%% file: 0_main.tex
\documentclass{nseJournal}

\usepackage{acronyms_vm}
\usepackage{recurring_equations}
\usepackage{packages_vm}
\usepackage{lineno}
\newcommand{\acov}{\eqt{\sigma_{\left(i,j\right),\left(k,l\right)}}}

\begin{document}

\title{Uncertainty Quantification for the Fission Matrix Method: A Rigorous Mathematical Framework and Computationally Efficient Alternatives} 

\addAuthor{\correspondingAuthor{Valerio Mascolino}}{a,b}
\correspondingEmail{valerio.mascolino@mines.edu}

\addAffiliation{a}{Nuclear Science and Engineering Program,\\Colorado School of Mines, 1500 Illinois St, Golden, CO 80401, USA}
\addAffiliation{b}{Mechanical Engineering Department,\\Colorado School of Mines, 1500 Illinois St, Golden, CO 80401, USA}

\addKeyword{Fission matrix}
\addKeyword{Uncertainty quantification}
\addKeyword{Delta method}
\addKeyword{JSI TRIGA}
\addKeyword{Monte Carlo}

\titlePage

\begin{abstract}
\input{1_abstract}
\end{abstract}

\input{2_body}

\pagebreak
\bibliographystyle{ans_js}                                                                           
\bibliography{references}

\end{document}

%% file: 1_abstract.tex
The fission matrix (FM) method recasts the neutron transport problem into matrix form, enabling fast, interpolation-based reactor calculations from a pre-computed database of Monte Carlo-derived coefficients. Despite its growing use, rigorous propagation of the underlying Monte Carlo statistical uncertainty through the FM method's eigenvalue and eigenvector has received comparatively little attention, with prior approaches typically assuming statistically independent coefficients without substantiating that assumption against the coefficients' actual covariance structure. This work derives a full-covariance Delta-method framework for propagating uncertainty to the FM eigenvalue, \keff, and fission source. The framework is evaluated, together with two lower-cost alternatives (a diagonal-covariance approximation and direct Monte Carlo resampling of the FM coefficients), against a detailed OpenMC model of the JSI TRIGA Mark-II reactor. For the all-rods-out reference configuration, all four uncertainty estimators (i.e., the rigorous full-covariance treatment, the diagonal approximation, resampling, and a fully independent empirical benchmark of 100 separate Monte Carlo realizations) agree to within 5.7\% for \keff\ and under 1\% for the fission source. The sign of the diagonal approximation's bias relative to the rigorous treatment flips between the two statistics levels, consistent with a change in the net correlation structure between FM coefficients, although it is not statistically resolved at the higher level and must be checked case by case. A global fixed-source FM generation strategy, evaluated directly, gives the same practical picture: the diagonal-covariance, resampling, and independent-run estimators agree within about 2\% and lie roughly 10\% below the full-covariance estimate, a gap attributed to sampling noise in the estimated covariance. In all cases, the FM coefficients' large off-diagonal covariance structure is close enough to symmetric to mostly cancel out when propagated to \keff\ and the fission source, rather than being small or systematically negative as previously assumed. Because interpolating and combining pre-computed database entries to represent configurations beyond those explicitly evaluated is a defining feature of practical FM-based tools, the same uncertainty estimators are also demonstrated for the FM method's control-rod interpolation and combination strategy (FM-CRd), validated here against an exact, fully-perturbed reference configuration, reproducing its \keff\ to within 60~pcm and its fission source shape to within 0.7\% relative error. This bias is about two standard deviations of the diagonal-covariance estimator's own propagated $\sigma_k$, a reasonable relative difference given that uncertainty. The diagonal-covariance and resampling uncertainty estimators continue to agree closely with each other for this case and remain conservative (by roughly a factor of two) with respect to the exact reference configuration's own statistical spread. Given this, together with the conservative direction of that gap and the prohibitive $\mathcal{O}(N_\text{cells}^4)$ memory cost of storing the full covariance matrix, the diagonal-covariance (i.e., statistically independent coefficients) Delta method is recommended as the practical default for propagating FM statistical uncertainty, with the full covariance calculation reserved for a one-time characterization of a given system and FM generation strategy.

%% file: 2_body.tex
\section{Introduction}
\label{sec:introduction}

Hybrid neutron transport methods offer an alternative to traditional approaches (deterministic, Monte Carlo) for the solution of the \gls{lbe} \cite{haghighatMonteCarloMethods2020}. Among these, the \gls{fm} method \cite{haghighatMonteCarloMethods2020,dufekFissionMatrixBased2009,carneyTheoryApplicationsFission2014} has received substantial attention in recent years due in part to the increased availability of computational resources, including the exponential increase in the number of CPUs and computer memory \cite{laureauTransientFissionMatrix2015,heCorrectionMethodRAPID2020,tophamIterativeFissionMatrix2020,rauStrategiesFastFission2021,pungercicVerificationNovelFuel2023,mascolinoNovelHybridDeterministic2024,mascolinoExperimentalValidation3D2024,dalingerAnalysisFissionMatrix2025,vitaCorrectionRatioApproach2026}. Modern implementations of the \gls{fm} method are based on the generation of large databases of \gls{fm} coefficients as a function of system-relevant parameters (e.g., control element positions and moderator temperature \cite{vitaCorrectionRatioApproach2026}). These coefficients can then be interpolated and combined using effective strategies to represent any combination of system parameters within the database validity range. For example, \gls{fm} coefficients can be generated for each control rod independently and then combined to account for the combined effect of all rods \cite{mascolinoValidationRAPIDsAlgorithm2021}. The \gls{fm} coefficients are accurately pre-calculated at reference configurations using any transport software. The generation of the coefficients using state-of-the-art Monte Carlo software (e.g., MCNP \cite{kuleszaMCNPCodeVersion2024}, Serpent \cite{leppanenStatusSerpentMonte2025}, OpenMC \cite{romanoOpenMCStateoftheartMonte2015}) is particularly effective, since the resulting coefficients are easy to calculate and to interpret physically \cite{mascolinoNovelHybridDeterministic2024}. Once a coefficients database has been pre-generated, the \gls{fm} method is capable of retaining the accuracy of the Monte Carlo method while achieving speedups of several orders of magnitude for any system configuration within the database validity range.

Given its accuracy and speed, the \gls{fm} method can play a prominent role in the analysis of next-generation nuclear reactors \cite{vitaFissionMatrixMethod2026a} to support design, optimization, and licensing. It lets reactor developers and deployers run hundreds of calculations in a short amount of time, for quick evaluation of design envelopes. The method has also found application beyond reactor cores, e.g., in spent-fuel storage and transport systems \cite{mascolinoEvaluationRAPIDUNF2017,mascolinoBenchmarkingRapidCode2018} and a subcritical multiplication pile \cite{roskoffExperimentalComputationalValidation2018}. Additionally, in its time-dependent formulation, the methodology could constitute the physics engine of a 3-D transport-based digital twin \cite{laureauTransientFissionMatrix2015,mascolinoValidationTransientFission2019}. Although the use of high-fidelity continuous-energy and -angle Monte Carlo methods for the pre-calculation of the coefficients reduces computational biases related to the discretization of the system, it also introduces statistical uncertainties into the method due to its stochastic nature. The calculation of uncertainties, however, has received comparatively little attention relative to the \gls{fm} method's other features and capabilities, despite evidence that Monte Carlo tallies can report deceptively low statistical uncertainties while still suffering from significant spatial undersampling \cite{strohDevelopmentPerformanceEvaluation2026}. A logical approach relies on sampling the coefficients of the matrix based on the statistical uncertainties of the coefficients, as computed by the Monte Carlo software during the pre-calculation phase \cite{mascolinoNovelHybridDeterministic2024,mascolinoDevelopmentBenchmarkingAdvanced2021}. However, this method is computationally very expensive, as it requires solving the \gls{fm} equations several times, one for each resampled set of \gls{fm} coefficients. Additionally, this method assumes fully statistically independent coefficients, an assumption believed to be conservative \cite{mascolinoExperimentalComputationalBenchmarking2019,mascolinoDevelopmentBenchmarkingAdvanced2021} on the reasoning that a shared, fixed pool of source particles forces an increase in one coefficient's tally to be matched, on average, by a decrease elsewhere, implying negative covariance between them. This reasoning is incomplete, however, since coefficients for either a single source region or the entire system are obtained from the same Monte Carlo calculation, which can also couple them positively. Additional work has analyzed an approach in which the coefficients of every calculation are assumed to be 100\% correlated, which makes for a more flexible approach to the uncertainty calculation without the need to perform multiple \gls{fm} calculations \cite{heDevelopmentUncertaintyAnalysis2022}.

To the best of the author's knowledge, none of these approaches has been demonstrated against a full propagation of uncertainties or has been substantiated by a rigorous calculation of statistical covariances between the coefficients. In this work, a full statistical covariance matrix and the correlation coefficients are characterized for a model of the JSI TRIGA Mark-II reactor \cite{snojHalfcenturyNuclearResearch2025}. Uncertainty for the criticality eigenvalue, \keff, is evaluated by propagating the covariance matrix using a linear perturbation approach based on the Delta method \cite{casellaStatisticalInference2024}. Prior \gls{uq} approaches for the \gls{fm} method, which ignore the computed covariances and rely only on the individual coefficient uncertainties reported by the Monte Carlo method, are compared against the Delta method. Since many \gls{fm} implementations generate the coefficients from fixed-source instead of criticality calculations \cite{mascolinoNovelHybridDeterministic2024,heDevelopmentUncertaintyAnalysis2022,balouzaComparingFissionMatrix2026}, the comparison is repeated for a global fixed-source generation strategy.

This article should serve as a helpful reference for nuclear scientists and engineers developing \gls{fm} algorithms to incorporate efficient yet reasonable \gls{uq} strategies into their analyses. First, an overview of the \gls{fm} method, as relevant to this work, is provided in Section \ref{sec:fm-method}. The computational model based on the JSI TRIGA Mark-II reactor is described in Section \ref{sec:model}, including the OpenMC \cite{romanoOpenMCStateoftheartMonte2015} model used for calculating the \gls{fm} coefficients and reference results, and the \gls{fm} spatial discretization and calculation process. The \gls{uq} analysis methodology is discussed in Section \ref{sec:methodology}, including the Delta method and simplified \gls{uq} techniques for the \gls{fm} method, as well as their extension to the interpolation and combination of \gls{fm} database entries via the \gls{fm}-\gls{crd} method. Section \ref{sec:results} presents and discusses the results, including covariances, correlation coefficients, and calculated uncertainties for the \gls{fm} method using the proposed methodologies. Section \ref{sec:conclusions} draws conclusions and provides recommendations for implementing \gls{uq} into \gls{fm} tools.

\section{The Fission Matrix method}
\label{sec:fm-method}
The \gls{fm} method consists of recasting the \gls{lbe} for neutrons into matrix form. The \gls{lbe} in its criticality form is given by \cite{bellNuclearReactorTheory1970}:
\begin{equation}
    \label{eq:lbe}
    \begin{aligned}
        \Sigma_t\psip
        & +\hat{\Omega}\cdot\nabla\psip
        = \\
         \quad & \displaystyle
        \int_{4\pi}d\Omega'\int_{0}^{+\infty}dE'\ \Sigma_s\left(\vec{r},E'\rightarrow E,\hat{\Omega}'\rightarrow \hat{\Omega}\right)\psipp\\
        & +\dfrac{1}{\keff}\dfrac{\chi(E)}{4\pi}\int_{4\pi}d\Omega'\int_{0}^{+\infty}dE'\ \nu\Sigma_f\left(\vec{r},E'\right)\psipp
    \end{aligned}
\end{equation}

The \gls{lbe} can be inverted and expressed in operator form as follows:
\begin{equation}
    \label{eq:lbe-operator-form}
    \hat{\mathcal{F}}\psip=\displaystyle\dfrac{1}{\keff}\dfrac{\chi(E)}{4\pi}\hat{\mathcal{F}}\hat{\mathcal{L}}^{-1}\hat{\mathcal{F}}\psip
\end{equation}
where the fission and transport operators, $\hat{\mathcal{F}}$ and $\hat{\mathcal{L}}$, can be expressed as:

\begin{equation}
    \label{eq:operators}
    \begin{cases}        \hat{\mathcal{F}}=\displaystyle\int_{4\pi}d\Omega'\int_0^{+\infty}dE'\nu\Sigma_f\left(\vec{r},E',\hat{\Omega}'\right)\\
        \hat{\mathcal{L}}=\displaystyle\Sigma_t+\hat{\Omega}\cdot\nabla-\int_{4\pi}d\Omega'\int_0^{+\infty}dE'\Sigma_s \left(\vec{r},E'\to E,\hat{\Omega}'\to\hat{\Omega}\right)
    \end{cases}
\end{equation}

Note that the $\hat{\mathcal{F}}\psip$ term that appears both in the left- and right-hand sides of the \gls{lbe} in operator form (Equation \ref{eq:lbe-operator-form}) yields the fission neutron source rate density, in units of $\left(\dfrac{\text{neutrons}}{\text{cm}^3\cdot\text{s}}\right)$.

Integrating over space and angle and discretizing the problem geometry into \nc fissionable regions yields the criticality form of the \gls{fm} equation:

\fmequation

In Equation \ref{eq:fmequation}, \fsv is referred to as the discretized fission source and each of its elements, $S_i$, yields the fission neutron source rate integrated over fissionable region $i$. \fma is the matrix form of the discretized operator that appears on the right-hand side of Equation \ref{eq:lbe-operator-form}. \matb{A} is referred to as the \textit{fission matrix} (\gls{fm}). Although the continuous form of the operator is mathematically intractable, since it includes the inverse of the transport operator, $\hat{\mathcal{L}}^{-1}$, the discretized form of the operator is actually very easy to compute using any transport software. In fact, from Equation \ref{eq:fmequation}, the fission matrix \matb{A} acts as a coupling operator for the fission source in every fissionable region. Specifically, each \gls{fm} coefficient, \aij, represents the number of neutrons born in region $i$ from fission events induced by a source neutron born in region $j$. Readers interested in the full formulation of the \gls{fm} method can refer to the plentiful scientific literature on the subject \cite{haghighatMonteCarloMethods2020,dufekFissionMatrixBased2009,carneyTheoryApplicationsFission2014,mascolinoNovelHybridDeterministic2024,mascolinoDevelopmentValidationNew2022,waltersRAPIDFissionMatrix2018}.

\begin{figure}[htbp!]
    \centering
    \includegraphics[width=.995\linewidth]{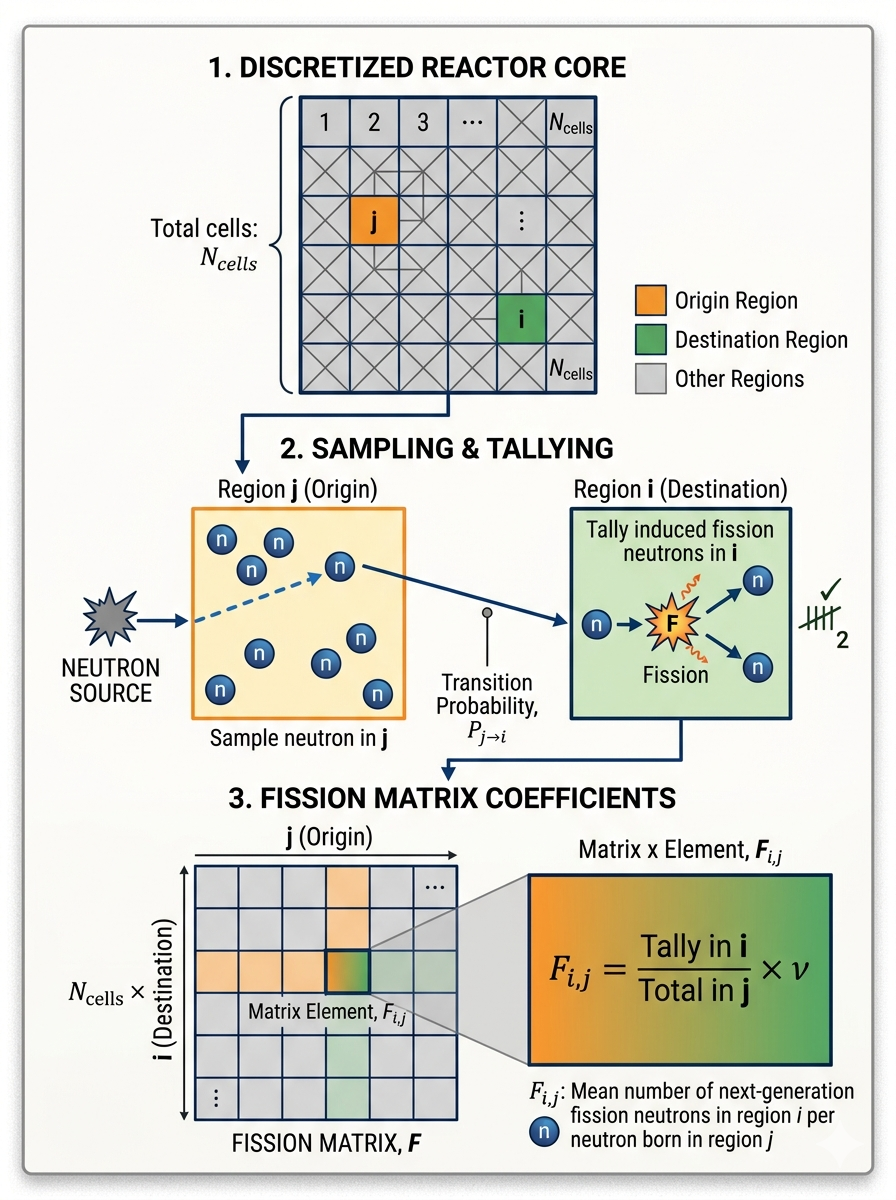}
    \caption{Schematic of the FM coefficients generation process.}
    \label{fig:fm-schematic}
\end{figure}

Although the generation of the \aij coefficients could easily be achieved using deterministic software, the most natural and accurate way to generate them is by means of a \acrfull{mc} software. In this work, the continuous-energy open-source software OpenMC \cite{romanoOpenMCMonteCarlo2013}, maintained by Argonne National Laboratory, was used to generate all of the \gls{fm} coefficients. The procedure to generate the coefficients involves dividing the model into \nc regions, tagging the source neutrons in OpenMC based on the region they were born in, $j$, and tallying the induced fission neutrons separately for every destination region, $i$. To properly normalize the \gls{fm} coefficients, a combination of OpenMC's filters and post-processing operations is required at this stage. Native support for this normalization within OpenMC itself is underway under a separate scope. A schematic for the generation of the \gls{fm} is shown in Figure \ref{fig:fm-schematic}.

\subsection{The Fission Matrix - Control Rod `deltas' method}
\label{sec:fm-crd}
One of the major strengths of the \gls{fm} method is its multi-stage nature \cite{haghighatMRTMethodologiesRealtime2016}. The coefficients can be pre-calculated uniformly for a set of system-relevant parameters (e.g., control rod positions) to generate a database. Then, for any user-input configuration within the database validity range, the coefficients can be interpolated and combined in seconds-to-minutes to solve the transport problem with accuracies that are comparable with the method used to generate the coefficients in the first place, e.g., continuous energy \gls{mc} calculations.

The \gls{fm}-\gls{crd} is a methodology for accounting for the effect of control rod insertion on the \gls{fm} coefficients, able to decouple the effect of a single control rod (or individual clusters moved together) and then obtain any combination of multiple control rods via interpolating and combining the limited \gls{fm} coefficient entries in the database. The method has been validated for the \jst reactor using both synthetic \cite{mascolinoDevelopmentValidationNew2022} and experimental \cite{mascolinoExperimentalValidation3D2024,mascolinoVerificationValidationRAPID2021} data, showing practically no loss in accuracy stemming from the interpolation and combination strategies.

If each of the four \glspl{cr} of the \jst reactor is identified by index $c$, then the impact of their insertion on the \gls{fm} coefficients can be calculated by the so-called \acrfull{crd}, as follows:

\begin{equation}
    \label{eq:crd}
    \matb{\text{CR}_{d,c}}(z_c)=\matb{A}^*_c(z_c)-\matb{A}_\text{ARO},\qquad c\in\left[1,N_\text{CR}\right]
\end{equation}

For each control rod, $c$, inserted at location $z_c$ (in steps), the \gls{crd} are calculated as the difference between the reference \gls{aro} \gls{fm} and the \gls{fm} evaluated with \textit{only} control rod $c$ inserted at location $z_c$. The \gls{cr} location is a continuous user input; however, the specific \glspl{fm} are only calculated at discrete control rod insertions, $z_{c,k}$ with $k\in\left[N_{z,\text{CR}}\right]$. To calculate $\matb{A}^*(z_c)$, the \gls{cr}-specific \gls{fm} database entries are first linearly interpolated independently, as follows:

\begin{equation}
    \matb{A}^*(z_c)=\dfrac{z_c-z_{c,k-1}}{z_{c,k}-z_{c,k-1}}\matb{A}(z_{c,k})+\dfrac{z_{c,k}-z_c}{z_{c,k}-z_{c,k-1}}\matb{A}(z_{c,k-1}),\quad
    \begin{cases}
        c\in\left[1,N_\text{CR}\right] \\
        k\in\left[1,N_{z,\text{CR}}\right]
    \end{cases}
\end{equation}

Finally, all the independently calculated $\text{CR}_{d,c}(z_c)$ can be combined to obtain the system-specific \gls{fm} for the user-specified case, $\matb{A}_\text{case}$:

\begin{equation}
    \matb{A}_\text{case}\left(z_1,z_2,..,z_{N_\text{CR}}\right)=\matb{A}_\text{ARO}+\displaystyle\sum_{c=1}^{N_\text{CR}}\text{CR}_{d,c}(z_c)
\end{equation}

The case-specific \gls{fm}, $\matb{A}_\text{case}$, is then solved for the criticality eigenvalue, \keff, and the fission source, \fsv, using Equation \ref{eq:fmequation}. Interpolating and combining separately pre-computed database entries in this way, rather than evaluating a fresh \gls{fm} for every configuration of interest, is what allows a single \gls{fm} database to cover a continuum of system states. This same strategy underlies most practical \gls{fm}-based tools and is envisioned for the method's kinetics extension via the \gls{tfm} \cite{laureauTransientFissionMatrix2015,mascolinoValidationTransientFission2019,mascolinoNovelHybridDeterministic2024}, to enable transport-based digital twins. As such, the \gls{fm}-\gls{crd} approach constitutes a natural test case for extending the \gls{uq} strategies explored in this work from a single, directly evaluated \gls{fm} to one assembled via database interpolation and combination.

\begin{figure}[htbp!]
    \centering
    \includegraphics[width=\linewidth]{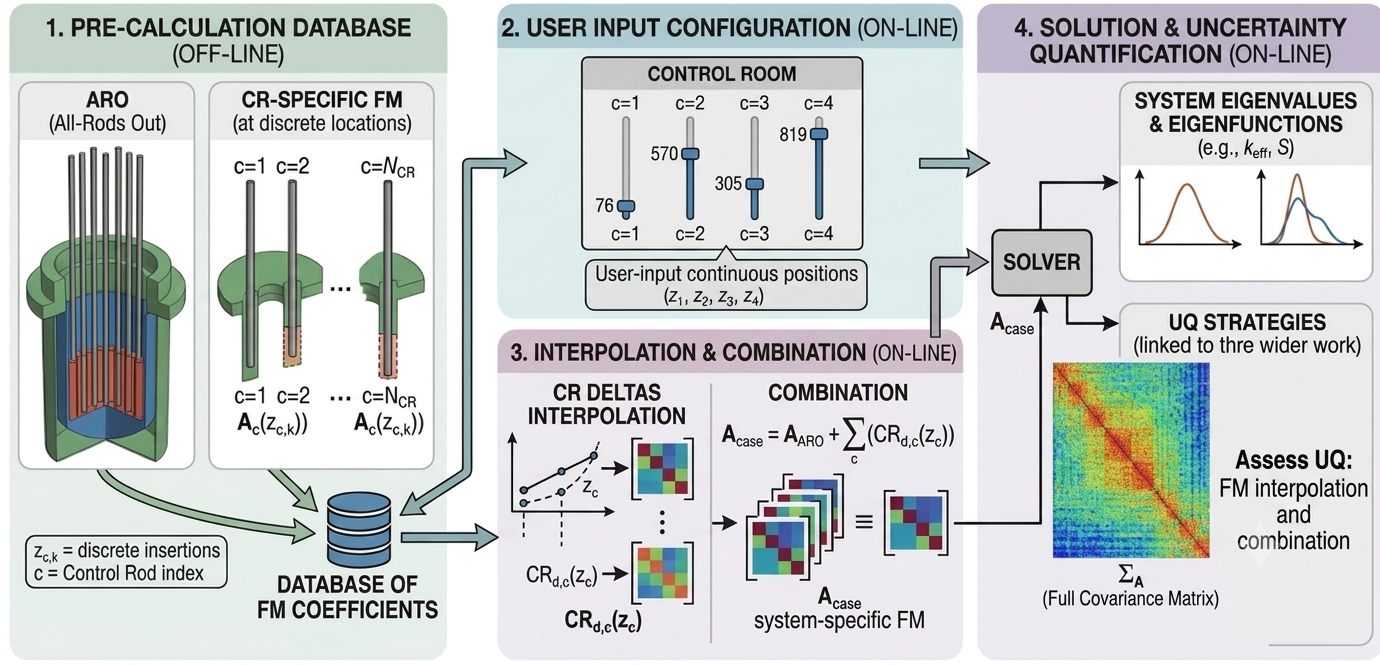}
    \caption{Workflow schematic for the FM-CRd method.}
    \label{fig:fm-crd}
\end{figure}

A workflow schematic of the \gls{fm}-\gls{crd} process is provided in Figure \ref{fig:fm-crd}. Additional detail about the method can be found in literature \cite{mascolinoExperimentalValidation3D2024,mascolinoDevelopmentValidationNew2022}.

\section{The JSI TRIGA Mark-II Reactor Computational Model}
\label{sec:model}
This section describes the computational model used to evaluate the various \gls{uq} methods for the fission matrix method: a recently validated OpenMC model of the JSI TRIGA Mark-II reactor (pending publication). Various runs were performed using core configuration 133, which contains all fresh fuel elements from the reactor's 1991 reconstruction \cite{jerajTRIGAMarkII1999}. This configuration was chosen to keep burnup-related effects, addressed separately by a dedicated \gls{fm}-based burnup methodology \cite{pungercicVerificationNovelFuel2023}, from affecting this evaluation. Several different sets of \gls{fm} coefficients were calculated using this model to directly compare the various \gls{uq} approaches. To understand the impact of \gls{fm} interpolation and combination on calculated uncertainties, the control rod positions were selected as the main dependency for the \gls{fm} coefficients. This reflects the \gls{fm}-\gls{crd} method's \cite{mascolinoDevelopmentValidationNew2022} extensive reliance on \gls{fm} interpolation strategies, as discussed in Section \ref{sec:fm-crd}. 

\begin{figure}[htbp!]
    \centering
    \includegraphics[width=1.0\linewidth]{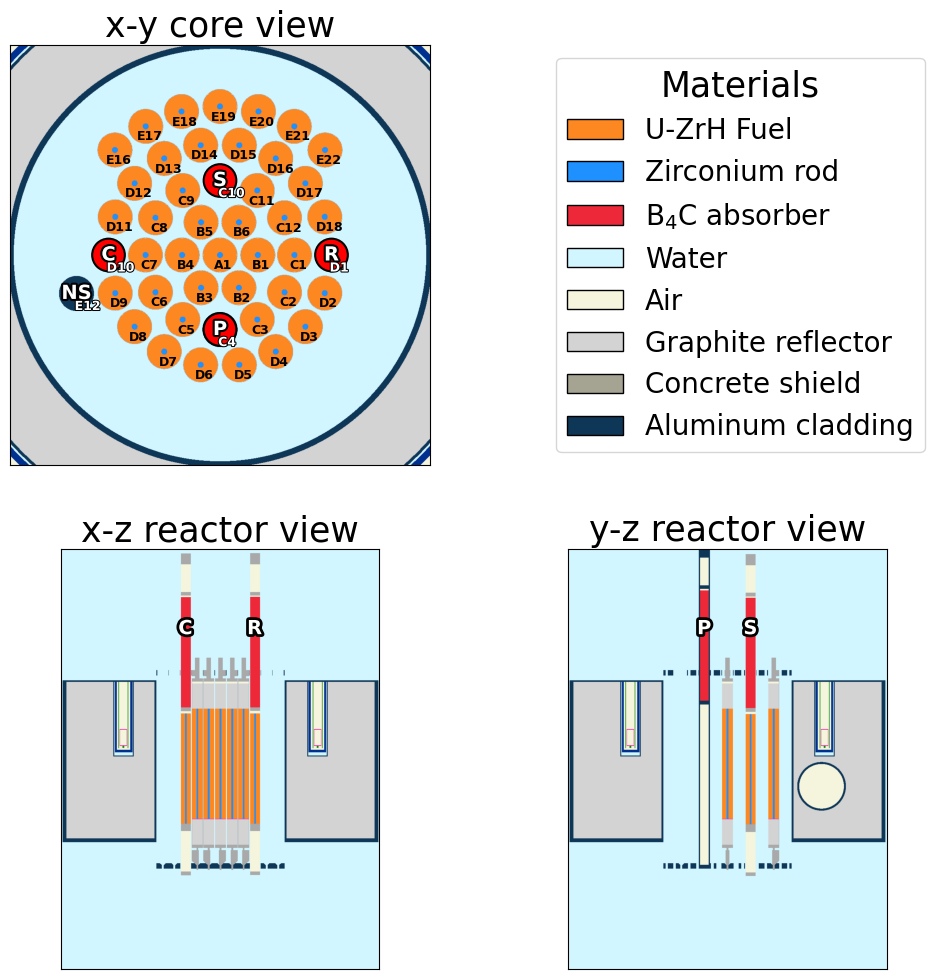}
    \caption{JSI TRIGA Mark-II reactor OpenMC model. The four control rods are identified in the various views according to their codes included in Table \ref{tab:cr_info}, and they are in the all-out position. Fuel rods in the x-y view show their location within the core lattice.}
    \label{fig:core}
\end{figure}

The \jst reactor is a 250-kW pool-type reactor located in Ljubljana, Slovenia. The reactor core features a circular lattice with 91 total locations in six concentric rings (A through F). Except for the central ring, A, which has only one core location, the other rings have positions in increasing multiples of 6 (e.g., 12 locations in ring C). Four control rods are used for reactivity control in the reactor. These control rods have different names and locations in the reactor core, summarized in Table \ref{tab:cr_info}. One of them, the transient (or pulse) rod, has an air follower and is used for pulse experiments. The other three have fuel followers that are almost identical to the active region of regular fuel rods. As such, when the fuel-followed control rods are extracted, they are replaced by fuel. Core configuration 133 also includes a neutron source element in the E-ring, used for reactor startup. In the OpenMC model, the neutron source geometry is modeled; however, no neutrons are sampled from it. The geometry of the \jst OpenMC model for core configuration 133 is shown in Figure \ref{fig:core}.

\newcolumntype{C}[1]{>{\centering\arraybackslash}p{#1}}
\begin{table}[htbp!]
    \centering
    \setlength{\tabcolsep}{4pt}
    \begin{tabular}{ l c c c c c}
        \toprule
         \multirow{2}{*}{\textbf{Name}} & \multirow{2}{*}{\textbf{Follower}} & \textbf{Core} & \multirow{2}{*}{\textbf{Codes}} & \textbf{Out pos.} & \textbf{In pos.} \\
          & & \textbf{Loc.} & & \textbf{(steps)} & \textbf{(steps)} \\
        \midrule
         Pulse/Transient & Air & C-04 & TRAN (P) & 0 & 900 \\
         Safety & Fuel & C-10 & SAFE (S) & 200 & 900 \\
         Regulating & Fuel & D-01 & REGU (R) & 181 & 900 \\
         Compensating/Shim & Fuel & D-10 & SHIM (C) & 200 & 900 \\
         \bottomrule
    \end{tabular}
    \caption{\jst control rods information. The all-out and all-in positions are provided in steps, corresponding to control room readings. The one-letter codes are consistent with Figure \ref{fig:core}.}
    \label{tab:cr_info}
\end{table}

\section{Methodology}
\label{sec:methodology}

To evaluate the various \gls{uq} methods, a rigorous mathematical framework was first established and used as the reference. The reference is based on the use of the Delta method \cite{casellaStatisticalInference2024}, a first-order Taylor expansion approach that can be used to calculate the propagation of variances and covariances to a matrix's eigenvalue and eigenvector. To use the method, covariances need to be calculated for the \gls{fm} coefficients. Several different sets of \gls{fm} coefficients were calculated using different configurations and statistical parameters. The specifics of the \gls{fm} calculation models, based on the \jst, are discussed in Section \ref{sec:fm-models}. An introduction to the Delta method and its application to the \gls{fm} method as well as the lower-cost \gls{uq} calculation alternatives are summarized in Section \ref{sec:uq-appr}, and the methodology for their evaluation against the Delta method is discussed.

\subsection{Fission Matrix Models}
\label{sec:fm-models}

\newcommand{\nz}{$N_\text{ax}$\xspace}

The chosen \gls{fm} discretization divides the \jst reactor into axial segments of each fuel rod. Core 133 has a total of 43 fuel rods, including the 3 control rod fueled followers. Two different discretizations are used for the number of axial segments, $N_\text{ax}$: 5 and 15. Different combinations of $N_\text{ax}$ and Monte Carlo criticality parameters (number of particles per batch, $\text{NPS}$; number of active cycles, $\text{NAC}$; and number of skipped cycles, $\text{NSK}$) were used to analyze the various \gls{uq} approaches. Specifically, Section \ref{sec:fm-modeling-uq} describes the models and assumptions used for the comparison of the Delta method with simplified \gls{uq} approaches, whereas Section \ref{sec:fm-modeling-crd} describes those used for the propagation of uncertainty using only the simplified approaches via the \gls{fm}-\gls{crd} method for control rod insertion, and Section \ref{sec:fm-modeling-fixed} describes the global fixed-source generation strategy used to test whether the covariance structure depends on how the coefficients are generated. A summary of all the \gls{fm} models used for this work is included in Section \ref{sec:fm-modeling-summary}.

\subsubsection{FM modeling for UQ approaches comparison}
\label{sec:fm-modeling-uq}
Previous \gls{fm} work related to the \jst reactor has successfully used $N_\text{ax}=15$ as a suitable division to obtain high-accuracy results \cite{mascolinoNovelHybridDeterministic2024,mascolinoExperimentalValidation3D2024,mascolinoDevelopmentValidationNew2022,mascolinoVerificationValidationRAPID2021,pungercicApplicationBRAPIDFission2022}. However, this posed a substantial limitation on the analysis planned for this work. The evaluation of sample covariances was needed to establish a rigorous \gls{uq} mathematical framework via the Delta method. However, since each \gls{fm} coefficient, \aij, can be correlated with any other value, $a_{k,l}$, the covariance matrix scales with $\nc^4$. Setting $N_\text{ax}=15$ gives $\nc=645$. Storing a covariance matrix of size $\nc^4=645^4\approx1.73\times10^{11}$ with double-precision floating points would have an approximate memory requirement of $\sim1.38$ TB, without considering any additional array for intermediate calculations or results. 

\begin{figure}[H]
    \centering
    \includegraphics[width=\linewidth]{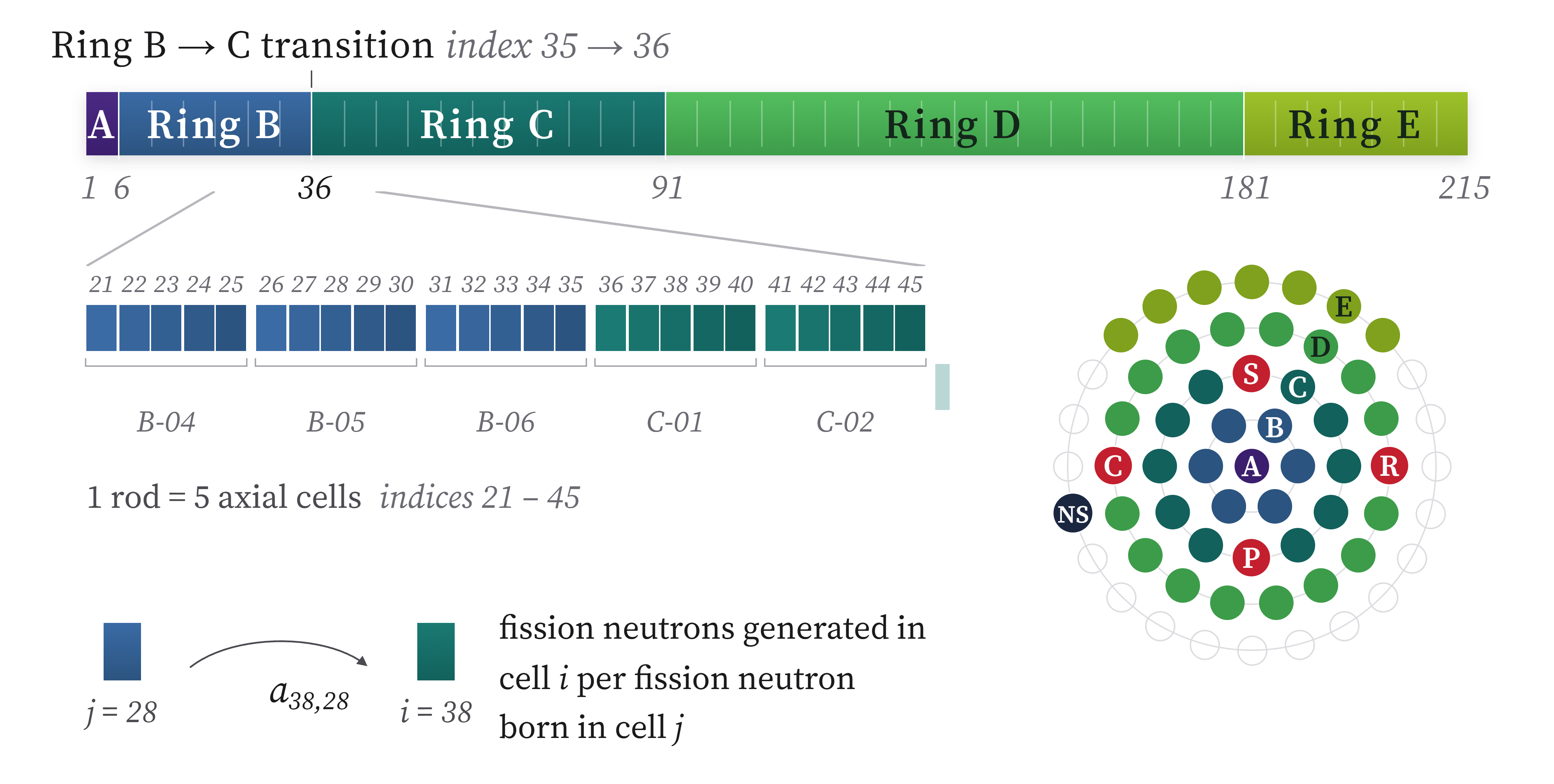}
    \caption{FM cells order in the $N_\text{ax}=5$, $N_\text{cells}=215$ TRIGA model.}
    \label{fig:fm-index-order}
\end{figure}

Moving from $N_\text{ax}=15$ to $N_\text{ax}=5$ reduces the size of the matrix by a factor $3^4=81$, resulting in a much more manageable covariance matrix requiring only 17.09 GB of RAM. This illustrates why explicitly storing the full covariance matrix of the \gls{fm} coefficients is impractical for any real-world application in which high-resolution 3-D distributions are needed. Hence, alternative and less resource-intensive \gls{uq} approaches are explored and evaluated, and their calculated uncertainty biases with respect to the rigorous mathematical model are assessed. The $N_\text{ax}=5$ fission matrix division and the order of the fission matrix cells is exemplified in Figure \ref{fig:fm-index-order}. Since the fission matrix coefficients of the CovHigh, CovLow, Indp, and CRd cases are obtained from criticality \gls{mc} calculations, the \glspl{fm} are generated using a converged fission source distribution within each fission region $i$. As such, even if some spatial resolution is lost in the process, the integrated values have the same level of accuracy. 

The covariance analysis was established on a highly statistically converged model (CovHigh) and then repeated for a case with significantly lower statistics (CovLow) to establish whether the simplified \gls{uq} methodologies remain valid irrespective of the statistical precision of the results. To do so, the calculations were repeated with a significantly smaller number of particles per each neutron generation in the OpenMC criticality calculations. This also enabled performing entirely independent \gls{mc} calculations to obtain truly independent coefficients and evaluate their calculated uncertainties against the Delta method.

\subsubsection{Control rods FM modeling using the CRd approach}
\label{sec:fm-modeling-crd}
As already noted in Section \ref{sec:fm-crd}, the \gls{fm}-\gls{crd} control-rod strategy is a concrete instance of the database interpolation and combination that underlies most practical \gls{fm} tools; this subsection therefore extends the \gls{uq} methodology compared in Section \ref{sec:fm-modeling-uq} to that interpolated-and-combined case. To carry the uncertainties through the \gls{fm} interpolation and combination strategies as part of the \gls{fm}-\gls{crd} method, the number of axial levels each fuel rod is divided into becomes important. In fact, the coarser the division, the higher the potential bias inserted by the \gls{fm}-\gls{crd} methodology \cite{mascolinoDevelopmentValidationNew2022}. As such, a second set of calculations was carried out with the previously demonstrated $N_\text{ax}=15$. Given the larger size of the database and the prohibitive memory requirements of the explicit covariance matrix, covariances were assumed to be 0 for all the \gls{fm}-\gls{crd} analyses (full statistical independence of the coefficients) and the alternative \gls{uq} strategies were used instead.

Equation \ref{eq:crd} and the combination above (Section \ref{sec:fm-crd}) can be expanded explicitly in terms of the nine independently-calculated \gls{fm} entering it: the shared $\matb{A}_\text{ARO}$ realization (subtracted once per rod but appearing only once as a random variable), and the two \gls{cr}-specific database entries bracketing each rod's test position. This gives:
\begin{equation}
    \label{eq:crd-expanded}
    \matb{A}_\text{CRd} = \left(1-N_\text{CR}\right)\matb{A}_\text{ARO} + \sum_{c=1}^{N_\text{CR}}\left[\dfrac{z_c-z_{c,k-1}}{z_{c,k}-z_{c,k-1}}\matb{A}(z_{c,k})+\dfrac{z_{c,k}-z_c}{z_{c,k}-z_{c,k-1}}\matb{A}(z_{c,k-1})\right]
\end{equation}
Since each of these nine \gls{fm} is obtained from a separate, independent OpenMC calculation, and each is itself treated under the diagonal-covariance (\textit{Independent}) assumption of Section \ref{sec:simplified_methods}, the coefficient-wise variance of $\matb{A}_\text{CRd}$ follows directly from Equation \ref{eq:crd-expanded} by squaring each linear coefficient:
\begin{equation}
    \label{eq:crd-cov-diag}
    \begin{aligned}
        \sigma^2_{i,j}\left(\matb{A}_\text{CRd}\right) = {} & \left(1-N_\text{CR}\right)^2\sigma^2_{i,j}\left(\text{ARO}\right) \\
        & + \sum_{c=1}^{N_\text{CR}}\left(\dfrac{z_c-z_{c,k-1}}{z_{c,k}-z_{c,k-1}}\right)^2\sigma^2_{i,j}\left(z_{c,k}\right) \\
        & + \sum_{c=1}^{N_\text{CR}}\left(\dfrac{z_{c,k}-z_c}{z_{c,k}-z_{c,k-1}}\right)^2\sigma^2_{i,j}\left(z_{c,k-1}\right)
    \end{aligned}
\end{equation}
where $\sigma^2_{i,j}(\text{ARO})$ and $\sigma^2_{i,j}(z_{c,k})$ denote the diagonal coefficient variance of the respective base \gls{fm}, as reported by OpenMC for that calculation. 

For the FM-CRd analysis, a random control rod configuration has been selected for testing so that the propagation of the uncertainties through the \gls{fm}-\gls{crd} interpolation approach discussed in Section \ref{sec:fm-crd} could be tested in limiting conditions. Additionally, the methodology requires selecting the number of discrete insertions of the control rods at which the \glspl{fm} are evaluated to build the \gls{fm}-\gls{crd} database. Based on previous JSI TRIGA analyses, this value was selected as $N_{z,\text{CR}}=16$. To test the \gls{fm}-\gls{crd} strategy at its fullest, a case was devised for which none of the control rods was exactly at the location at which the database was evaluated. The selected configuration is summarized in Table \ref{tab:cr_pos}.

\begin{table}[H]
    \centering
    \begin{tabular}{c p{220pt} c}
        \toprule
        \textbf{CR} & \textbf{FM database entries locations ($N_{z,\text{CR}}=16$)} & \textbf{Test location} \\
        \midrule
        P   &   [0, \textbf{56}, \textbf{112}, 168, 225, 281, 337, 393, 450, 506, 562, 618, 675, 731, 787, 843, 900]    & 76\\
        \midrule
        S   &   [200, 243, 287, 331, 375, 418, 462, 506, \textbf{550}, \textbf{593}, 637, 681, 725, 768, 812, 856, 900] & 570 \\
        \midrule
        R   &   [181, 225, \textbf{270}, \textbf{315}, 360, 405, 450, 495, 540, 585, 630, 675, 720, 765, 810, 855, 900] & 305 \\
        \midrule
        C   &   [200, 243, 287, 331, 375, 418, 462, 506, 550, 593, 637, 681, 725, 768, \textbf{812}, \textbf{856}, 900] & 819 \\
        \bottomrule
    \end{tabular}
    \caption{Control rod positions in the \jst FM-CRd database. The FM method uncertainties are calculated at the configuration identified by the test location. The database entries involved for the interpolation of the matrix using the FM-CRd method are shown in boldface.}
    \label{tab:cr_pos}
\end{table}

The table shows the positions at which the \gls{fm} matrices are evaluated based on a CR insertion performed in $N_{z,\text{CR}}=16$ steps, as well as the location of the test case used to perform the \gls{uq} analysis. Effectively, only the runs that are identified in boldface needed to actually be performed to calculate the \gls{fm} coefficients needed for the test case, together with the \gls{aro} baseline configuration required to evaluate $\matb{A}_\text{ARO}$ in Equation \ref{eq:crd}, for a total of nine \gls{fm} calculations. Additionally, the combined case was also evaluated explicitly to test the effectiveness of the FM-CRd method and compare uncertainties with and without the interpolation and combination of the coefficients necessary for the \gls{fm}-\gls{crd} method, resulting in a total of ten \gls{fm} evaluations needed for this analysis.

\subsubsection{Global fixed-source FM calculations}
\label{sec:fm-modeling-fixed}
The FixedSrc case tests whether the \gls{fm} covariance structure depends on how the coefficients are generated. In the criticality-based cases above, the coefficients come from a self-consistent fission source that is resampled from generation to generation. Here, they come instead from a fixed-source calculation. The same \gls{aro} configuration and $N_\text{ax}=5$ discretization ($\nc=215$) as CovHigh and CovLow are used, which keeps $\bm{\Sigma}_A$ tractable (Section \ref{sec:fm-modeling-uq}).

All $\nc$ fuel cells are active simultaneously in a single OpenMC fixed-source run. Each cell hosts its own independent source of unit strength. The spatial distribution of the source is assumed to be uniform over the fuel cell, whereas the energy is sampled from the Watt fission spectrum for U-235. The emission is isotropic. This is analogous to many applications of the \gls{fm} method in which the coefficients of a given source region, $j$, are generated via independent fixed-source calculations \cite{haghighatMonteCarloMethods2020}. Here, all source regions are simply run together. Since every particle is assigned to one of the \nc\ sources by an independent random draw, each cell receives $\text{NPS}/\nc\approx4{,}651$ source particles per batch on average, and the actual number fluctuates from batch to batch. Secondary fission neutrons are not transported, so each tally scores only the neutrons born from induced fissions caused directly by the source neutrons. As discussed in Section \ref{sec:covariances}, the raw tally is then already a valid \gls{fm} realization, and no renormalization is applied.

The production run uses $\text{NPS}=10^6$ particles per batch and 215 batches, all of them active (Table \ref{tab:fm-calc-summary}). No inactive batches are needed since there is no fission source to converge, and no fission source is carried over between batches. A statepoint is saved after every batch. As in Section \ref{sec:covariances}, forward-differencing the statepoints yields $M=215$ realizations of the raw tally, which are used for the \textit{Covariances}, \textit{Independent}, and \textit{Resample} estimators of Section \ref{sec:estimators-compared}.

The \textit{Runs} estimator uses 100 additional, fully independent replicate runs of the same configuration, each with a different random seed. Each replicate consists of two batches of $1.075\times10^6$ particles and saves only its final statepoint. Its $2.15\times10^6$ particles are exactly $1/100$ of the production run's $2.15\times10^8$, so the 100 replicates together carry the same particle budget as the production run.

\subsubsection{Summary of FM models and calculations}
\label{sec:fm-modeling-summary}
\begin{figure}[htbp]
    \centering
    \includegraphics[width=\linewidth]{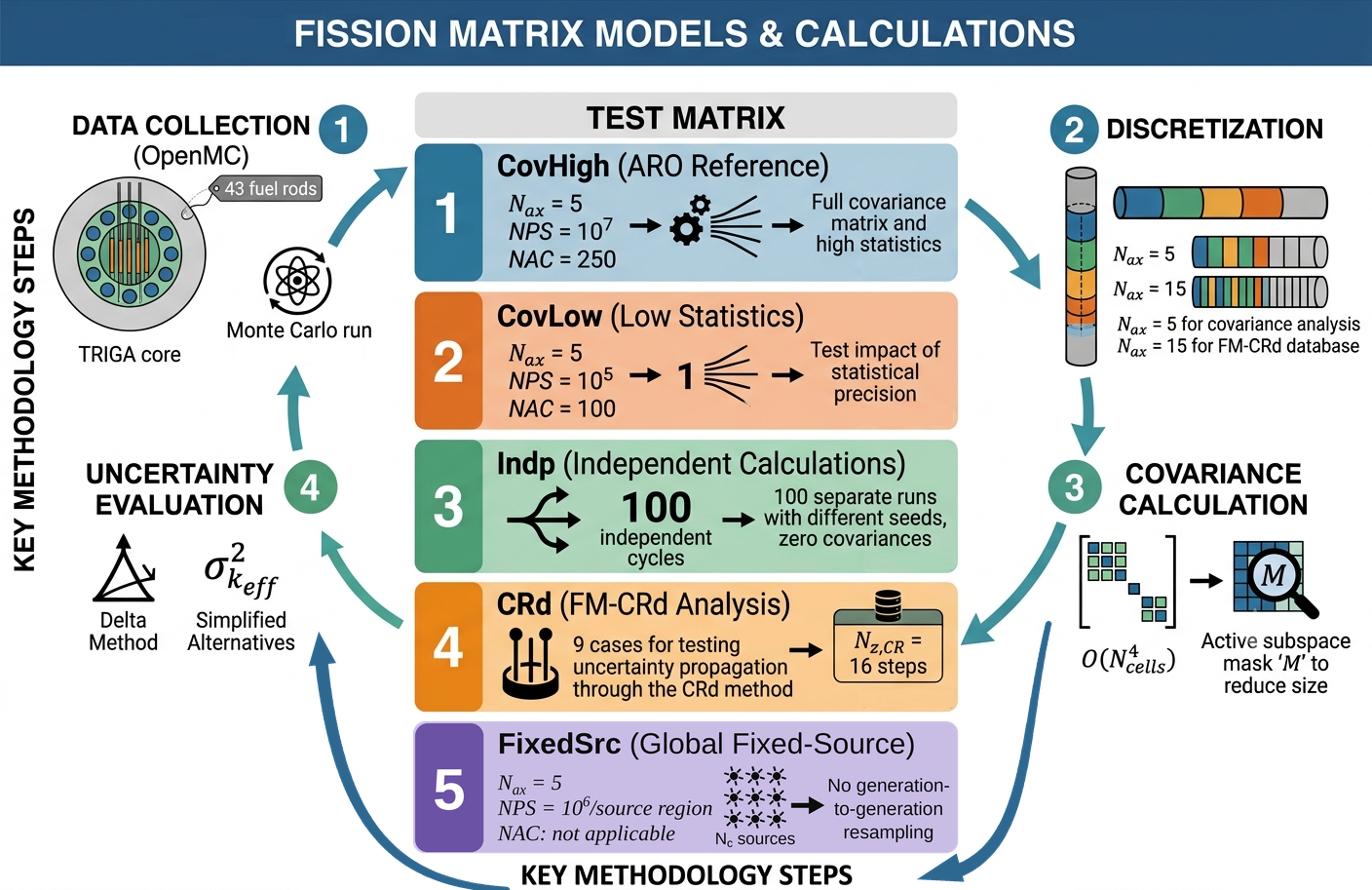}
    \caption{Summary of \gls{fm} calculations performed for this work.}
    \label{fig:fm-calcs}
\end{figure}

The fission matrix models  run for this work are summarized in the infographic in Figure \ref{fig:fm-calcs}. The main features of each of the different models utilized are summarized in Table \ref{tab:fm-calc-summary}.

\begin{table}[htbp]
    \centering
    \begingroup
    \renewcommand{\baselinestretch}{1}
    \setlength{\tabcolsep}{4pt}
    \small
    \begin{tabular}{c c c c c c p{170pt}}
        \toprule
        \textbf{ID} & \textbf{Name} & $N_\text{ax}$ & \textbf{NPS} & \textbf{NAC} & $z_{\text{CR}_c}$ & \textbf{Description} \\
        \midrule
        1   & CovHigh & 5     & $10^7$ & 250  & ARO       &
            Reference case for which the full covariance matrix is calculated with high statistics. The rigorous Delta method and simplified \gls{uq} strategies are compared against the same set of data.\\
        \midrule
        2   & CovLow  & 5     & $10^5$ & 100  & ARO       &
            Analogous to case 1, but the NPS is reduced by a factor of 100 and NAC is brought down to 100 to test the impact of statistical precision on the simplified \gls{uq} methodologies and how they compare to the Delta method.\\
        \midrule
        3   & Indp    & 5     & $10^5$ & 1    & ARO       &
            100 different calculations were run with different seeds. Each realization of \aij is completely independent, ensuring zero covariances. This case is compared to case 2, since all the other parameters are identical. \\
        \midrule
        4   & CRd     & 15    & $10^6$ & 250  & Various   &9 cases (see Table \ref{tab:cr_pos}) are run to test the propagation of uncertainty through the \gls{fm}-\gls{crd} method using simplified \gls{uq} methods. 100 additional replicates of the exact, fully-perturbed configuration (one active batch each) provide the \textit{Runs} benchmark. \\
        \midrule
        5   & FixedSrc & 5    & $10^6$ & 215  & ARO       &
            Global fixed-source calculation: all \nc\ fuel cells are simultaneously active, independent, equal-strength sources, rather than one self-consistent source resampled generation-to-generation. This production run provides the covariances. 100 additional replicates (2 batches of $1.075\times10^6$ particles each) provide the \textit{Runs} benchmark. \\
        \bottomrule
    \end{tabular}
    \endgroup
    \caption{Test matrix for the \gls{fm}-\gls{uq} analysis.}
    \label{tab:fm-calc-summary}
\end{table}

\subsection{Uncertainty Quantification Approaches for the Fission Matrix Method}
\label{sec:uq-appr}
This work establishes and compares different methods for \gls{uq} of the \gls{fm} method. The rigorous Delta method approach based on first-order Taylor expansion and linear perturbation is discussed in Section \ref{sec:delta}. The method requires the evaluation of \gls{fm} covariances, not readily available from the output of any Monte Carlo software. The approach used for evaluating the \gls{fm} covariances is described in Section \ref{sec:covariances}. Finally, the Delta method is quantitatively compared to simplified \gls{uq} approaches historically used for the \gls{fm} method, discussed in Section \ref{sec:simplified_methods}.

\subsubsection{The Delta Method formulation}
\label{sec:delta}
To establish a rigorous mathematical baseline for the uncertainty, a first-order multivariate Taylor series expansion (the Delta Method) is formulated \cite{casellaStatisticalInference2024}. To map the covariances of the \gls{fm} coefficients to the fundamental eigenvalue $k_{\text{eff}}$, an unperturbed eigen-decomposition is executed on the expected value of the fission matrix, $\mathbf{\bar{A}}$:
\begin{equation}
    \label{eq:base_problem}
    \mathbf{\bar{A}}\cdot\mathbf{S}_0 = k_0 \mathbf{S}_0, 
    \qquad 
    {\mathbf{S_0^*}}^T\cdot \mathbf{\bar{A}} = k_0{\mathbf{S_0^*}}^T
\end{equation}
where $\mathbf{S}_0$ and $\mathbf{S_0^*}$ represent the right and left fundamental eigenvectors, respectively. $\mathbf{S}_0$ is normalized such that $\sum_{i=1}^{\nc}S_{0,i}=1$, consistent with its physical interpretation as a unit-normalized fission source distribution; $\mathbf{S}_0^*$ is then normalized to satisfy the bi-orthogonal relation ${\mathbf{S}_0^*}^T \mathbf{S}_0 = 1$. The superscript $T$ indicates the transpose operator. Since both normalizations must hold for every perturbed configuration, their derivatives vanish identically: $\mathbf{1}^T\delta\mathbf{S}=0$ and ${\mathbf{S}_0^*}^T\delta\mathbf{S}+{\delta\mathbf{S}^*}^T\mathbf{S}_0=0$, where $\mathbf{1}\in\mathbb{R}^{\nc}$ denotes the all-ones vector. The first of these identities constrains $\delta\mathbf{S}$ alone, and is used below (Equation \ref{eq:delta-s-final} onward) to resolve the eigenvector perturbation $\delta\mathbf{S}$ uniquely. The second only fixes the scale of the perturbed left eigenvector, $\delta\mathbf{S}^*$, and places no further constraint on $\delta\mathbf{S}$. In the context of the \gls{fm}, the right- and left-eigenvector can be physically interpreted as the fission source, \fsv, and the fission importance function, $\mathbf{S}^*$, respectively \cite{mascolinoNovelHybridDeterministic2024}. 

The solution to a perturbed state, characterized by \gls{fm} $\mathbf{\tilde{A}}$, is described by the following equation:
\begin{equation}
    \label{eq:perturbed}
    \mathbf{\tilde{A}}\cdot\mathbf{\tilde{S}}=\tilde{k}\mathbf{\tilde{S}}
\end{equation}

Assuming linear perturbations, we can rewrite the perturbed terms as the sum of the unperturbed functions and a small perturbation component:
\begin{equation}
    \begin{cases}
        \mathbf{\tilde{A}}=\bar{\mathbf{A}}+\mathbf{\delta A} \\
        \tilde{k}=k_0+\delta k \\
        \mathbf{\tilde{S}}=\mathbf{S_0}+\mathbf{\delta S}
    \end{cases}
\end{equation}

Equation \ref{eq:perturbed} becomes:
\begin{equation}
    \left(\mathbf{\bar{A}}+\delta\mathbf{A}\right)\cdot\left(\mathbf{S_0}+\delta\mathbf{S}\right)=\left(k_0+\delta k\right)\left(\mathbf{S}_0+\mathbf{\delta S}\right)
\end{equation}

Expanding the above equation, using the identity from Equation \ref{eq:base_problem}, and neglecting the second-order terms $\delta\mathbf{A}\cdot\delta\mathbf{S}$ and $\delta k\,\delta\mathbf{S}$, we obtain:
\begin{equation}
    \label{eq:branch}
    \mathbf{\bar{A}}\cdot\mathbf{\delta S}+\mathbf{\delta A}\cdot\mathbf{S_0}=k_0\mathbf{\delta{S}}+\delta k\mathbf{S_0}
\end{equation}

Left-multiplying Equation \ref{eq:branch} by ${\mathbf{S_0^*}}^T$ and using the left-eigenvector relation of Equation \ref{eq:base_problem}, the terms in $\delta\mathbf{S}$ cancel identically, so no constraint on $\delta\mathbf{S}$ is needed. The bi-orthogonal normalization ${\mathbf{S_0^*}}^T\mathbf{S_0}=1$ then yields:
\begin{equation}
    \label{eq:delta-k}
    \delta k=\mathbf{S_0^*}^T\cdot\mathbf{\delta\bar{A}}\cdot \mathbf{S_0}=\displaystyle \sum_{i=1}^{\nc} \sum_{j=1}^{\nc} S^*_{0,i} \delta a_{i,j}S_{0,j}
\end{equation}

Finally, we can calculate the sensitivity coefficient of $k$ as a function of the variation of each coefficient, \aij:

\begin{equation}
    \label{eq:sensitivity}
    J_{k;i,j}=\dfrac{\partial k}{\partial\aij}=S^*_{0,i}S_{0,j}
    \qquad\rightarrow\qquad
    \mathbf{J_k} = \mathbf{S_0^*} \otimes \mathbf{S_0}
\end{equation}

where $J_{k; i,j}$ is the generic element of the Jacobian vector \cite{magnusMatrixDifferentialCalculus2019}, $\mathbf{J_k} \in \mathbb{R}^{\nc^2\times1}$, and $\otimes$ indicates the Kronecker product. The fundamental eigenpair of a non-negative \gls{fm} is real (Perron-Frobenius theorem), so all Jacobians in this section are real. The Kronecker form follows from the vectorization identity $\text{vec}\left(\mathbf{u}\mathbf{v}^T\right)=\mathbf{u}\otimes\mathbf{v}$ applied row-major throughout this section, i.e., consistent with the array-storage (C-order) convention used by the numerical implementation, so that $\delta A_{i,j}$ corresponds to entry $(i-1)\nc+j$ of $\text{vec}(\delta\mathbf{A})$. Each of the sensitivity coefficients in Equation \ref{eq:sensitivity} represents the effective (i.e., importance-weighted) contribution of the fission source to the value of \keff. The mathematically rigorous variance of the eigenvalue $k_{\text{eff}}$ using the Delta method is then evaluated via the `sandwich' product \cite{jcgmEvaluationMeasurementData2008}:

\begin{equation}
    \label{eq:var-keff}
    \sigma^2_\Delta(k_{\text{eff}}) \approx  \mathbf{J}^T \bm{\Sigma}_A \mathbf{J}
\end{equation}
where $\bm{\Sigma}_A$ is the \gls{fm} covariance matrix, built from the individual coefficient covariances of Equation \ref{eq:cov-def}, as discussed in Section \ref{sec:covariances}. The calculation of the uncertainties of the fission source, \matb{S} (the right eigenvector $\mathbf{S_0}$, as stated above) is conceptually analogous, but requires additional care. Unlike the scalar eigenvalue $k$, the eigenvector $\mathbf{S}$ is only determined up to the null space of the perturbed operator, and resolving that ambiguity correctly is essential to obtaining a valid result. Starting from Equation \ref{eq:branch}, we rearrange it as:
\begin{equation}
    \label{eq:delta-s-rearranged}
    (\mathbf{\bar{A}}-k_0\mathbf{I})\delta \mathbf{S}=\delta k\mathbf{S_0}-\mathbf{\delta A}\cdot\mathbf{S_0}
\end{equation}

We define $\mathbf{M}=\mathbf{\bar{A}}-k_0\mathbf{I}$. $\mathbf{M}$ is a singular matrix, with $\det(\mathbf{M})=0$, and therefore does not admit a rigorous inverse. The Moore-Penrose pseudo-inverse, $\mathbf{M^+}$, is used instead \cite{penroseGeneralizedInverseMatrices1955}. Since $\bar{\mathbf{A}}$ is, in general, not symmetric, $\mathbf{M}$'s right and left null spaces differ and $\mathbf{M}^+$ can only be applied directly to a right-hand side that already lies in the range of $\mathbf{M}$, i.e., a vector $\mathbf{v}$ satisfying ${\mathbf{S_0^*}}^T\mathbf{v}=0$. The right-hand side of Equation \ref{eq:delta-s-rearranged} does satisfy this condition, and using the expression for $\delta k$ of Equation \ref{eq:delta-k} yields:

\begin{equation}
    {\mathbf{S_0^*}}^T\left(\delta k\,\mathbf{S_0}-\mathbf{\delta A}\cdot\mathbf{S_0}\right)=\delta k\left({\mathbf{S_0^*}}^T\mathbf{S_0}\right)-{\mathbf{S_0^*}}^T\mathbf{\delta A}\,\mathbf{S_0}=\delta k-\delta k=0
\end{equation}
so $\mathbf{M}^+$ may be applied directly to Equation \ref{eq:delta-s-rearranged} in full, retaining the $\delta k\,\mathbf{S_0}$ term that a naive application of the pseudo-inverse would otherwise incorrectly discard. This term does not vanish, since $\mathbf{S_0}\notin\ker\left(\mathbf{M^+}\right)=\text{span}\left(\mathbf{S_0^*}\right)$ in general:
\begin{equation}
    \label{eq:delta-s-particular}
    \delta\mathbf{S}_\text{p}=\mathbf{M^+}\left(\delta k\,\mathbf{S_0}-\mathbf{\delta A}\cdot\mathbf{S_0}\right)=\delta k\,\mathbf{M^+}\mathbf{S_0}-\mathbf{M^+}\mathbf{\delta A}\,\mathbf{S_0}
\end{equation}
Equation \ref{eq:delta-s-particular} is \textit{a} solution of Equation \ref{eq:delta-s-rearranged}, but not necessarily the physical one: since $\mathbf{S_0}\in\ker(\mathbf{M})$, $\delta\mathbf{S}_\text{p}+c\,\mathbf{S_0}$ solves Equation \ref{eq:delta-s-rearranged} equally well for any scalar $c$. The pseudo-inverse does select one member of this family. Since $\text{range}\left(\mathbf{M^+}\right)=\text{range}\left(\mathbf{M}^T\right)=\ker(\mathbf{M})^\perp$, it returns the minimum-norm solution, which satisfies $\mathbf{S_0}^T\delta\mathbf{S}_\text{p}=0$. This is not the gauge required here. The required gauge is the sum-normalization of $\mathbf{S_0}$ introduced above, namely $\mathbf{1}^T\delta\mathbf{S}=0$, with $\mathbf{1}\in\mathbb{R}^{\nc}$ the all-ones vector. The two gauges differ because $\bar{\mathbf{A}}$ (and hence $\mathbf{M}$) is not symmetric in general, and $\mathbf{S_0}$ is not proportional to $\mathbf{1}$. The minimum-norm solution is therefore corrected by removing the component along $\mathbf{S_0}$ that violates the sum-normalization, which gives the unique, physically consistent solution:
\begin{equation}
    \delta\mathbf{S}=\delta\mathbf{S}_\text{p}-\mathbf{S_0}\left(\mathbf{1}^T\delta\mathbf{S}_\text{p}\right)=\left(\mathbf{I}-\mathbf{S_0}\mathbf{1}^T\right)\delta\mathbf{S}_\text{p}
\end{equation}
Defining the gauge-corrected pseudo-inverse $\hat{\mathbf{M}}^+\equiv\left(\mathbf{I}-\mathbf{S_0}\mathbf{1}^T\right)\mathbf{M^+}$, each of whose columns sums to zero by construction, this becomes:
\begin{equation}
    \label{eq:delta-s-final}
    \delta\mathbf{S}=\delta k\,\hat{\mathbf{M}}^+\mathbf{S_0}-\hat{\mathbf{M}}^+\mathbf{\delta A}\,\mathbf{S_0}
\end{equation}

By applying the properties of the Kronecker product and vectorization, and substituting $\delta k={\mathbf{S_0^*}}^T\mathbf{\delta A}\,\mathbf{S_0}$ (Equation \ref{eq:delta-k}) so that Equation \ref{eq:delta-s-final} is expressed entirely as a linear map of $\mathbf{\delta A}$, we obtain:
\begin{equation}
    \delta\mathbf{S}=-\left[\hat{\mathbf{M}}^+\left(\mathbf{I}-\mathbf{S_0}{\mathbf{S_0^*}}^T\right)\otimes\mathbf{S_0}^T\right]\text{vec}\left(\mathbf{\delta A}\right)
\end{equation}

With $\hat{\mathbf{M}}^+$ in hand, the Jacobian matrix for the fission source, $\mathbf{J_S}\in\mathbb{R}^{\nc\times\nc^2}$, can be calculated as follows:

\begin{equation}
    \label{eq:jacobian-source}
    J_{S_l;i,j}=\dfrac{\partial S_l}{\partial\aij}=S_{0,j}\left[S^*_{0,i}\left(\hat{\mathbf{M}}^+\mathbf{S_0}\right)_l-\hat{M}^+_{l,i}\right]
\end{equation}
or in matrix notation:

\begin{equation}
    \label{eq:jacobian-source-matrix}
    \mathbf{J_S}=-\left[\hat{\mathbf{M}}^+\left(\mathbf{I}-\mathbf{S_0}{\mathbf{S_0^*}}^T\right)\right]\otimes\mathbf{S_0}^T
\end{equation}
As in Equation \ref{eq:sensitivity}, the Kronecker form of Equation \ref{eq:jacobian-source} follows the row-major vectorization convention.

Finally, the covariance matrix for the source can be calculated using the sandwich product:
\begin{equation}
    \label{eq:var-source}
    \mathbf{\Sigma_S}\approx \mathbf{J_S}\mathbf{\Sigma_A}\mathbf{J_S}^T
\end{equation}

Since each column of $\hat{\mathbf{M}}^+$ sums to zero, $\mathbf{1}^T\mathbf{J_S}=\mathbf{0}^T$, and therefore $\mathbf{1}^T\bm{\Sigma}_S\mathbf{1}=0$ exactly. The matrix $\bm{\Sigma}_S$ is singular by construction, with rank at most $\nc-1$. The same holds for the diagonal-covariance approximation of Section \ref{sec:simplified_methods}. This provides a consistency check of the implementation. It also clarifies what $\text{diag}\left(\bm{\Sigma}_S\right)^{1/2}$ represents: the uncertainty of the source shape under the fixed sum-normalization. Since the per-cell errors sum to zero, they are negatively correlated on average.

The accuracy of the first-order treatment of $\delta\mathbf{S}$ depends on the spectral gap of $\bar{\mathbf{A}}$. The smallest non-zero singular value of $\mathbf{M}$ is of the order of $k_0-|k_1|$, where $k_1$ is the second-largest eigenvalue. The eigenvector sensitivity is therefore amplified by roughly $1/(1-\text{DR})$, with $\text{DR}=|k_1|/k_0$ the dominance ratio. For the models of this work, $\text{DR}=0.742$ (CovLow), $0.742$ (CovHigh), and $0.746$ (FixedSrc), which corresponds to an amplification of about $3.9$. This amplification is modest, so the linearization is adequate for the statistical uncertainties considered here. This is corroborated by \textit{Resample}, which uses no linearization and agrees closely with the diagonal-covariance Delta method (Section \ref{sec:res-uncertainty}).

All of the Delta method calculations above are implemented into a Python script and use widely available and optimized mathematical libraries such as NumPy \cite{harrisArrayProgrammingNumPy2020} and SciPy \cite{virtanenSciPy10Fundamental2020}.

\subsubsection{Evaluation of Fission Matrix Covariances}
\label{sec:covariances}
Let $\mathbf{T}_m \in \mathbb{R}^{\nc\times\nc}$ denote the realization of the raw fission matrix tally extracted from an OpenMC statepoint after active generation $m$, where \nc represents the number of spatial mesh cells. Since OpenMC only retains the sum and squared sum of tallies at any active generation but not its realization, a forward-difference operator is applied across adjacent statepoints to isolate the net stochastically accumulated tallies strictly within a discrete batch/generation:
\begin{equation}
    \mathbf{T}_m = \mathbf{T}^\Sigma_m - \mathbf{T}^\Sigma_{m-1}, \quad \mathbf{T}^\Sigma_0\equiv\mathbf{0},\quad \forall m \in\left[1,M\right]
\end{equation}
where $\mathbf{T}^\Sigma_m=\sum_{m'=1}^{m}\mathbf{T}_{m'}$ denotes OpenMC's own cumulative tally sum reported at statepoint $m$, and $M$ is the total number of active generations simulated (100 for CovLow, 250 for CovHigh, per Table \ref{tab:fm-calc-summary}) and corresponds to the number of independent realizations of the raw fission matrix, $\mathbf{T}$.

Due to the fission matrix generation process in OpenMC, the actual physical fission matrix coefficients, $a_{i,j}^{(m)}$, for the $m$-th realization are calculated by scaling the raw tally entries $T_{i,j}^{(m)}$ to obtain a value that is normalized by the number of fission neutrons born, i.e., the fission source, in cell $j$ \cite{balouzaComparingFissionMatrix2026,paineNeuralNetworkEnabledThermal29}:
\begin{equation}
    \label{eq:renorm-crit}
    a_{i,j}^{(m)} = T_{i,j}^{(m)} \times\frac{\displaystyle\sum_{k=1}^{N}\sum_{l=1}^{N} T_{k,l}^{(m)}}{\displaystyle\sum_{l=1}^{N} T_{j,l}^{(m)}}=T_{i,j}^{(m)}\times C_{j}
\end{equation}
where the numerator of the scaling factor $C_j$ represents the grand total of all fission neutrons generated across the entire geometry, and the denominator represents the tally originating from cell $j$, integrated over all destination cells $l$. Note, however, that for the global fixed-source calculation of Table \ref{tab:fm-calc-summary} (case 5), the renormalization of Equation \ref{eq:renorm-crit} is unnecessary. Each of the \nc\ fuel cells is assigned its own, independently declared unit source strength, all active simultaneously within a single run, rather than a single self-consistent fission source resampled from the previous batch's fission bank. OpenMC normalizes each source's tally contribution by that source's own declared strength regardless of how many Monte Carlo particles are actually realized from it in a given batch, so the scaling factor $C_j$ of Equation \ref{eq:renorm-crit} reduces to 1 identically and $a_{i,j}^{(m)}=T_{i,j}^{(m)}$ directly, with $\mathbf{T}_m$ itself already a valid realization of the fission matrix. 

 The expected value of the fission matrix, $\bar{\mathbf{A}}= \mathbb{E}(A)$ is defined as:
\begin{equation}
    \label{eq:mean-A}
    \mathbf{\bar{A}} = \frac{1}{M}\sum_{m=1}^{M} \mathbf{A}^{(m)}
\end{equation}

Since the raw fission matrix tallies, $T_{i,j}$, are obtained from the same Monte Carlo calculation and are scored based on the same simulated particles and random events, the coefficients $a_{i,j}$ are statistically correlated and have non-zero covariances. Each couple of coefficients, $a_{i,j}$ and $a_{k,l}$, first has a per-realization sample covariance, computed from the $M$ realizations via the standard unbiased estimator\ \cite{casellaStatisticalInference2024}:

\begin{equation}
    \label{eq:cov-per-realization}
    \widehat{\sigma}^{(1)}_{(i,j),(k,l)} = \frac{1}{M-1}\sum_{s=1}^{M} \left(a_{i,j}^{(s)} - \bar{a}_{i,j}\right)\left(a_{k,l}^{(s)} - \bar{a}_{k,l}\right)
\end{equation}

However, \keff\ and the fission source \matb{S} are evaluated at the sample mean fission matrix, $\mathbf{\bar A}$ (Equation \ref{eq:renorm-crit}), not at any single realization $a_{i,j}^{(m)}$: the quantity that actually propagates through the Delta-method sandwich products of Equations \ref{eq:var-keff} and \ref{eq:var-source} is therefore the covariance of that mean, not of a single realization. Since $\mathbf{\bar A}$ averages $M$ realizations, this is smaller than the per-realization covariance by a factor of $M$:

\begin{equation}
    \label{eq:cov-def}
    \text{Cov}(a_{i,j}, a_{k,l}) =\acov= \frac{1}{M}\,\widehat{\sigma}^{(1)}_{(i,j),(k,l)} = \frac{1}{M(M-1)}\sum_{s=1}^{M} \left(a_{i,j}^{(s)} - \bar{a}_{i,j}\right)\left(a_{k,l}^{(s)} - \bar{a}_{k,l}\right)
\end{equation}

 The individual covariances, \acov, can be built into a covariance matrix, denoted $\bm{\displaystyle\Sigma}_A$, of size $\nc^2\times\nc^2$. It is clear that, even for a reasonably sized \gls{fm} of size $\nc=1{,}000=10^3$, the size of the covariance matrix becomes incredibly large at $10^{12}$ values. If double-precision floating-point numbers are used for a fully dense covariance matrix, this results in 8 TB memory requirement just to store it. This is highly impractical, and, as such, the coefficients' covariances cannot be used directly in real-time \gls{fm} applications for the purpose of \gls{uq}, justifying the analysis and evaluation of alternative and more efficient \gls{uq} approaches against the impractical, mathematically rigorous reference.

Storing $\bm{\Sigma}_A$ is not strictly required by the Delta method itself. Since Equation \ref{eq:cov-def} is a sample covariance, the sandwich products of Equations \ref{eq:var-keff} and \ref{eq:var-source} reduce exactly (to roundoff) to sums over the $M$ realizations:
\begin{equation}
    \label{eq:streaming}
    \sigma^2_\Delta(\keff)=\frac{1}{M(M-1)}\sum_{m=1}^{M}\left(\delta k^{(m)}\right)^2,\qquad
    \bm{\Sigma}_S=\frac{1}{M(M-1)}\sum_{m=1}^{M}\delta\mathbf{S}^{(m)}{\delta\mathbf{S}^{(m)}}^T
\end{equation}
where $\delta k^{(m)}={\mathbf{S_0^*}}^T\left(\mathbf{A}^{(m)}-\bar{\mathbf{A}}\right)\mathbf{S_0}$ and $\delta\mathbf{S}^{(m)}=-\hat{\mathbf{M}}^+\left(\mathbf{I}-\mathbf{S_0}{\mathbf{S_0^*}}^T\right)\left(\mathbf{A}^{(m)}-\bar{\mathbf{A}}\right)\mathbf{S_0}$. This requires $\mathcal{O}(M\nc^2)$ time and $\mathcal{O}(\nc^2)$ memory, instead of $\mathcal{O}(\nc^4)$ memory. It does, however, require the $M$ per-batch realizations $\mathbf{A}^{(m)}$. OpenMC only retains cumulative tallies, so these realizations are only available if a statepoint is saved after every batch, as described above. A typical \gls{fm} database, on the other hand, retains only the mean coefficients and the per-coefficient uncertainties reported by the Monte Carlo code. In this work, $\bm{\Sigma}_A$ is built explicitly because its structure is itself characterized (Section \ref{sec:res-cov-structure}). The impractical cost discussed here therefore applies to the explicit covariance matrix and to the per-batch data needed to estimate it.

\subsubsection{Simplified Low-Cost Alternative UQ Methods}
\label{sec:simplified_methods}

While the Delta method framework outlined above provides a mathematically exact first-order representation of uncertainty, the construction and storage of $\bm{\Sigma}_A$ scales at $\mathcal{O}(N_\text{cells}^4)$, presenting a severe computational bottleneck for large-scale geometries. To alleviate this burden, two simplified alternative methodologies are proposed and evaluated.

\paragraph{Diagonal Covariance Matrix Approximation}
The first simplified approach assumes that the correlations between distinct \gls{fm} matrix coefficients are negligible compared to their localized variances. Under this assumption, $\bm{\Sigma}_A$ is reduced to a strictly diagonal matrix where $\sigma_{(i,j), (k,l)} = 0$ for $(i,j) \neq (k,l)$. The eigenvalue variance calculated using Equation \ref{eq:var-keff} simplifies to a simple linear vector dot-product:
\begin{equation}
    \label{eq:var-keff-ind}
    \sigma^2(k_{\text{eff}}) \approx
    \displaystyle \sum_{i=1}^{\nc}\sum_{j=1}^{\nc} \left( \frac{\partial k_{\text{eff}}}{\partial a_{i,j}} \right)^2 \sigma^2_{i,j}
    =
    \sum_{i=1}^{\nc}\sum_{j=1}^{\nc} \left(S^*_{0,i}S_{0,j}\right)^2 \sigma^2_{i,j}
\end{equation}
reducing the computational scaling from $\mathcal{O}(\nc^4)$ to $\mathcal{O}(\nc^2)$. Similarly, the fission source variances using Equation \ref{eq:var-source} reduce to:
\begin{equation}
    \label{eq:var-source-ind}
    \sigma^2(S_l)\approx \sum_{i=1}^{\nc}\sum_{j=1}^{\nc}\left(J_{S_l;i,j}\right)^2\sigma^2_{i,j}
    =
    \sum_{i=1}^{\nc}\sum_{j=1}^{\nc}\left\{S_{0,j}\left[S^*_{0,i}\left(\hat{\mathbf{M}}^+\mathbf{S_0}\right)_l-\hat{M}^+_{l,i}\right]\right\}^2\sigma^2_{i,j}
\end{equation}
using the gauge-corrected Jacobian $J_{S_l;i,j}$ of Equation \ref{eq:jacobian-source}.

\paragraph{Direct Independent Sampling of Fission Matrix Coefficients}
The second simplified approach entirely bypasses the calculation of coefficient gradients and covariance matrices. Under the assumption of statistically independent \gls{fm} coefficients, it samples each coefficient from a Gaussian distribution, based on the central limit theorem \cite{haghighatMonteCarloMethods2020}, as follows:

\begin{equation}
    f(a^{(m)}_{i,j})=\dfrac{1}{\sqrt{2\pi\sigma^2_{i,j}}}\cdot\exp\left[-\dfrac{(a^{(m)}_{i,j}-\bar{a}_{i,j})^2}{2\sigma_{i,j}^2}\right],\qquad m\in[1,M]
\end{equation}
where $a^{(m)}_{i,j}$ is the $m$-th sample of the \aij \gls{fm} coefficient, $\bar{a}_{i,j}$ and $\sigma_{i,j}$ are the mean value and statistical uncertainty of the coefficient as calculated by OpenMC. After sampling the \gls{fm} coefficients, $M$ values of $k_{eff}^{(m)}$ and $\matb{S}^{(m)}$ are then calculated using Equation \ref{eq:fmequation}. Average values and standard deviations are simply calculated as the sample average and uncertainty of the \keff and source samples, $\keff^{(m)}$ and $\mathbf{S}^{(m)}$, as follows:

\begin{equation}
    \begin{cases}
        \bar{k}_\text{eff}=\displaystyle\dfrac{1}{M}\sum_{m=1}^M\keff^{(m)}\\
        s_{\keff}=\displaystyle\sqrt{\dfrac{\sum_{m=1}^M\left(\keff^{(m)}-\bar{k}_\text{eff}\right)^2}{M-1}}
    \end{cases},
    \qquad
    \begin{cases}
        \bar{S}_i=\displaystyle\dfrac{1}{M}\sum_{m=1}^M S_i^{(m)}\\
        s_{S_i}=\displaystyle\sqrt{\dfrac{\sum_{m=1}^M\left(S_i^{(m)}-\bar{S}_i\right)^2}{M-1}}
    \end{cases}
\end{equation}

This method limits the memory footprint strictly to the size of two $\nc\times\nc$ \gls{fm} matrices at any given time.

\subsection{Evaluation Metrics}
\label{sec:metrics}

The Delta method and the two simplified alternatives of Section \ref{sec:simplified_methods} are compared against each other and against independent statistical evidence using the metrics defined below. All metrics are computed identically for every case in Table \ref{tab:fm-calc-summary}, and their values are reported and discussed in Section \ref{sec:results}.

\subsubsection{Uncertainty estimators}
\label{sec:estimators-compared}
Four estimators of $\sigma(\keff)$ and $\sigma(\fsv)$ are compared throughout:
\begin{itemize}
    \item \textit{Covariances}: the rigorous Delta method of Section \ref{sec:delta}, using the full \gls{fm} covariance $\bm{\Sigma}_A$ propagated through Equations \ref{eq:var-keff} and \ref{eq:var-source} (i.e., $\sigma_\Delta(\keff)$ and $\text{diag}\left(\bm{\Sigma}_S\right)^{1/2}$).
    \item \textit{Independent}: the diagonal-covariance approximation of Equations \ref{eq:var-keff-ind} and \ref{eq:var-source-ind}, i.e., the Delta method evaluated as if all the \gls{fm} coefficients $\aij$ were statistically independent.
    \item \textit{Resample}: the direct-sampling approach of Section \ref{sec:simplified_methods}, i.e., the sample standard deviation of $M=10{,}000$ values of $\keff^{(m)}$ (or $S_l^{(m)}$) obtained by independently sampling every \gls{fm} coefficient from its own Gaussian distribution and re-solving Equation \ref{eq:fmequation} in full, with no linearization.
    \item \textit{Runs}: the empirical standard error of the mean of \keff and \fsv across $M=100$ fully independent OpenMC realizations, i.e., the sample standard deviation across the realizations divided by $\sqrt{M}$. It is therefore the statistical uncertainty of the 100-run average. The realizations are: those of the Indp case for CovLow, and the replicate runs of the CRd and FixedSrc cases for those cases (Table \ref{tab:fm-calc-summary}). It is not available for CovHigh. Since each realization is a complete, independently-seeded Monte Carlo calculation, the covariance between realizations is 0 by construction, while any correlation among the coefficients of one realization is captured implicitly in the run-to-run spread. As such, \textit{Runs} represents an absolute reference limited only by the selected sample size, $M=100$.
\end{itemize}

Each alternative estimator is reported as a ratio to \textit{Covariances}, giving both Table \ref{tab:res-keff-sigma} (\keff) and Table \ref{tab:res-source-sigma} (fission source) one common reference. For \textit{Independent}, this ratio is the whole point of the diagonal-covariance approximation: it shows how much ignoring the off-diagonal terms changes the result. \textit{Runs} has zero cross-realization covariance by construction: each realization uses an independent seed and at most two batches (Table \ref{tab:fm-calc-summary}), so there is no significant batch-to-batch (inter-cycle) correlation within a realization, and none between realizations either. Correlations among the coefficients of a single realization are still captured implicitly in the run-to-run spread, so \textit{Runs} makes no diagonal-covariance assumption. Its ratio to \textit{Covariances} therefore tests the Delta-method result against an assumption-free reference, whereas the ratios of \textit{Independent} and \textit{Resample} also include the effect of assuming a diagonal covariance. \textit{Resample} also assumes a diagonal covariance matrix since sampling \gls{fm} coefficients independently cannot reproduce any cross-coefficient covariance. Its ratio to \textit{Covariances} combines that assumption's effect with the extra difference between Monte Carlo resampling and the Delta method's linearization.

\subsubsection{Point estimates}
For \keff, OpenMC's native track-length estimator is read directly from the simulation output, with its own reported statistical uncertainty, and is used throughout. The choice of the OpenMC track-length estimator is the most consistent with the one used for tallying the \gls{fm} coefficients themselves. It is compared against the \gls{fm}-derived eigenvalue obtained by solving Equation \ref{eq:fmequation} for $\bar{\mathbf{A}}$ (i.e., $k_0$ of Equation \ref{eq:base_problem}); the difference between the two, in pcm ($10^5\times\Delta\keff$), serves as a numerical-consistency check between the fission-matrix eigen-solve and OpenMC's own criticality estimators. For the fission source, the peak (space-integrated, unit-normalized) value and the \gls{rsu} are compared between the OpenMC tally and the \gls{fm} solution:
\begin{equation}
    \label{eq:rsu}
    \text{RSU}(\mathbf{S}) = \frac{\sum_{i=1}^{\nc}\sigma_{S,i}}{\sum_{i=1}^{\nc}S_i} = \sum_{i=1}^{\nc}\sigma_{S,i}
\end{equation}
the second equality following from the unit-sum normalization of \fsv. The agreement between the OpenMC-tallied and \gls{fm}-solved source \textit{shapes} (as opposed to their uncertainties) is quantified by a normalized, source-weighted mean absolute deviation, labeled ``NRMSE'' for consistency with the accompanying numerical results:
\begin{equation}
    \label{eq:nrmse}
    \text{NRMSE} = \frac{\displaystyle\sum_{i=1}^{\nc}\left|S_i^{\text{OpenMC}}-S_i^{\text{FM}}\right|}{\displaystyle\sum_{i=1}^{\nc}S_i^{\text{OpenMC}}}
\end{equation}

\subsubsection{FM covariance structure}
The correlation matrix associated with $\bm{\Sigma}_A$ is
\begin{equation}
    \rho_{(i,j),(k,l)}=\frac{\sigma_{(i,j),(k,l)}}{\sigma_{i,j}\sigma_{k,l}}
\end{equation}
Two diagnostics summarize how much of that correlation structure sits off the diagonal. The first is the off-diagonal energy fraction, corresponding to the fraction of the correlation matrix's squared Frobenius norm carried by its off-diagonal entries, as follows:
\begin{equation}
    \eta=\left\|\bm{\rho}^{\text{off-diag}}\right\|_F^2\Big/\left\|\bm{\rho}\right\|_F^2
\end{equation}
The second is the largest off-diagonal correlation magnitude, $\max_{(i,j)\neq(k,l)}\left|\rho_{(i,j),(k,l)}\right|$. The practical consequence of dropping that off-diagonal structure on the evaluated variance estimators is quantified by the diagonal-only \keff variance error:
\begin{equation}
    \label{eq:matrix-error}
    \epsilon_k = \frac{\left|\sigma^2_\Delta(\keff)-\sigma^2_{\text{diag}}(\keff)\right|}{\sigma^2_\Delta(\keff)}
\end{equation}
where $\sigma^2_\text{diag}(\keff)$ is the \textit{Independent} estimator's variance.

Every off-diagonal pair of \gls{fm} coefficients, $\text{Cov}(a_{i,j},a_{k,l})$ with $(i,j)\neq(k,l)$, is further classified into one of two physically distinct groups: \textit{same-source-column} pairs ($j=l$, $i\neq k$) and \textit{different-source-column} pairs ($j\neq l$). The \textit{same-source-column} pairs are expected to experience correlation (and, hence, non-zero covariances) due to the fact that they are competing for a shared pool of source particles coming from the same region, $j=l$. This is consistent with the same shared-particle-pool reasoning previously used to argue that \gls{fm} coefficient covariances are negative in general \cite{mascolinoExperimentalComputationalBenchmarking2019,mascolinoDevelopmentBenchmarkingAdvanced2021}, restricted here to the specific pairs for which it actually applies. As such, this effect is expected to be negative. The \textit{different-source-column} pairs, on the other hand, are dominated by inter-cycle correlation of the fission source between Monte Carlo generations. This effect can be due to local, generation-wise, fission source fluctuations that may affect all of the coefficients in a nearby region and, hence, may carry a positive correlation. This effect may carry into successive generations and strongly depends on the convergence of the fission source. Each group's mean covariance and the fraction of its pairs that are negative/positive are reported to characterize the sign structure of $\bm{\Sigma}_A$. Both mechanisms stem from the specific tallying approach used to generate the \gls{fm} coefficients in this work (Section \ref{sec:fm-method}), which relies on eigenvalue (criticality) calculations where the fission source is resampled generation-to-generation. Other \gls{fm} generation strategies use fixed-source calculations instead \cite{mascolinoNovelHybridDeterministic2024,heDevelopmentUncertaintyAnalysis2022,balouzaComparingFissionMatrix2026}: since the fission source is not resampled between batches, the positive fission-source-convergence correlation described above cannot occur, so the same-source-column competition mechanism is expected to be the only one that survives, driven by the \aij competing for the same source-particle budget for any fixed $j$; different-source-column pairs, with no analogous shared mechanism linking them, are expected to be statistically independent. This work evaluates such a case directly (FixedSrc) and finds that the predicted sign holds only on average: the same-source-column covariance is intrinsically small enough, relative to what a single Monte Carlo run's batch-to-batch statistics can resolve, that it cannot be distinguished empirically from zero without either an analytic (rather than sampled) treatment or an impractically large number of batches (Section \ref{sec:res-fixed-source}).

Finally, a per-coefficient diagonal-dominance diagnostic, in the same units as $\sigma_{i,j}$, quantifies whether a given \gls{fm} coefficient's own variance or the raw sum of its covariance with every other coefficient is larger in magnitude:
\begin{equation}
    \label{eq:cov-dominance}
    D_{i,j} = \frac{\displaystyle\sum_{(k,l)\neq(i,j)}\text{Cov}(a_{i,j},a_{k,l})}{\sigma_{i,j}}
\end{equation}
Its sign shows whether coefficient $(i,j)$'s net correlation with the rest of the matrix leans positive or negative; its magnitude, compared against the coefficient's own standard deviation $\sigma_{i,j}$, shows whether that net correlation is small or large relative to it. This raw sum is not weighted by any estimator's Jacobian, so a large $D_{i,j}$ does not by itself imply a large net effect on a propagated quantity such as $\sigma^2_\Delta(k_\text{eff})$ (Equation \ref{eq:var-keff}): once every coefficient's covariance row is weighted by the Jacobian and summed, extensive sign cancellation can leave the net effect far smaller than $D_{i,j}$ alone would suggest (Section \ref{sec:res-cov-structure}). The analogous per-cell ratio for the fission source standard deviation, $\sigma_{S,l}^{\text{diag}}/\sigma_{S,l}$ (\textit{Independent}-to-\textit{Covariances}), is averaged over all \nc cells, and the number of cells for which the full-covariance $\sigma_{S,l}$ is larger, respectively smaller, than the diagonal-only $\sigma_{S,l}^{\text{diag}}$ is tabulated.

\subsubsection{FM-CRd validation}
The \gls{fm}-\gls{crd} method of Section \ref{sec:fm-crd} is validated by combining the \gls{aro} baseline with the single-rod-perturbation \gls{fm} evaluations of Table \ref{tab:cr_pos}, linearly weighted toward each rod's actual test position, into $\mathbf{A}_\text{CRd}$, the weighted sum of the already-renormalized per-run \glspl{fm}. $\mathbf{A}_\text{CRd}$ is solved via Equation \ref{eq:fmequation} and compared against the exact, fully-perturbed reference solution (all four rods at their real positions simultaneously) using the same pcm eigenvalue difference as above and a source-weighted relative error:
\begin{equation}
    \label{eq:crd-source-error}
    \text{Source rel. error} = \frac{\displaystyle\sum_{i=1}^{\nc}\left|S_i^{\text{CRd}}-S_i^{\text{exact}}\right|}{\displaystyle\sum_{i=1}^{\nc}S_i^{\text{exact}}}
\end{equation}

\section{Results}
\label{sec:results}

Results are presented for the two long-run Delta-method cases of Table \ref{tab:fm-calc-summary} (CovLow, $\text{NPS}=10^5$; CovHigh, $\text{NPS}=10^7$), the Indp case used as the empirical \textit{Runs} benchmark, the FixedSrc global fixed-source case, and the \gls{fm}-\gls{crd} validation case. Reported uncertainties are one-standard-deviation values, in pcm ($10^{-5}\Delta\keff$) for \keff\ and as the source \gls{rsu} of Equation \ref{eq:rsu} for \fsv.

\subsection{Reference k-effective and Fission Source Results}
\label{sec:res-reference}

Table \ref{tab:res-keff} compares OpenMC's own \keff\ (Section \ref{sec:metrics}) against the eigenvalue obtained by solving Equation \ref{eq:fmequation} for the mean fission matrix $\bar{\mathbf{A}}$. For both cases, the \gls{fm}-derived \keff\ agrees with OpenMC's \keff\ to within 1 pcm, well inside the OpenMC-reported statistical uncertainty even for the higher-statistics CovHigh case (3 pcm), confirming that the \gls{fm} eigen-solve reproduces OpenMC's own criticality result without introducing a discretization bias at $N_\text{ax}=5$. This is expected when obtaining the \gls{fm} coefficients directly from a criticality calculation instead of a series of fixed-source calculations, as done in previous work \cite{waltersRAPIDFissionMatrix2018}. Section \ref{sec:res-fixed-source} evaluates this point-estimate agreement directly for a global fixed-source calculation of this same reactor. The fission matrix uncertainties presented in Table \ref{tab:res-keff} are obtained with the analytical Delta method, and agree with the $\sigma_k$ obtained by OpenMC to the displayed precision (both round to 48 pcm for CovLow and 3 pcm for CovHigh, though the underlying raw values differ by a few percent, since OpenMC's own reported $\sigma_k$ is itself rounded in its log output).

\begin{table}[htbp!]
    \centering
    \setlength{\tabcolsep}{3pt}
    \begin{tabular}{l c c c c c}
        \toprule
        \multirow{2}{*}{\textbf{Case}} & \multicolumn{2}{c}{OpenMC} & \multicolumn{2}{c}{Fission Matrix} & \textbf{Diff.}\\
         & \keff & \boldmath$\sigma_k$ \textbf{(pcm)} & \keff & \boldmath$\sigma_k$ \textbf{(pcm)} & \textbf{(pcm)} \\
        \midrule
        CovLow  & 1.00846 & 48 & 1.00845 & 48 & $-1$ \\
        CovHigh & 1.00780 & 3  & 1.00779 & 3  & $-1$ \\
        \bottomrule
    \end{tabular}
    \caption{Reference \keff\ values: OpenMC's \keff\ (with its own reported $\sigma_k$), the eigenvalue of the mean \gls{fm} $\bar{\mathbf{A}}$, the rigorous Delta-method $\sigma_k$ (\textit{Covariances}, Section \ref{sec:estimators-compared}), and the difference between the \gls{fm}-derived and OpenMC \keff.}
    \label{tab:res-keff}
\end{table}

Table \ref{tab:res-source} compares the OpenMC-tallied fission source with the \gls{fm}-solved source \fsv. The two source shapes agree closely: the normalized deviation of Equation \ref{eq:nrmse} is below $2\times10^{-5}$ for CovLow and below $1.1\times10^{-6}$ for CovHigh, and the peak source values match to the reported precision in both cases. As expected, the \gls{rsu} of the OpenMC-tallied source improves by roughly an order of magnitude between the CovLow and CovHigh statistics levels, consistent with the $100\times$ increase in \text{NPS} and $2.5\times$ increase in \text{NAC} between the two cases (Table \ref{tab:fm-calc-summary}). Once again, the \gls{fm} uncertainties needed to calculate the RSU (Equation \ref{eq:rsu}) are obtained using the Delta method.

\begin{table}[htbp!]
    \centering
    \begin{tabular}{l l c c}
        \toprule
        \textbf{Method} & \textbf{Qty} & \textbf{CovLow} & \textbf{CovHigh} \\
        \midrule
        \multirow{2}{*}{OpenMC} & Max. & $8.02\times10^{-3}$ & $7.96\times10^{-3}$ \\
         & RSU & $6.53\times10^{-3}$ & $4.15\times10^{-4}$ \\
        \midrule
        \multirow{2}{*}{FM} & Max. & $8.02\times10^{-3}$ & $7.96\times10^{-3}$ \\
         & RSU & $6.51\times10^{-3}$ & $4.14\times10^{-4}$ \\
        \midrule
        \multicolumn{2}{c}{NRMSE}         & $1.60\times10^{-5}$ & $1.07\times10^{-6}$ \\
        \bottomrule
    \end{tabular}
    \caption{Reference fission-source values: peak (unit-normalized) source and \gls{rsu} from the OpenMC tally and from the \gls{fm} solution of $\bar{\mathbf{A}}$, and the normalized source deviation (NRMSE, Equation \ref{eq:nrmse}) between the two.}
    \label{tab:res-source}
\end{table}

\subsection{Delta-Method and Simplified Uncertainty Estimates}
\label{sec:res-uncertainty}

Table \ref{tab:res-keff-sigma} compares the rigorous \textit{Covariances} estimate of $\sigma(\keff)$ against \textit{Independent}, \textit{Resample}, and \textit{Runs}, all reported relative to \textit{Covariances} (see Section \ref{sec:estimators-compared}). For CovLow, the \textit{Independent} estimator underestimates $\sigma(\keff)$ by 5.7\% relative to \textit{Covariances}. The empirical \textit{Runs} benchmark relies on neither a covariance nor a linearization assumption, though its realizations are uncorrelated by construction. It sits close to \textit{Independent}, 4.1\% below \textit{Covariances}, supporting the diagonal covariance matrix approximation's accuracy at this statistics level for \keff. \textit{Resample}, run at $M=10{,}000$ Monte Carlo samples to keep its own sampling noise well below the effect being measured, tracks \textit{Independent} closely, 5.7\% below \textit{Covariances} -- expected, since both estimators target the same diagonal-covariance variance, one analytically and one by direct Monte Carlo resampling. For CovHigh, the \textit{Independent}-to-\textit{Covariances} ratio flips sign: \textit{Independent} now \textit{overestimates} $\sigma(\keff)$ by 1.5\%, and \textit{Resample} again tracks it closely, overestimating by 1.4\%. This sign flip in the \textit{Independent}-to-\textit{Covariances} ratio between the two statistics regimes (i.e., below 1 for CovLow, above 1 for CovHigh) is consistent with a change in the net sign of the \gls{fm} covariance's off-diagonal terms, due to the reducing impact of the fission-source-generation correlation described in Section \ref{sec:metrics}. Section \ref{sec:res-cov-structure} discusses this mechanism further and shows that the CovHigh point estimate's sign specifically is not tightly resolved at the available statistics. \textit{Runs} is unavailable for CovHigh, since the Indp case of Table \ref{tab:fm-calc-summary} was only run at the CovLow statistics level to maintain a manageable wall-clock execution time.

\begin{table}[htbp!]
    \centering
    \begin{tabular}{l c c}
        \toprule
        \textbf{$\sigma$ estimator} & \textbf{CovLow} & \textbf{CovHigh} \\
        \midrule
        Covariances   & 48     & 3     \\
        \midrule
        Independent   & 46     & 3     \\
        Ind./Cov.        & 0.943& 1.015 \\
        \midrule
        Resample      & 45     & 3     \\
        Res./Cov.       & 0.943  & 1.014 \\
        \midrule
        Runs          & 46     & --    \\
        Runs./Cov.      & 0.959  & --    \\
        \bottomrule
    \end{tabular}
    \caption{$\sigma_k$ (pcm) estimated via \textit{Covariances} (rigorous Delta method), \textit{Independent} (diagonal-covariance Delta method), \textit{Resample} (Monte Carlo resampling), and \textit{Runs} (100 independent OpenMC realizations, CovLow only), all reported relative to \textit{Covariances}, as defined in Section \ref{sec:estimators-compared}.}
    \label{tab:res-keff-sigma}
\end{table}

Table \ref{tab:res-source-sigma} repeats the comparison for the fission-source \gls{rsu}, with \textit{Independent}, \textit{Resample}, and \textit{Runs} again all reported relative to \textit{Covariances}, matching Table \ref{tab:res-keff-sigma}'s convention (Section \ref{sec:estimators-compared}). Unlike \keff, every estimator here overestimates the rigorous \textit{Covariances} result, or matches it almost exactly: \textit{Independent} is 0.5\% high for CovLow and 0.02\% low for CovHigh, \textit{Resample} is 0.5\% high for CovLow and 0.03\% low for CovHigh, and \textit{Runs} is only 0.6\% high for CovLow.

\begin{table}[htbp!]
    \centering
    \begin{tabular}{l c c}
        \toprule
        \textbf{RSU ($\sigma_S$) estimator} & \textbf{CovLow} & \textbf{CovHigh} \\
        \midrule
        Covariances   & $6.51\times10^{-3}$ & $4.14\times10^{-4}$ \\
        \midrule
        Independent   & $6.54\times10^{-3}$ & $4.14\times10^{-4}$ \\
        Ind./Cov.     & 1.005     & 1.000\\
        \midrule
        Resample      & $6.54\times10^{-3}$ & $4.14\times10^{-4}$ \\
        Res./Cov.     & 1.005     & 1.000     \\
        \midrule
        Runs          & $6.55\times10^{-3}$ & --        \\
        Runs./Cov.    & 1.006     & --        \\
        \bottomrule
    \end{tabular}
    \caption{Fission-source \gls{rsu} estimated via the same four approaches as Table \ref{tab:res-keff-sigma}, defined in Section \ref{sec:estimators-compared}.}
    \label{tab:res-source-sigma}
\end{table}

\subsection{FM Covariance Structure}
\label{sec:res-cov-structure}

Table \ref{tab:res-cov-structure} quantifies how much of the \gls{fm} correlation structure lies off the diagonal. For both cases, essentially all of the correlation matrix's Frobenius-norm energy sits off-diagonal ($\eta=99.78\%$ for CovLow, $99.46\%$ for CovHigh), and the largest off-diagonal correlation magnitude reaches unity for CovLow (a pair of coefficients that are, to the shown precision, perfectly correlated) and 0.573 for CovHigh. Despite this, dropping the off-diagonal structure entirely (the \textit{Independent} approximation) changes the propagated \keff\ variance by 11.0\% for CovLow and 3.0\% for CovHigh (Equation \ref{eq:matrix-error}). 

\begin{table}[htbp!]
    \centering
    \begin{tabular}{l c c c}
        \toprule
        \textbf{Case} & \textbf{Off-diag. energy fraction, } $\eta$ & \boldmath$\max|\rho|$ & \boldmath$\epsilon_k$ \textbf{(Eq. \ref{eq:matrix-error})} \\
        \midrule
        CovLow  & 0.9978 & 1     & 11.0\% \\
        CovHigh & 0.9946 & 0.573 & 3.0\%  \\
        \bottomrule
    \end{tabular}
    \caption{Off-diagonal structure of the \gls{fm} correlation matrix $\bm{\rho}$ and its impact on the \keff\ variance.}
    \label{tab:res-cov-structure}
\end{table}

Every off-diagonal \gls{fm} covariance pair was classified into the same-source-column (competition, $j=l$) and different-source-column (inter-cycle, $j\neq l$) groups of Section \ref{sec:metrics} and summarized exactly, not sampled, over the full population (Table \ref{tab:res-sign-structure}). For both cases the sign split is close to even (59.4\%/38.4\% negative/positive overall for CovLow, 50.2\%/49.8\% for CovHigh). The vastly larger different-column (inter-cycle) group, dominated by batch-to-batch correlation of the fission source between Monte Carlo generations \cite{mervinUncertaintyUnderpredictionMonte2014}, closely tracks the overall split. This is expected, since it makes up $\approx99.5\%$ of all off-diagonal pairs for $\nc=215$. Critically, the \textit{net} sign of the overall mean flips between the two cases: positive for CovLow ($1.20\times10^{-17}$) despite a numerical majority of negative pairs, and negative for CovHigh ($-1.13\times10^{-20}$) despite a near-even split. This happens because the rarer but individually larger inter-cycle-correlated pairs outweigh the more numerous but individually smaller competition-correlated pairs, by differing margins in each case. A net positive mean means the diagonal-only approximation underestimates the true, fully correlated variance. A net negative mean means it is mildly conservative instead. This mirrors the sign flip of the \textit{Independent}-to-\textit{Covariances} ratio already observed in Table \ref{tab:res-keff-sigma}. That ratio's sign, however, is not tightly resolved at the batch counts available here. Bootstrap-resampling each case's Monte Carlo batches (with replacement, at the same $M$ used for that case -- 100 for CovLow, 250 for CovHigh -- 30 resamples each) reproduces the reported sign in 25/30 resamples for CovLow (83\%) but only 17/30 for CovHigh (57\%). A simple chronological split of each case's batches into two halves flips the sign entirely for both. CovLow's bias direction is therefore reasonably well supported by the available statistics; CovHigh's is close to indistinguishable from batch-to-batch noise at $M=250$. This does not overturn the underlying practical conclusion; if anything it reinforces it. The direction of the diagonal approximation's bias is not fixed a priori, can be unresolved even at a single, otherwise well-characterized statistics level, and must be checked (and, ideally, bootstrapped) per case. This is the central practical justification for the full Delta-method covariance propagation of Section \ref{sec:delta} over the diagonal shortcut.

\begin{table}[htbp!]
    \centering
    \setlength{\tabcolsep}{3.5pt}
    \begin{tabular}{l l c c c c}
        \toprule
        \textbf{Case} & \textbf{Group} & \textbf{Mean} & \textbf{\% neg.} & \textbf{\% pos.} & \textbf{$n$ pairs} \\
        \midrule
        \multirow{3}{*}{CovLow}
         & Overall          & $1.20\times10^{-17}$  & 59.37 & 38.38 & $2.137\times10^{9}$ \\
         & Same column ($j=l$)       & $2.24\times10^{-14}$  & 58.36 & 39.41 & $9.892\times10^{6}$ \\
         & Different column ($j\ne l$)  & $-9.20\times10^{-17}$ & 59.37 & 38.37 & $2.127\times10^{9}$ \\
        \midrule
        \multirow{3}{*}{CovHigh}
         & Overall          & $-1.13\times10^{-20}$ & 50.16 & 49.84 & $2.137\times10^{9}$ \\
         & Same column ($j=l$)       & $9.13\times10^{-17}$  & 48.03 & 51.97 & $9.892\times10^{6}$ \\
         & Different column ($j\ne l$)  & $-4.36\times10^{-19}$ & 50.17 & 49.83 & $2.127\times10^{9}$ \\
        \bottomrule
    \end{tabular}
    \caption{Sign structure of the off-diagonal \gls{fm} covariance, classified by same-source-column (competition) vs.\ different-source-column (inter-cycle) pairs.}
    \label{tab:res-sign-structure}
\end{table}

\begin{figure}[htbp!]
    \centering
    \includegraphics[width=\linewidth]{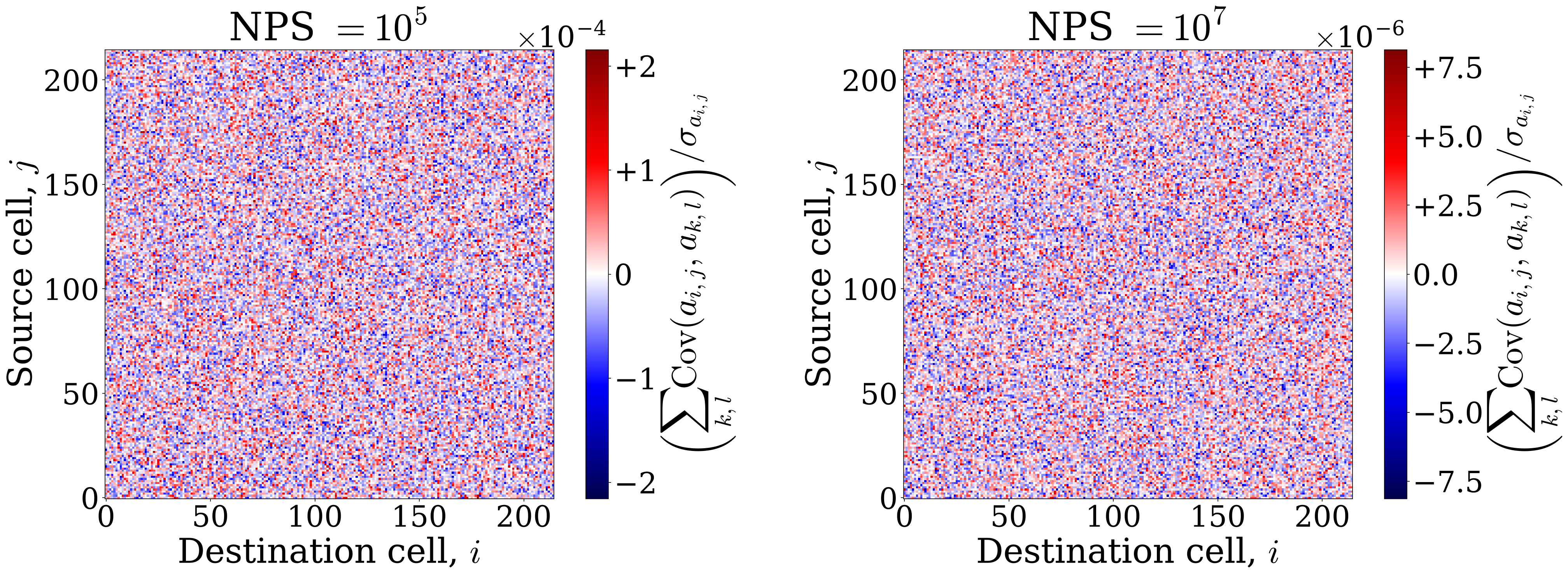}
    \caption{Diagonal-dominance diagnostic $D_{i,j}$ (Equation \ref{eq:cov-dominance}) for CovLow (left) and CovHigh (right), in the same $(i,j)$ layout as Figure \ref{fig:jacobian-k-pins-zoom}.}
    \label{fig:cov-dominance}
\end{figure}

Figure \ref{fig:cov-dominance} maps the per-coefficient diagonal-dominance diagnostic $D_{i,j}$ of Equation \ref{eq:cov-dominance} across the full $\nc\times\nc$ index grid, for both statistics levels. Both panels read as unstructured, symmetric speckle around zero rather than any smooth spatial pattern. This mirrors the near-coin-flip sign split of Table \ref{tab:res-sign-structure} carrying over to the per-coefficient level: 22,336 of 46,225 coefficients (48.3\%) are net positive for CovLow, 23,191 (50.2\%) for CovHigh (Table \ref{tab:res-cov-dominance}). The magnitude of $D_{i,j}$ is not small. Its population mean absolute value, $3.83\times10^{-5}$ (CovLow) and $1.44\times10^{-6}$ (CovHigh), actually \textit{exceeds} the coefficients' own mean standard deviation $\sigma_{i,j}$ when the latter is computed self-consistently from the same covariance matrix ($1.46\times10^{-6}$ for CovLow, $9.29\times10^{-8}$ for CovHigh), by a factor of $\approx26$ and $\approx15$, respectively. The fully dimensionless per-coefficient ratio $D_{i,j}/\sigma_{i,j}$, comparing each coefficient's aggregate off-diagonal covariance directly to its own variance, has a median of $33$ (CovLow) and $19$ (CovHigh), and exceeds unity for all but $2.4\%$ (CovLow) and $4.2\%$ (CovHigh) of the $\nc^2$ coefficients. This means a typical \gls{fm} coefficient's aggregate covariance with the other $\nc^2-1$ coefficients, not its own variance, dominates its row, often by more than an order of magnitude.

This row-wise dominance does not, however, translate into an equally large effect on the propagated \keff\ variance. Splitting Equation \ref{eq:var-keff}'s sandwich product into diagonal and off-diagonal contributions, and summing each coefficient's off-diagonal terms over its row, the magnitudes of these row-wise contributions add up to $2.60\times10^{-6}$ (CovLow) and $6.21\times10^{-9}$ (CovHigh): about $12.6\times$ and $7.5\times$ $\sigma^2_\text{diag}(k_\text{eff})$ itself. Summing the magnitudes of every individual off-diagonal term instead gives values roughly $140\times$ and $150\times$ larger, respectively. But the covariance's sign structure is close to a 50/50 mix (Table \ref{tab:res-sign-structure}), so these row-wise contributions largely cancel each other, and what actually survives in the signed sum is only $2.56\times10^{-8}$ (CovLow) and $-2.42\times10^{-11}$ (CovHigh): a cancellation factor (absolute row-wise sum over signed sum) of roughly $100\times$ and $260\times$, respectively. That surviving residual is exactly the modest, but non-trivial, $\epsilon_k$ of Table \ref{tab:res-cov-structure} (11.0\%/3.0\%). $D_{i,j}$ and $\epsilon_k$ therefore answer different questions: $D_{i,j}$ shows that a coefficient's raw, Jacobian-unweighted covariance with the rest of the matrix genuinely dominates its own variance; $\epsilon_k$ shows that once that same covariance is weighted by the \keff\ Jacobian and summed over the full $\nc^2\times\nc^2$ population, it mostly cancels, leaving only a modest net effect on the propagated variance. This near-total, near-50/50 cancellation is the underlying reason the \textit{Independent} estimator tracks \textit{Covariances} so closely in practice (5.7\% for CovLow, 1.5\% for CovHigh, Table \ref{tab:res-keff-sigma}). The individual coefficients are, by every pairwise and row-wise measure used in this section, very far from statistically independent, but that dependence is close enough to symmetric that assuming zero covariance introduces only a small error in the propagated \keff\ and source uncertainty. This is the quantitative basis for recommending the diagonal-covariance \textit{Independent} estimator as the practical default (Section \ref{sec:conclusions}), not evidence that the coefficients are, in any literal sense, close to independent.

\begin{figure}[htbp!]
    \centering
    \includegraphics[width=\linewidth]{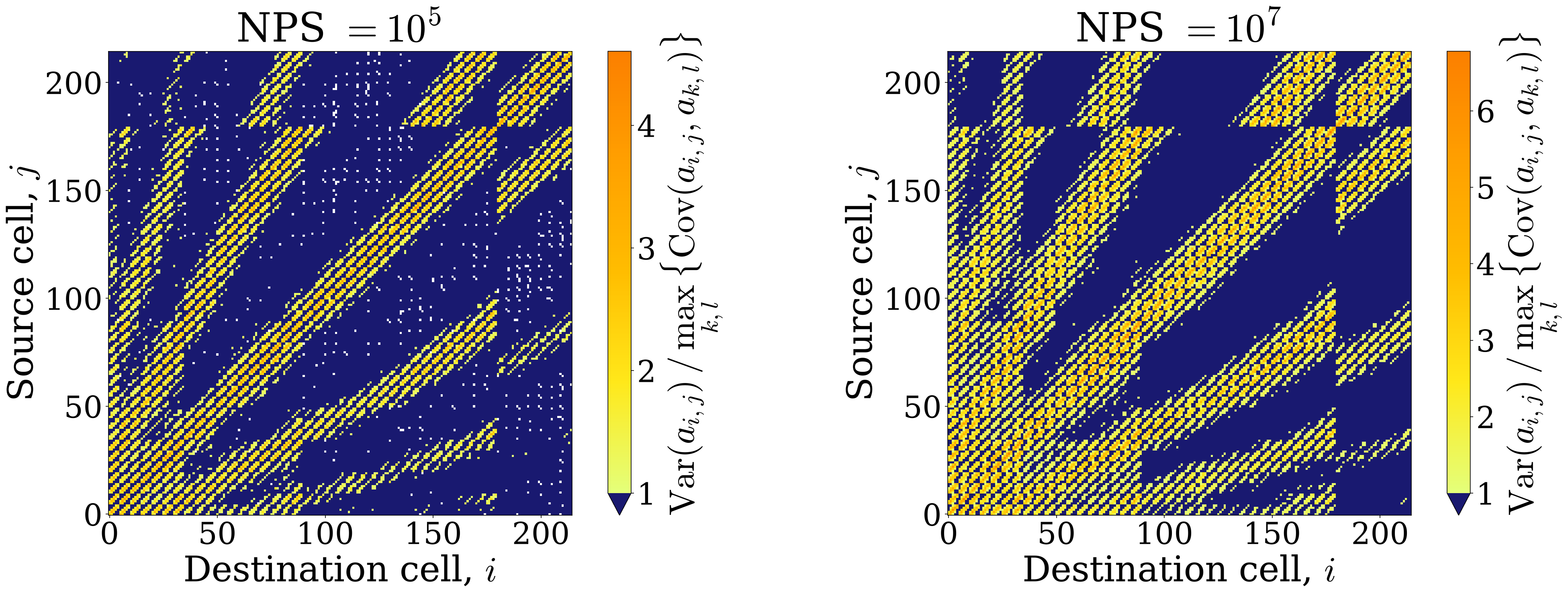}
    \caption{Ratio of each \gls{fm} coefficient's own variance to the largest covariance it has with any other coefficient, for CovLow (left) and CovHigh (right). Values below 1 (the largest single covariance exceeds the coefficient's own variance) are shown in dark navy.}
    \label{fig:variance-max-cov}
\end{figure}

The dominance is row-wise: it arises from a large number of small, similarly-signed covariance terms accumulating across the row, rather than from any single dominant pairwise term. This also explains the structure visible in Figure \ref{fig:variance-max-cov}, which maps each \gls{fm} coefficient's own variance relative to the \textit{largest single} covariance it has with any other coefficient (Section \ref{sec:metrics}). That ratio exceeds unity (colored) along most of the grid, and drops below unity (the largest single covariance exceeds the coefficient's own variance) only along a comparatively sparse set of diagonal-adjacent bands. Those bands align with the rod-major, axial-minor \gls{fm} index ordering of Figure \ref{fig:fm-index-order}. The strongest \textit{single} pairwise covariances still arise between axially- or radially-adjacent cells rather than between arbitrary, unrelated pairs.

\begin{table}[htbp!]
    \centering
    \begin{tabular}{l c c}
        \toprule
        \textbf{Quantity} & \textbf{CovLow} & \textbf{CovHigh} \\
        \midrule
        $n_+$ (Eq.\ \ref{eq:cov-dominance} $>0$)                     & 22{,}336              & 23{,}191              \\
        $n_-$ (Eq.\ \ref{eq:cov-dominance} $<0$)                     & 23{,}365              & 23{,}034              \\
        mean $|D|$                                    & $3.83\times10^{-5}$  & $1.44\times10^{-6}$  \\
        mean $\sigma_{i,j}$ (self-consistent)          & $1.46\times10^{-6}$  & $9.29\times10^{-8}$  \\
        mean $|D|$ / mean $\sigma_{i,j}$                & 26.3                  & 15.5                  \\
        median $|D_{i,j}/\sigma_{i,j}|$                & 32.9                  & 18.7                  \\
        frac.\ coefficients w/ $|D_{i,j}/\sigma_{i,j}|<1$ & 2.4\%               & 4.2\%                 \\
        mean Ind./Cov. (source $\sigma$)               & 1.009                 & 1.002                 \\
        $n$ cells: smaller $\sigma_{S}$ w/ full cov.   & 118                   & 108                   \\
        $n$ cells: larger $\sigma_{S}$ w/ full cov.    & 97                    & 107                   \\
        \bottomrule
    \end{tabular}
    \caption{Diagonal-dominance diagnostic $D_{i,j}$ (Equation \ref{eq:cov-dominance}), summarized over all 46,225 \gls{fm} coefficients, and the analogous \textit{Independent}-to-\textit{Covariances} (Section \ref{sec:estimators-compared}) source-$\sigma$ ratio, summarized over all 215 fission-source cells.}
    \label{tab:res-cov-dominance}
\end{table}

\begin{figure}[htbp!]
    \centering
    \includegraphics[width=\linewidth]{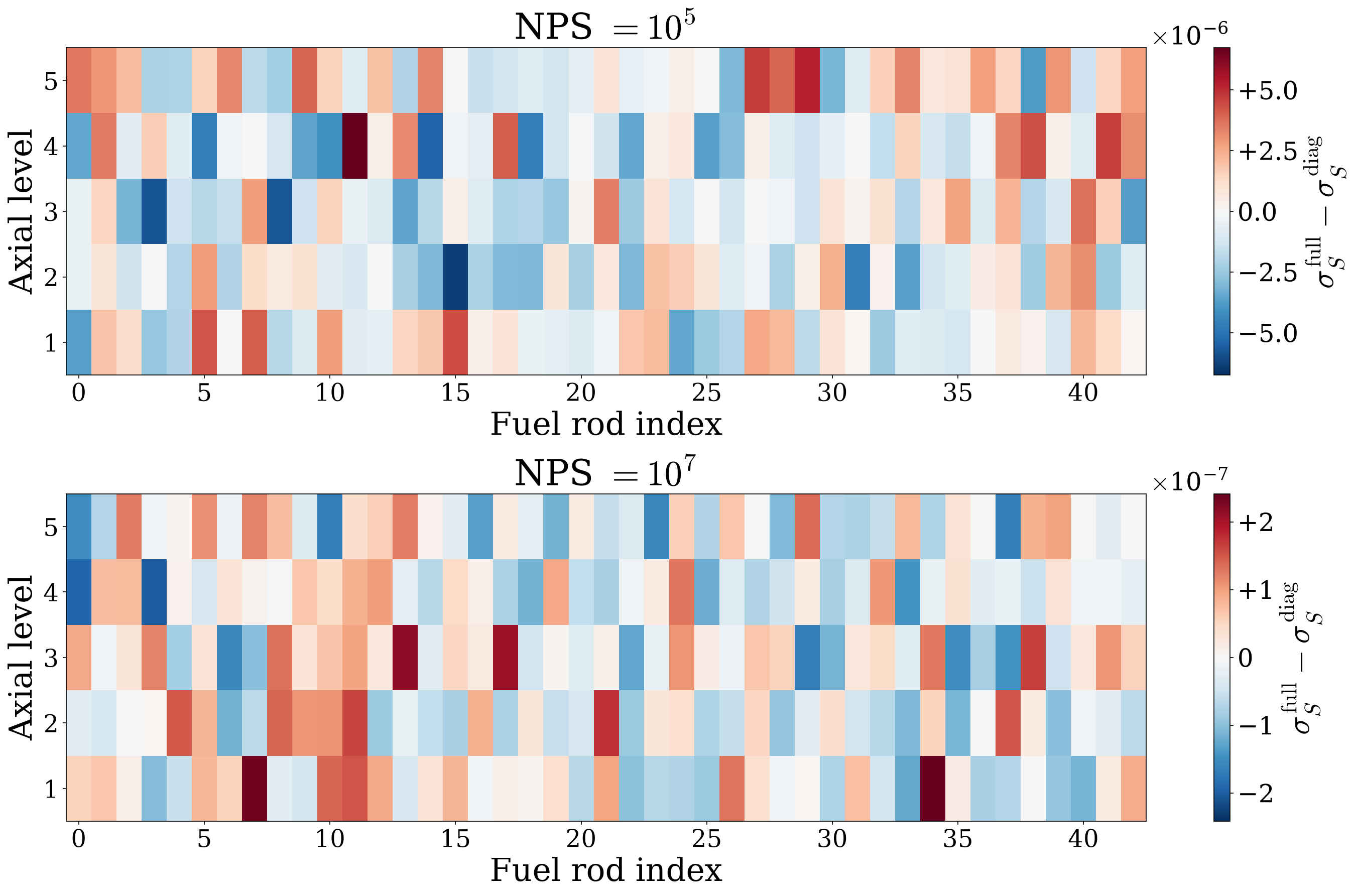}
    \caption{Difference between the full-covariance and diagonal-only fission-source uncertainty, $\sigma_S^\text{full}-\sigma_S^\text{diag}$, for CovLow (top) and CovHigh (bottom), reshaped into each cell's (fuel rod, axial level) position.}
    \label{fig:source-uncertainty-diff}
\end{figure}

On the fission source itself, Figure \ref{fig:source-uncertainty-diff} maps $\sigma_S^\text{full}-\sigma_S^\text{diag}$ across all 215 fuel cells (43 rods $\times$ 5 axial levels) for each statistics level, using an independent, symmetric color scale per case since CovLow and CovHigh differ by roughly an order of magnitude. The two approximations disagree cell by cell rather than in one uniform direction: 118 of 215 cells (54.9\%) show a smaller full-covariance $\sigma_S$ for CovLow, 108 of 215 (50.2\%) for CovHigh (Table \ref{tab:res-cov-dominance}), with no evident concentration in any particular radial or axial region. This is in line with the same near-random sign structure found for the \gls{fm} covariance itself. The per-cell average of this ratio (mean Ind./Cov. of 1.009 for CovLow, 1.002 for CovHigh) is close to, but not identical to, the sum-of-stds \textit{Ind./Cov. ratio} of Table \ref{tab:res-source-sigma} (1.005/0.9998): the two differ because the per-cell average weights every cell equally, whereas the sum-of-stds ratio is implicitly weighted toward the highest-source cells.

\subsection{Spatial Sensitivity Structure}
\label{sec:res-spatial}

\begin{figure}[htbp!]
    \centering
    \includegraphics[width=\linewidth]{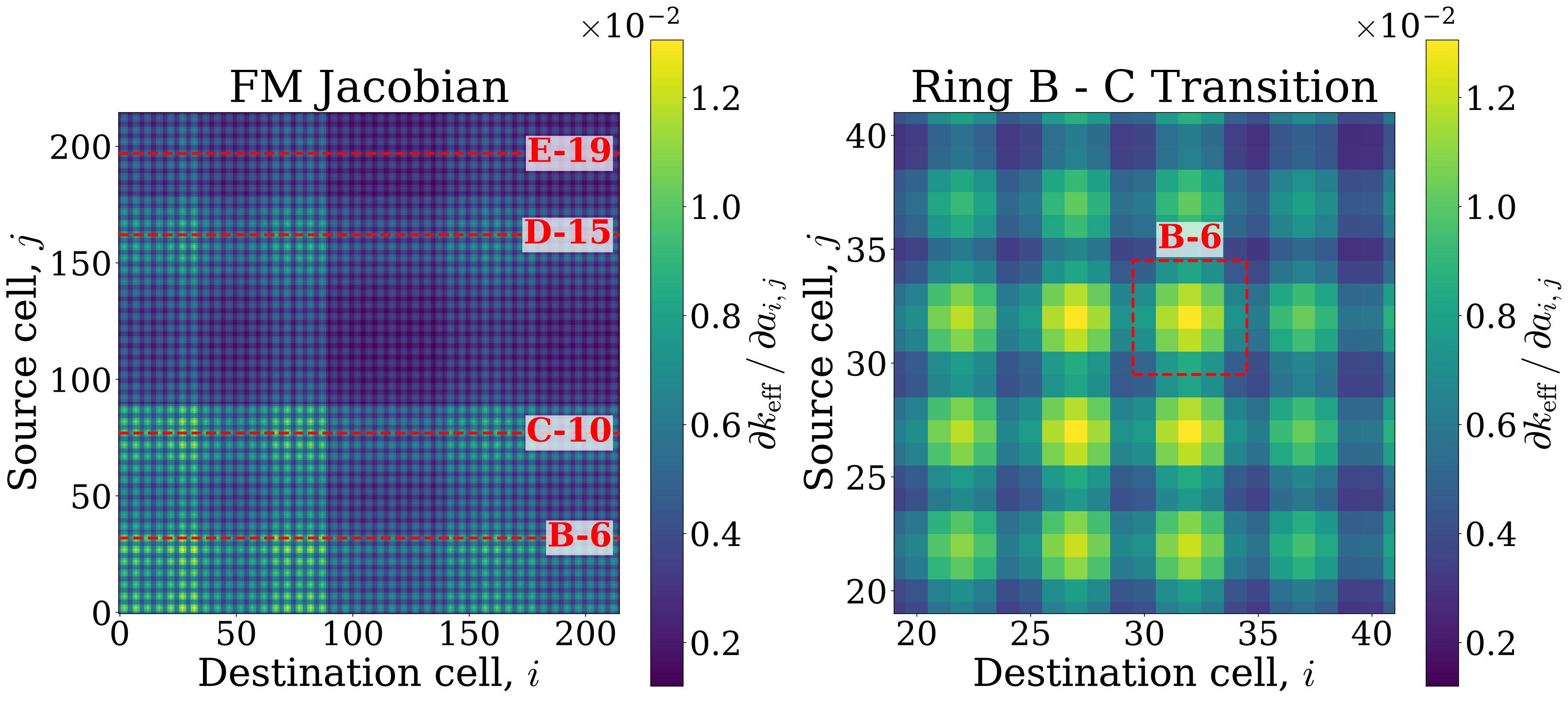}
    \caption{\keff\ sensitivity coefficients $\partial\keff/\partial\aij$ (Equation \ref{eq:sensitivity}), CovHigh. Left: full map, with the axial-center row of pins B-6, C-10, D-15, and E-19 marked. Right: zoomed into the ring B--C transition (indices 19--41), with the five cells of rod B-6 (indices 30--34) enclosed by the red dashed square.}
    \label{fig:jacobian-k-pins-zoom}
\end{figure}

Figure \ref{fig:jacobian-k-pins-zoom} (left panel) plots the \keff\ sensitivity coefficients $\partial\keff/\partial\aij=S^*_{0,i}S_{0,j}$ of Equation \ref{eq:sensitivity} for CovHigh, with the axial-center row of four representative pins spanning rings B--E marked. As a rank-one outer product of the fission source and its adjoint, the map factors into a row-wise pattern (set by $S_{0,j}$) and a column-wise pattern (set by $S^*_{0,i}$): sensitivity is highest when both the source cell $j$ and the destination cell $i$ fall within the central rings (indices below approximately 90, i.e., rings A--C of Figure \ref{fig:fm-index-order}), and drops sharply once either index falls in the outer rings, consistent with fissions born in the well-thermalized core center contributing most strongly to \keff\ regardless of where the induced fission is ultimately counted. Each marked pin row sits on a local sensitivity peak within its ring, matching the expectation that a pin's axial center carries the highest local importance along its fuel column (relative to its own top and bottom, where axial leakage reduces it). The right panel zooms into the ring B--C transition (indices 19--41). There, the sensitivity map's lattice-like block structure is directly visible: one bright cell per fuel pin, repeating with the 5-cell-per-rod axial spacing of the \gls{fm} index ordering. The red dashed square encloses the five axial cells of rod B-6.

\begin{figure}[htbp!]
    \centering
    \includegraphics[width=\linewidth]{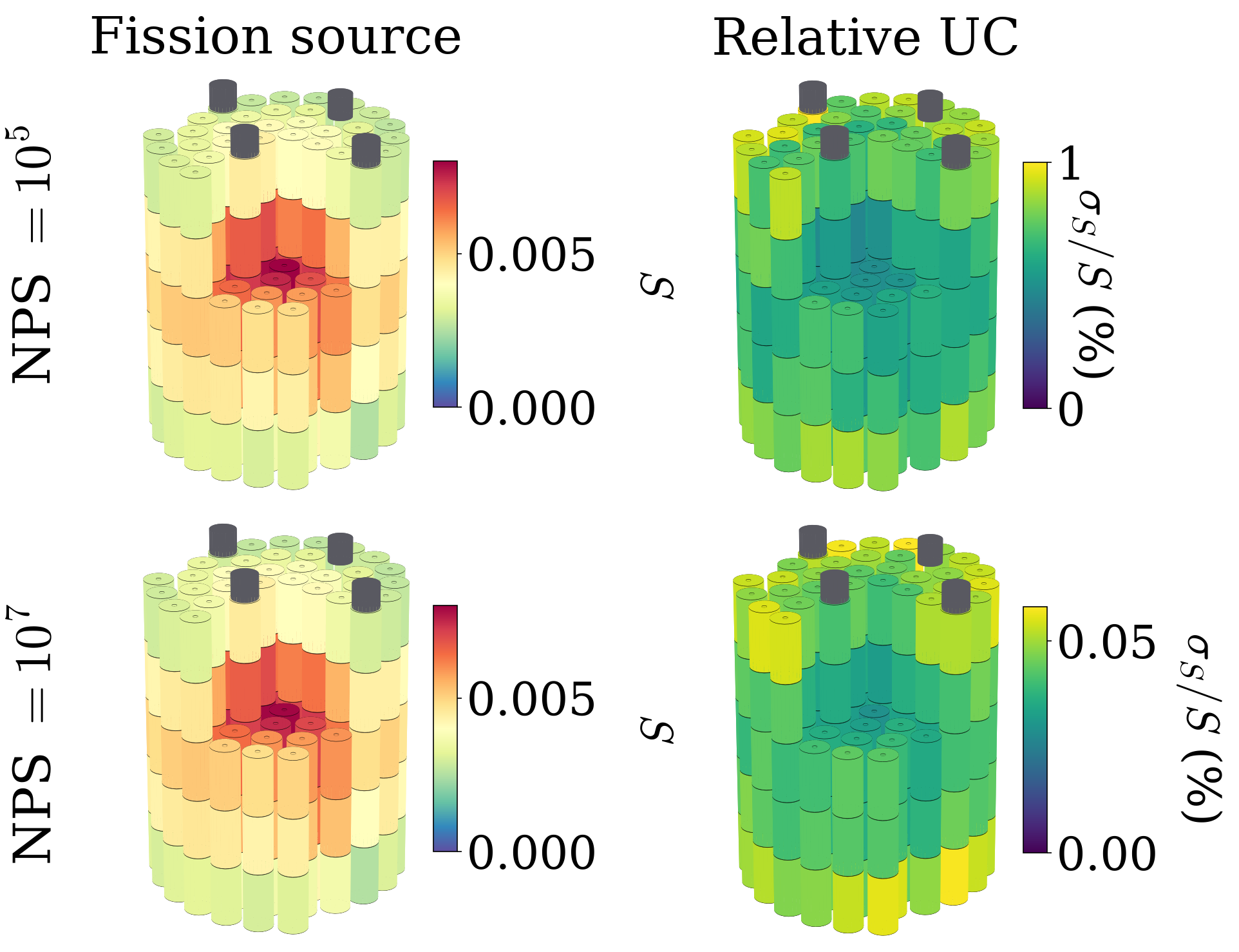}
    \caption{Fission source (left) and its relative uncertainty (right) on the \jst\ core geometry, for CovLow (top) and CovHigh (bottom).}
    \label{fig:source-3d}
\end{figure}

Figure \ref{fig:source-3d} shows the fission source and its relative uncertainty, $\sigma_S/|S|$, mapped onto the actual \jst\ core geometry for both statistics levels. The source shape itself, shown on one shared color scale, is essentially identical between CovLow and CovHigh, as expected since it is a property of the reactor configuration rather than of the Monte Carlo statistics: it peaks in the central fuel rods and decreases smoothly toward the core periphery. The relative uncertainty (each case shown on its own scale, since CovHigh's is roughly an order of magnitude smaller than CovLow's) is comparatively uniform across the bulk of the core, with a modest increase toward the top axial level and the radial periphery, in keeping with the lower absolute fission source, and correspondingly lower tally counts, in those cells.

\subsection{Independent Realization Cross-Check}
\label{sec:res-independent}

\begin{figure}[htbp!]
    \centering
    \includegraphics[width=\linewidth]{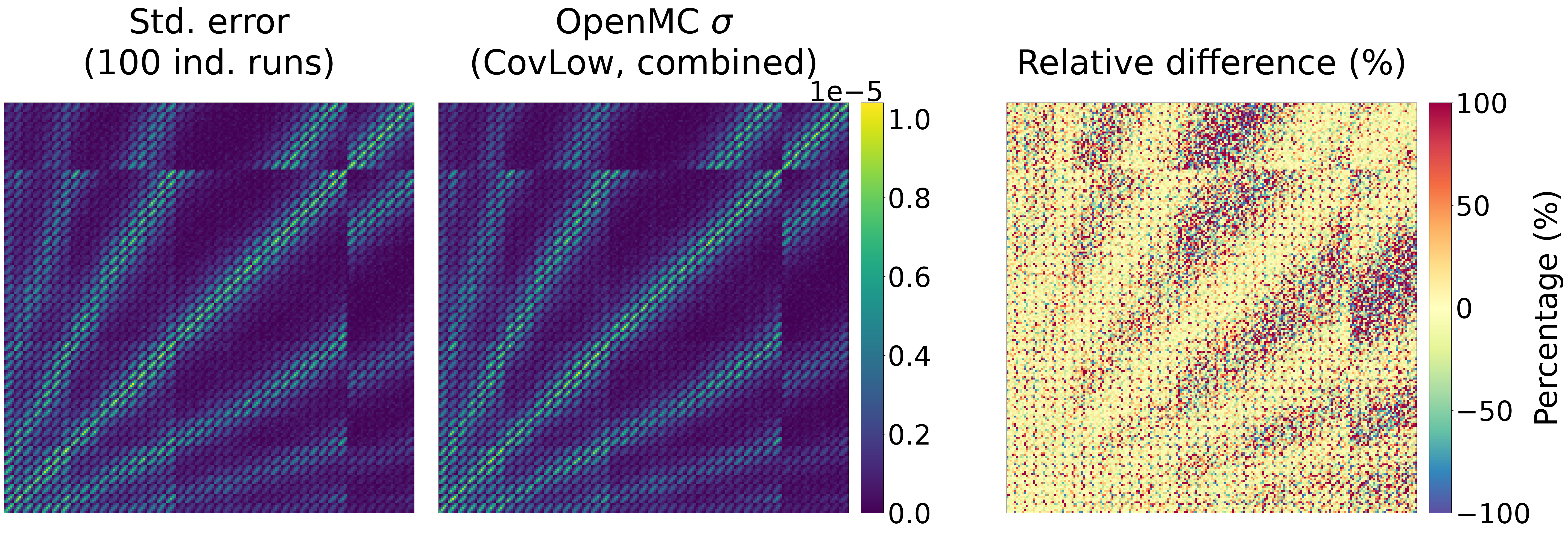}
    \caption{Raw-tally standard error of the mean from the 100 independent OpenMC realizations (left) vs.\ OpenMC's own combined-batch standard deviation from the reference CovLow run (center), and their pairwise relative difference (right).}
    \label{fig:independent-runs}
\end{figure}

Figure \ref{fig:independent-runs} compares, element by element, the standard error of the raw tally (the sample standard deviation across the 100 independent OpenMC realizations of the Indp case, divided by $\sqrt{100}$) against OpenMC's own internally-reported combined-batch standard deviation from the reference run underlying CovLow. The two uncertainty maps are visually indistinguishable, both reproducing the same diagonal-banded spatial pattern seen throughout this work, while the pairwise relative difference is unstructured noise spanning roughly $\pm100\%$. This is expected: each pixel compares two independently-estimated uncertainties from only $M=100$ samples each, so even elements with identical true uncertainty can differ substantially pixel-by-pixel due to sampling noise in the ratio itself. This qualitative agreement supports the \textit{Runs} comparison of Table \ref{tab:res-keff-sigma}: OpenMC's own internally-computed combined-batch statistics agree with a fully independent-realization estimate of the same underlying uncertainty.

\subsection{Global Fixed Source Generation Uncertainty}
\label{sec:res-fixed-source}

Sections \ref{sec:res-reference}--\ref{sec:res-independent} characterized the \gls{fm} covariance structure exclusively for coefficients generated from a criticality (eigenvalue) calculation, in which the fission source is resampled generation-to-generation from the previous batch's fission bank. This section evaluates the FixedSrc case of Table \ref{tab:fm-calc-summary}, in which every one of the \nc\ fuel cells is instead an independent, equal-strength neutron source, all active simultaneously within a single run and never resampled (Section \ref{sec:covariances}). This tests whether that difference in generation strategy changes the \gls{fm} covariance structure and the resulting diagonal-approximation bias, as anticipated in Section \ref{sec:metrics}. The FixedSrc calculation is described in Section \ref{sec:fm-modeling-fixed}.

Solving Equation \ref{eq:fmequation} for the mean FixedSrc fission matrix gives $\keff=0.99587$, compared to the $\keff=1.00779$ obtained from the criticality-based CovHigh case (Table \ref{tab:res-keff}) for the identical \gls{aro}, $N_\text{ax}=5$ geometry. Table \ref{tab:res-fixed-source-keff} summarizes both values with their \textit{Covariances} $\sigma_k$. The resulting difference is $-1192$ pcm. This discrepancy is expected since the fixed external source's spatial and spectral shape differs from the criticality case's self-consistent, converged fission source. Particularly, with $N_\text{ax}=5$, the assumption that the fission source is uniform over the axial direction introduces a significant bias. This bias has been characterized in detail in previous \gls{fm} analyses and is outside the scope of this work, which focuses on the propagated statistical uncertainty. As such, it is kept separate from the $\sigma_k$ comparison that follows, in the same way Section \ref{sec:res-crd-uq} separates the \gls{fm}-\gls{crd} interpolation bias from its propagated uncertainty.

\begin{table}[htbp!]
    \centering
    \begin{tabular}{l c c}
        \toprule
        \textbf{Case} & \keff & \boldmath$\sigma_k$ \textbf{(pcm)} \\
        \midrule
        CovHigh  & 1.00779 & 3  \\
        FixedSrc & 0.99587 & 12 \\
        \midrule
        Diff. (FixedSrc $-$ CovHigh), pcm & \multicolumn{2}{c}{$-1192$} \\
        \bottomrule
    \end{tabular}
    \caption{\gls{fm}-derived \keff\ and rigorous \textit{Covariances} $\sigma_k$ for the CovHigh and FixedSrc cases, and the \keff\ difference between them. Both cases use the \gls{aro}, $N_\text{ax}=5$ geometry.}
    \label{tab:res-fixed-source-keff}
\end{table}

Table \ref{tab:res-fixed-source-uq} reports $\sigma_k$ and the source \gls{rsu} for FixedSrc using the same four estimators applied to the criticality cases (\textit{Covariances}, \textit{Independent}, \textit{Resample}, \textit{Runs}; Section \ref{sec:estimators-compared}). \textit{Independent}, \textit{Resample}, and \textit{Runs} all sit \textit{below} \textit{Covariances} here (0.914, 0.916, and 0.896 for $\sigma_k$; 0.959, 0.959, and 0.998 for the source \gls{rsu}). This is opposite to the ``always conservative'' expectation for fixed-source generation raised in Section \ref{sec:metrics}. Critically, \textit{Independent}, \textit{Resample}, and \textit{Runs} all agree closely with \textit{one another} (within 2.0\% for $\sigma_k$, 4.0\% for the source \gls{rsu}), despite \textit{Runs} making no covariance assumption (Section \ref{sec:estimators-compared}) while \textit{Independent} and \textit{Resample} assume a diagonal covariance. This rules out a failure of the diagonal approximation itself as the explanation, since an assumption-free, fully independent benchmark reaches essentially the same answer. The discrepancy is therefore concentrated in \textit{Covariances}, not in the cheap estimators it is meant to validate, motivating the covariance-structure comparison below.

\begin{table}[htbp!]
    \centering
    \begin{tabular}{l c c}
        \toprule
        \textbf{$\sigma$ estimator} & \boldmath$\sigma_k$ \textbf{(pcm)} & \textbf{Source RSU} \\
        \midrule
        Covariances   & 12    & $1.76\times10^{-3}$ \\
        \midrule
        Independent   & 11    & $1.69\times10^{-3}$ \\
        Ind./Cov.     & 0.914 & 0.959 \\
        \midrule
        Resample      & 11    & $1.69\times10^{-3}$ \\
        Res./Cov.     & 0.916 & 0.959 \\
        \midrule
        Runs          & 11    & $1.76\times10^{-3}$ \\
        Runs./Cov.    & 0.896 & 0.998 \\
        \bottomrule
    \end{tabular}
    \caption{$\sigma_k$ (pcm) and source \gls{rsu} for the FixedSrc case, estimated via \textit{Covariances} (rigorous Delta method), \textit{Independent} (diagonal-covariance Delta method), \textit{Resample} (Monte Carlo resampling, $M=10{,}000$, without the criticality-only renormalization of Equation \ref{eq:renorm-crit}, per Section \ref{sec:covariances}), and \textit{Runs} (100 independent replicate realizations, Table \ref{tab:fm-calc-summary}), all reported relative to \textit{Covariances} as in Table \ref{tab:res-keff-sigma}.}
    \label{tab:res-fixed-source-uq}
\end{table}

Table \ref{tab:res-fixed-source-cov-structure} repeats the covariance-structure diagnostics of Section \ref{sec:res-cov-structure} for FixedSrc, alongside the CovLow/CovHigh values of Tables \ref{tab:res-cov-structure} and \ref{tab:res-cov-dominance} for direct comparison. Unlike those two cases, where the reported quantities approximate $\text{Cov}(a_{i,j},a_{k,l})$ via the raw tally covariance, FixedSrc's values are exact, since $a_{i,j}=T_{i,j}$ identically here (Section \ref{sec:covariances}). Almost every diagnostic is larger in magnitude for FixedSrc than for either criticality case: the diagonal-dominance diagnostic's population mean, $\text{mean}\,|D|$ (Equation \ref{eq:cov-dominance}), is 30 times CovLow's and nearly 800 times CovHigh's. This difference is expected given the lower particle count and, hence, statistics. The propagated-variance error $\epsilon_k$ (Equation \ref{eq:matrix-error}) is 16.5\%, larger than either criticality case's 11.0\%/3.0\%. The sign split, however, remains a near-even coin flip for both the same-column and different-column groups (50.1\%/49.9\% and 51.4\%/48.6\% negative/positive, respectively, out of 9.89 million and 2.13 billion pairs), just as it was for the criticality cases (Table \ref{tab:res-sign-structure}).

The same row-wise-dominance-versus-cancellation picture found for CovLow/CovHigh holds for FixedSrc as well. FixedSrc's mean $|D|$ exceeds its own self-consistent mean $\sigma_{i,j}$ by a factor of $17.1$, comparable to the criticality cases' $26.3\times$/$15.5\times$; the fully dimensionless per-coefficient ratio has a median of $20.2$ and drops below unity for only $3.7\%$ of coefficients, again in the same range as CovLow's $2.4\%$ and CovHigh's $4.2\%$. As before, this large row-wise dominance mostly cancels once weighted by the \keff\ Jacobian and summed: the absolute row-wise off-diagonal contribution to $\sigma^2_\Delta(\keff)$ is $9.51\times10^{-8}$ (about $8\times$ $\sigma^2_\text{diag}(k_\text{eff})$) but only $2.35\times10^{-9}$ survives in the signed sum, a cancellation factor of $\approx40\times$. This is smaller than CovLow's $\approx102\times$ or CovHigh's $\approx257\times$, and tracks FixedSrc's larger $\epsilon_k$ (16.5\% vs.\ 11.0\%/3.0\%): less cancellation among the Jacobian-weighted off-diagonal terms leaves more of the raw row-wise dominance to survive into the propagated \keff\ variance. Even so, that weaker cancellation still keeps the practical effect small: \textit{Independent} and \textit{Covariances} agree to within 8.6\% for $\sigma_k$ for FixedSrc (Table \ref{tab:res-fixed-source-uq}), comparable to the largest gap already found for the criticality cases (5.7\%, Table \ref{tab:res-keff-sigma}). The pattern is consistent across all three cases characterized in this work: substantial row-wise covariance dominance, largely canceled by the near-50/50 sign split once weighted by the propagation Jacobian and summed, leaving the diagonal-covariance \textit{Independent} approximation a reliable practical stand-in for the costly explicit full-covariance calculation.

\begin{table}[htbp!]
    \centering
    \begin{tabular}{l c c c}
        \toprule
        \textbf{Quantity} & \textbf{CovLow} & \textbf{CovHigh} & \textbf{FixedSrc} \\
        \midrule
        Off-diag.\ energy fraction, $\eta$        & 0.9978                & 0.9946                & 0.9954                \\
        $\max|\rho|$                                & 1                     & 0.573                 & 0.928                 \\
        $\epsilon_k$ (Eq.\ \ref{eq:matrix-error})   & 11.0\%                & 3.0\%                 & 16.5\%                \\
        mean $|D|$ (Eq.\ \ref{eq:cov-dominance})    & $3.83\times10^{-5}$   & $1.44\times10^{-6}$   & $1.14\times10^{-3}$   \\
        mean $\sigma_{i,j}$ (self-consistent)        & $1.46\times10^{-6}$   & $9.29\times10^{-8}$   & $6.64\times10^{-5}$   \\
        mean $|D|$ / mean $\sigma_{i,j}$              & 26.3                  & 15.5                  & 17.1                  \\
        median $|D_{i,j}/\sigma_{i,j}|$              & 32.9                  & 18.7                  & 20.2                  \\
        frac.\ coefficients w/ $|D_{i,j}/\sigma_{i,j}|<1$ & 2.4\%             & 4.2\%                 & 3.7\%                 \\
        $n_+$ (Eq.\ \ref{eq:cov-dominance} $>0$)     & 22{,}336              & 23{,}191              & 23{,}136              \\
        $n_-$ (Eq.\ \ref{eq:cov-dominance} $<0$)     & 23{,}365              & 23{,}034              & 23{,}089              \\
        Off-diag.\ contrib.\ to $\sigma^2_\Delta(\keff)$, abs.\ row-wise sum & $2.60\times10^{-6}$ & $6.21\times10^{-9}$ & $9.51\times10^{-8}$ \\
        Off-diag.\ contrib.\ to $\sigma^2_\Delta(\keff)$, signed   & $2.56\times10^{-8}$ & $-2.42\times10^{-11}$ & $2.35\times10^{-9}$ \\
        Cancellation factor (abs.\ row-wise/signed) & $102\times$           & $257\times$           & $40\times$            \\
        Same-column mean cov.                       & $2.24\times10^{-14}$  & $9.13\times10^{-17}$  & $2.28\times10^{-11}$  \\
        Diff.-column mean cov.                      & $-9.20\times10^{-17}$ & $-4.36\times10^{-19}$ & $-9.50\times10^{-14}$ \\
        \bottomrule
    \end{tabular}
    \caption{FM covariance-structure diagnostics of Sections \ref{sec:res-cov-structure} and \ref{sec:metrics} for FixedSrc, alongside the criticality-case values of Tables \ref{tab:res-cov-structure}, \ref{tab:res-sign-structure}, and \ref{tab:res-cov-dominance}. The self-consistent $\sigma_{i,j}$, per-coefficient dominance ratio, and cancellation-factor rows are the same diagnostics introduced for CovLow/CovHigh above, computed identically for FixedSrc. The row-wise sum adds the magnitudes of each coefficient's net off-diagonal contribution to $\sigma^2_\Delta(\keff)$, and the signed row is the plain sum of those contributions. The $\sigma_{i,j}$-based statistics exclude coefficients with exactly zero variance.}
    \label{tab:res-fixed-source-cov-structure}
\end{table}

This larger apparent magnitude is not evidence of a stronger true correlation; it is better explained by how severely underdetermined the covariance estimate is at the available statistics. $\bm{\Sigma}_A$ is estimated here from only $M=215$ batch realizations of a $\nc^2\times\nc^2=46{,}225\times46{,}225$ object, the same order of underdetermination as the criticality cases ($M=100$/$250$), but for a system where, as argued in Section \ref{sec:metrics}, the true covariance is expected to be smaller still: with no fission-bank resampling, different-source-column pairs have no shared mechanism linking them at all, and the same-source-column (competition) mechanism is bounded by a simple multinomial argument. For a fixed source column $j$ realizing $N_j$ Monte Carlo particles in a batch ($N_j\approx\text{NPS}/\nc\approx4{,}651$ for FixedSrc), each particle contributes to at most one destination cell's tally, so $\{T_{i,j}\}_i$ is a multinomial-type allocation across destinations for that batch. The covariance between two categories of a multinomial distribution is standard \cite{johnsonDiscreteMultivariateDistributions1997,casellaStatisticalInference2024}. Normalizing the counts by $N_j$ gives:
\begin{equation}
    \label{eq:multinomial-cov}
    \text{Cov}(a_{i,j},a_{k,j}) \approx -\frac{a_{i,j}\,a_{k,j}}{N_j}, \qquad i\neq k,
\end{equation}
an intrinsically tiny, noise-free quantity: evaluating Equation \ref{eq:multinomial-cov} for the two most strongly-coupled cells sharing a source gives $\approx-5\times10^{-7}$, and for more weakly-linked pairs, several orders of magnitude smaller still ($10^{-9}$--$10^{-11}$). Different-source-column pairs are expected to be exactly independent by this same argument, since they are computed from disjoint sets of Monte Carlo particles with no shared history; OpenMC's own multi-source sampling, however, assigns each particle to one of the \nc\ sources via an independent random draw rather than a fixed count per source, so a small residual anti-correlation of order $1/\nc\approx0.005$ arises from every source column drawing from the same total per-batch particle budget. Both effects are several orders of magnitude below the sample-covariance estimator's own noise floor at $M=215$: this is consistent with, and offered as the explanation for, the near-coin-flip sign split and larger apparent $D_{i,j}$/$\epsilon_k$ values of Table \ref{tab:res-fixed-source-cov-structure}, rather than a genuine departure from the negative-competition mechanism expected in Section \ref{sec:metrics}.

Directly confirming Equation \ref{eq:multinomial-cov}'s prediction empirically, rather than accepting it analytically, would require the sample-covariance estimator's standard error to fall below these tiny true values. For a given coefficient pair, this standard error scales as $1/\sqrt{M}$, so resolving even the two most strongly-coupled cells above at 95\% confidence needs approximately $6{,}200$ batches, about 29 times the $M=215$ used here; more weakly-linked pairs need up to several million. Reaching that many independent batches within reasonable wall-clock time is not simply a matter of running the same calculation longer: it would require restructuring how OpenMC accumulates fission-matrix tallies in the first place, from per-batch-averaged statistics (Equation \ref{eq:renorm-crit}) to true per-particle statistics, so that each of the $\text{NPS}\times\text{NAC}$ individual particle histories, rather than each batch, becomes one independent sample. Such a change to OpenMC's own tallying implementation is beyond the scope of this work and is proposed as future work in Section \ref{sec:conclusions}.

None of this changes the practical conclusion, however. Table \ref{tab:res-fixed-source-uq} shows \textit{Covariances}, \textit{Independent}, \textit{Resample}, and \textit{Runs} agreeing to within 10.4\% for $\sigma_k$ and 4.1\% for the source \gls{rsu} -- larger than, but the same order as, the largest spread already found between estimators for the criticality cases (5.7\%, Table \ref{tab:res-keff-sigma}), and still small enough not to change the practical recommendation. Whether or not the small, theoretically-expected same-column covariance of Equation \ref{eq:multinomial-cov} can be rigorously confirmed at achievable statistics therefore does not materially change the propagated uncertainty estimate for this generation strategy: the diagonal-covariance \textit{Independent} method remains a safe, practical default regardless of whether the \gls{fm} coefficients are generated from a criticality or a global fixed-source calculation.

\subsection{FM-CRd Uncertainty Propagation}
\label{sec:res-crd-uq}

Having established in Section \ref{sec:res-uncertainty} that the diagonal-covariance and resampling estimators closely track the rigorous Delta method when the \gls{fm} is evaluated directly, this section turns to the \gls{fm}-\gls{crd} case introduced in Section \ref{sec:fm-modeling-crd}. With this approach, the configuration is instead assembled from interpolated and combined database entries, the mode of use that motivates most practical \gls{fm}-based tools in the first place. Table \ref{tab:res-crd} validates the \gls{fm}-\gls{crd} combination $\mathbf{A}_\text{CRd}$ of Section \ref{sec:metrics} against the exact, fully-perturbed reference (all four rods simultaneous at random positions different from those used to obtain the \gls{fm} coefficients), using only the \gls{aro} baseline and the eight single-rod-perturbation \gls{fm} evaluations of Table \ref{tab:cr_pos} (boldface). $\mathbf{A}_\text{CRd}$ reproduces the `exact' \keff\ (i.e., using a \gls{fm} calculation with the rods at the exact test locations from Table \ref{tab:cr_pos}) to within 60 pcm and the exact source shape to within 0.7\% relative error. This comparison validates the \gls{fm}-\gls{crd} point-estimate accuracy itself and is consistent with the more detailed \gls{fm}-\gls{crd} validation work \cite{mascolinoDevelopmentValidationNew2022}. As noted in Section \ref{sec:fm-modeling-crd}, propagating uncertainty through the \gls{fm}-\gls{crd} interpolation and combination steps relies on the simplified, zero-covariance \gls{uq} approaches of Section \ref{sec:simplified_methods} rather than the full Delta-method covariance, given the prohibitive size of the explicit $\bm{\Sigma}_A$ at $N_\text{ax}=15$.

\begin{table}[htbp!]
    \centering
    \begin{tabular}{l c c c}
        \toprule
        \textbf{Variant} & \keff & \textbf{Diff. (pcm)} & \textbf{Src.\ err.\ (\%)} \\
        \midrule
        $\mathbf{A}_\text{CRd}$     & 0.95398 & $-60$ & 0.653 \\
        Exact            & 0.95458 & 0     & 0     \\
        \bottomrule
    \end{tabular}
    \caption{\gls{fm}-\gls{crd} validation: $\mathbf{A}_\text{CRd}$ (the sum of already-renormalized single-rod \glspl{fm}) against the exact, fully-perturbed reference. ``Diff.'' and ``Src.\ err.'' are $\mathbf{A}_\text{CRd}$'s difference from the exact \keff\ (pcm) and source relative error (\%), respectively.}
    \label{tab:res-crd}
\end{table}

\begin{figure}[H]
    \centering
    \includegraphics[width=1\linewidth]{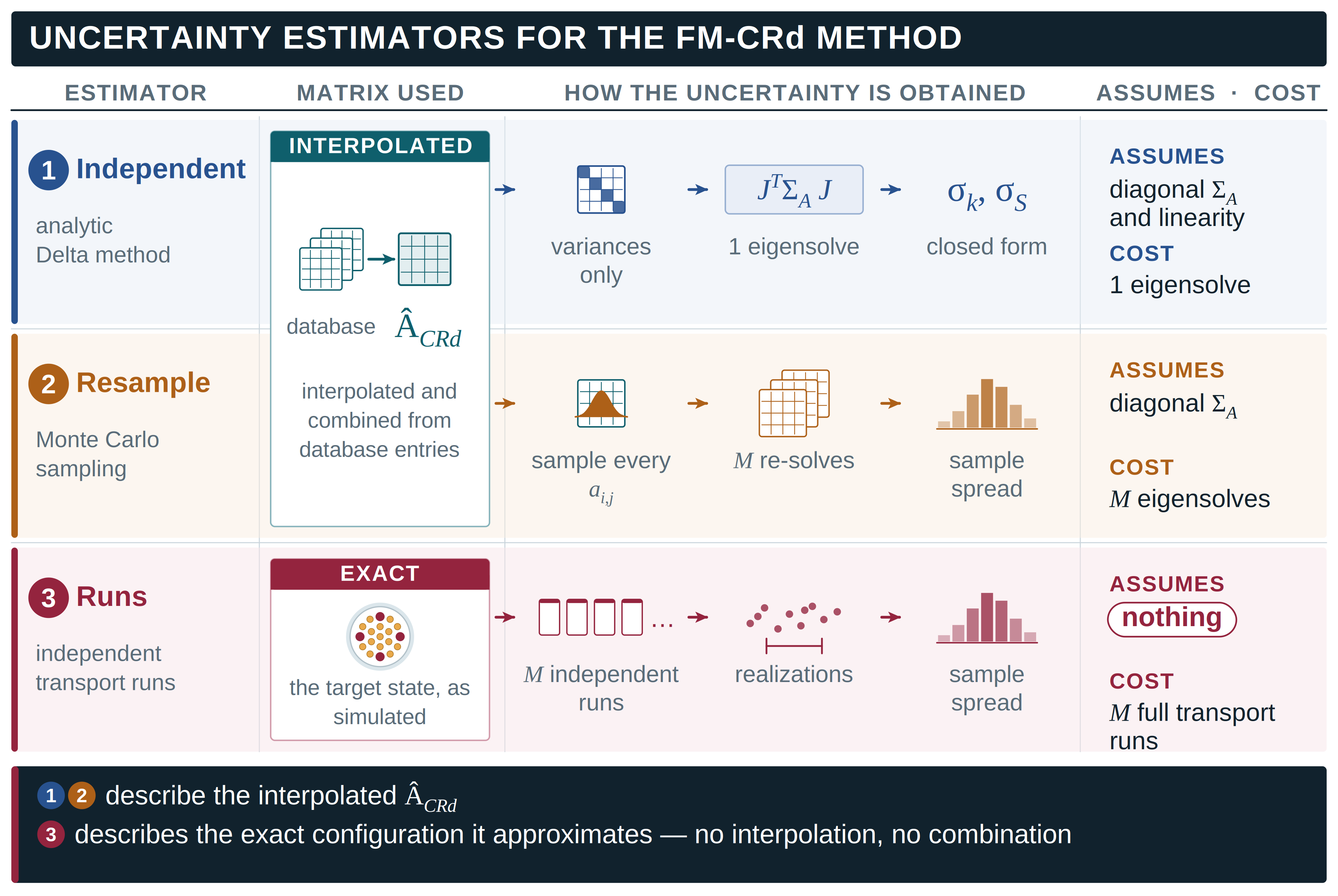}
    \caption{Schematic comparison of the three uncertainty estimators evaluated for the \gls{fm}-\gls{crd} target configuration (all $N_\text{CR}$ rods at arbitrary, between-node positions)}
    \label{fig:crd-estimators}
\end{figure}

Beyond the point-estimate validation of Table \ref{tab:res-crd}, the \gls{fm}-\gls{crd} combination's propagated uncertainty is evaluated using the same \textit{Independent}, \textit{Resample}, and \textit{Runs} estimators of Section \ref{sec:estimators-compared}. As already noted in Section \ref{sec:fm-modeling-crd}, no rigorous \textit{Covariances} estimate is available for this case: the full Delta-method covariance $\bm{\Sigma}_A$ is intractable to store explicitly at $N_\text{ax}=15$ ($\nc=645$, so $\bm{\Sigma}_A$ would require storing $\nc^4\approx1.73\times10^{11}$ entries). \textit{Runs} is therefore used as the reference instead: unlike \textit{Independent} and \textit{Resample}, which both \textit{assume} a diagonal \gls{fm} covariance, \textit{Runs} has zero cross-realization covariance \textit{by construction} (Section \ref{sec:estimators-compared}), making it the most rigorous benchmark available in the absence of \textit{Covariances}. \textit{Resample} perturbs the nine base \glspl{fm} (one for \gls{aro}, and two per single \gls{cr} axial interpolation) needed to build $\mathbf{A}_\text{CRd}$ via Equation \ref{eq:crd}. Note that in the FM-CRd case, \textit{Runs} is chosen as the empirical standard error of the mean across $M=100$ independent OpenMC realizations of the exact, fully-perturbed configuration validated in Table \ref{tab:res-crd} (all four \glspl{cr} simultaneously at their real operating positions). Each realization run at the same $\text{NPS}=10^6$ as the production \gls{fm}-\gls{crd} case of Table \ref{tab:fm-calc-summary} but with only a single active batch and an independent seed. Unlike the \gls{aro} case of Section \ref{sec:res-uncertainty}, where \textit{Runs} and \textit{Independent} estimate the uncertainty of the exact same, directly-evaluated \gls{fm}, here \textit{Runs} instead measures the true statistical uncertainty of the averaged \keff\ and \fsv\ for the exact, non-interpolated configuration itself. This approach entirely bypasses the \gls{fm}-\gls{crd} interpolation and combination machinery that \textit{Independent} and \textit{Resample} propagate uncertainty through. Close agreement between \textit{Runs} and \textit{Independent} therefore validates more than the diagonal-covariance assumption alone: it would indicate that carrying that assumption through the \gls{fm}-\gls{crd} interpolation and combination steps introduces no additional uncertainty beyond what a direct, non-interpolated calculation of the same configuration already exhibits. Figure \ref{fig:crd-estimators} summarizes the differences between the three estimators for these \gls{fm}-\gls{crd} cases.

Table \ref{tab:res-crd-uq} reports $\sigma_k$ and the source \gls{rsu} (Equation \ref{eq:rsu}) for all three estimators, with \textit{Runs} as the reference for the reasons given above. \textit{Resample} agrees with \textit{Independent} to within 6.9\% for $\sigma_k$ and 5.8\% for the source \gls{rsu}. These results are comparable to the largest \textit{Independent}-to-\textit{Covariances} discrepancy observed for the \gls{aro} case (5.7\% for $\sigma_k$, Table \ref{tab:res-keff-sigma}), and consistent with \textit{Resample}'s additional resampling noise on top of the diagonal-covariance assumption it shares with \textit{Independent} (Section \ref{sec:estimators-compared}). 

The \textit{Ind./Runs} and \textit{Res./Runs} ratios in Table \ref{tab:res-crd-uq} are markedly farther from unity than the close, few-percent agreement between \textit{Independent} and the rigorous \textit{Covariances} estimator observed for the directly-evaluated \gls{aro} case (Table \ref{tab:res-keff-sigma}): \textit{Independent} exceeds \textit{Runs} by a factor of 1.98 for $\sigma_k$ and 2.41 for the source \gls{rsu}, and \textit{Resample} by a factor of 1.84 and 2.27, respectively. This gap has a specific, identifiable origin in Equation \ref{eq:crd-cov-diag} itself, not in a failure of either low-cost estimator. That equation, derived in Section \ref{sec:fm-modeling-crd} directly from Equation \ref{eq:crd-expanded}'s expansion of the \gls{fm}-\gls{crd} combination, gives the \gls{aro} baseline's own coefficient variance a leading weight of $(1-N_\text{CR})^2$: for the $N_\text{CR}=4$ simultaneously-perturbed rods of this test case, that is $(1-4)^2=9$, a ninefold amplification relative to the $\pm1$ weight a directly-evaluated \gls{fm} would carry.

Evaluating Equation \ref{eq:crd-cov-diag} numerically for this test case shows this is not a minor correction: the \gls{aro} term alone accounts for 79\% of \textit{Independent}'s total propagated $\sigma_k^2$ and source \gls{rsu}$^2$ here, with the eight interpolated single-rod-perturbation terms of Equation \ref{eq:crd-cov-diag}'s remaining two sums together contributing only the other 21\%. The \gls{aro} run's own, unamplified statistical noise ($\sigma_k\approx8.7$ pcm, backed out from its $26.0$ pcm amplified contribution) is in fact smaller than \textit{Runs}' 15 pcm. Note that the two are not compared at equal statistics: \textit{Runs} totals 100 active batches of $10^6$ particles, whereas each base \gls{fm} entering \textit{Independent} and \textit{Resample} has 250. If it scaled as $1/\sqrt{N_\text{batches}}$, the \textit{Runs} $\sigma_k$ at equal statistics would be about $15\sqrt{100/250}\approx9.5$ pcm, comparable to the \gls{aro} run's 8.7 pcm. This widens the gap between \textit{Independent} and \textit{Runs} without changing the conclusion. This corresponds specifically to the $(1-N_\text{CR})^2=9$ weight of Equation \ref{eq:crd-cov-diag}, not an unusually noisy underlying calculation, that inflates \textit{Independent}'s total to 29 pcm. \textit{Runs}, a single direct calculation of the target configuration, carries no such amplification.

\textit{Independent} and \textit{Resample} continuing to agree closely with \textit{each other} (within 6.9\% for $\sigma_k$ and 5.8\% for the source \gls{rsu}, as noted above) despite \textit{Resample} using no linearization at all confirms this amplification is a structural property of the \gls{fm}-\gls{crd} combination itself (Equation \ref{eq:crd-cov-diag}). Since the \gls{aro} weight grows as $(1-N_\text{CR})^2$, the same effect is expected to grow correspondingly for configurations combining more than four rods simultaneously. This gap is nonetheless a favorable outcome in practice: since the real system \textit{Runs} samples does not itself combine or interpolate, \textit{Independent} and \textit{Resample} overestimating it rather than underestimating it means the cheap, practical UQ estimate does not under-report the true uncertainty of the configuration it approximates. The diagonal-covariance approximation therefore remains a safe, if here less tight, choice for this more complex, interpolated use case.

\begin{table}[htbp!]
    \centering
    \begin{tabular}{l c c}
        \toprule
        \textbf{$\sigma$ estimator} & \boldmath$\sigma_k$ \textbf{(pcm)} & \textbf{Source RSU} \\
        \midrule
        Runs          & 15    & $3.25\times10^{-3}$ \\
        \midrule
        Independent   & 29    & $7.84\times10^{-3}$ \\
        Ind./Runs     & 1.98  & 2.41 \\
        \midrule
        Resample      & 27    & $7.38\times10^{-3}$ \\
        Res./Runs     & 1.84  & 2.27 \\
        \bottomrule
    \end{tabular}
    \caption{$\sigma_k$ (pcm) and source \gls{rsu} (Equation \ref{eq:rsu}) for the \gls{fm}-\gls{crd} case, estimated via \textit{Runs} (100 independent OpenMC realizations of the exact, fully-perturbed configuration, used as reference in place of the unavailable \textit{Covariances}), \textit{Independent} (diagonal-covariance Delta method), and \textit{Resample} (Monte Carlo resampling of the nine base \glspl{fm}), as defined in Section \ref{sec:estimators-compared}.}
    \label{tab:res-crd-uq}
\end{table}

\begin{figure}[htbp!]
    \centering
    \includegraphics[width=0.7\linewidth]{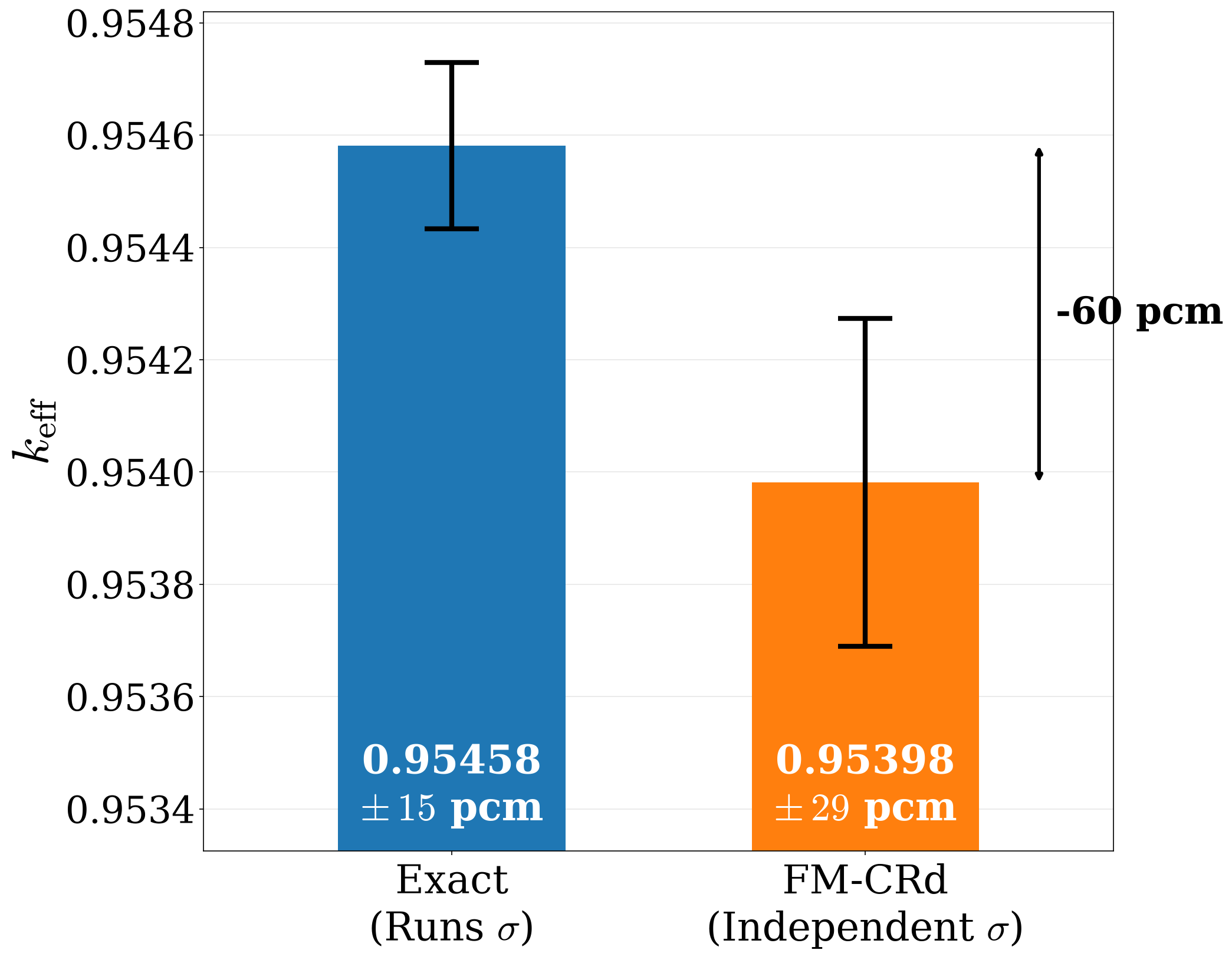}
    \caption{\keff\ of the exact, fully-perturbed configuration, with error bars showing the $\sigma_k$ of the \textit{Runs} estimator (15 pcm), and of the \gls{fm}-\gls{crd} combined case $\mathbf{A}_\text{CRd}$, with error bars showing the $\sigma_k$ of the \textit{Independent} estimator (29 pcm). The vertical axis is truncated.}
    \label{fig:crd-keff-comparison}
\end{figure}

The $60$ pcm \keff\ bias of Table \ref{tab:res-crd} and the $\sigma_k$ values of Table \ref{tab:res-crd-uq} characterize two distinct error sources. They should not be conflated. The bias is deterministic: it comes from the \gls{fm}-\gls{crd} interpolation and combination steps themselves, and would be present even with zero Monte Carlo noise. The $\sigma_k$ values are statistical: they come from the finite number of Monte Carlo histories used to compute the contributing \gls{fm} coefficients. \textit{Runs} is not the right reference for connecting these two quantities. As noted above, it measures the exact configuration's own statistical spread, not the \gls{fm}-\gls{crd} combination's propagated uncertainty. Its role here is instead to confirm that \textit{Independent} does not underestimate the true uncertainty of the interpolated case. It does not: \textit{Independent}'s 29 pcm exceeds \textit{Runs}' 15 pcm. The relevant comparison for the bias is therefore \textit{Independent}'s own $\sigma_k$. The $60$ pcm bias is approximately two standard deviations of \textit{Independent}'s 29 pcm. This is a reasonable relative difference given the propagated uncertainty. Figure \ref{fig:crd-keff-comparison} illustrates this comparison: it shows the exact configuration's \keff\ with the $\sigma_k$ of \textit{Runs}, next to the \gls{fm}-\gls{crd} combined \keff\ with the $\sigma_k$ of \textit{Independent}. It supports the effectiveness of the \gls{fm}-\gls{crd} linear combination approach, which has also been shown to have the potential for further improvement via a finer $N_{z,\text{CR}}$ discretization \cite{mascolinoDevelopmentValidationNew2022}.

\section{Conclusions and Future Work}
\label{sec:conclusions}

This work established a rigorous, full-covariance Delta-method framework for propagating Monte Carlo statistical uncertainty through the fission matrix (\gls{fm}) method's criticality eigenvalue and fission source. This rigorous analytical framework is used to benchmark two substantially cheaper alternatives: the diagonal-covariance (\textit{Independent}) Delta method and direct Monte Carlo resampling of the \gls{fm} coefficients (\textit{Resample}) against both the rigorous covariance treatment and a fully independent empirical benchmark (\textit{Runs}), for a detailed OpenMC model of the \jst reactor.

The full-covariance \textit{Covariances} estimator is not systematically the largest, or the smallest, of the four. For the \gls{aro} reference configuration, dropping the off-diagonal covariance terms (as in the \textit{Independent} approximation) \textit{underestimates} the rigorous $\sigma_k$ by 5.7\% at the lower (CovLow) statistics level, but \textit{overestimates} it by 1.5\% at the higher (CovHigh) level (Table \ref{tab:res-keff-sigma}). This sign flip is consistent with a change in the net sign of the off-diagonal \gls{fm} covariance itself, from a net-positive mean at CovLow to a net-negative mean at CovHigh (Table \ref{tab:res-sign-structure}), even though the population of individual coefficient pairs splits close to evenly between positive and negative correlation at both statistics levels (Section \ref{sec:res-cov-structure}). This point estimate's sign is well supported by a batch-bootstrap check for CovLow but is close to indistinguishable from batch-to-batch noise for CovHigh (Section \ref{sec:res-cov-structure}), so it should not be read as a firmly established trend at either statistics level individually. The qualitative conclusion that the diagonal approximation's bias direction is not fixed a priori, and must be checked per case, holds regardless. Two distinct physical mechanisms compete to set that sign. Coefficients sharing the same source region $j$ (i.e., the same \gls{fm} column) compete for that region's fixed pool of source particles and are therefore negatively correlated -- the mechanism previously proposed, on the same shared-particle-pool reasoning, to argue that \gls{fm} coefficient covariances are negative in general and that the independence assumption is therefore conservative \cite{mascolinoExperimentalComputationalBenchmarking2019,mascolinoDevelopmentBenchmarkingAdvanced2021}. Coefficients belonging to different source columns, on the other hand, are instead coupled by inter-cycle correlation of the fission source shape between successive Monte Carlo generations \cite{mervinUncertaintyUnderpredictionMonte2014}, an effect that can carry either sign locally and can trend positive as the fission source progressively converges.

This different-source-column group makes up $\approx99.5\%$ of all off-diagonal pairs at $\nc=215$ (Section \ref{sec:res-cov-structure}) and has no analogue in that earlier argument, which considered only the particle-competition mechanism; accounting for it shows the true picture is not uniformly negative covariance, but a population of individual pairs split close to evenly between positive and negative correlation (Table \ref{tab:res-sign-structure}). Critically, that near-50/50 split -- not a small true covariance, as its raw row-wise magnitude is in fact large relative to a coefficient's own variance (Section \ref{sec:res-cov-structure}) -- is what very largely cancels once weighted by the \keff\ Jacobian and summed across the full coefficient population, by a factor of $\approx100$--$260\times$ for the criticality cases characterized here, and by a smaller but still substantial $\approx40\times$ for FixedSrc (Table \ref{tab:res-fixed-source-cov-structure}), consistent with that case's larger propagated-uncertainty error. The propagated-uncertainty error from assuming independence (Table \ref{tab:res-cov-structure}) is therefore real but modest for a qualitatively different reason than previously proposed: not because the true \gls{fm} covariance is small or systematically negative, but because its large, genuinely non-independent structure is close enough to symmetric to mostly cancel out. Since inter-cycle correlation is a direct consequence of resampling the fission source generation-to-generation \textit{within a single criticality calculation}, it is specific to that generation strategy. The assumption from previous work \cite{mascolinoNovelHybridDeterministic2024,heDevelopmentUncertaintyAnalysis2022, balouzaComparingFissionMatrix2026} using fixed-source \gls{fm} generation instead, in which each source region's coefficients are obtained from either local or global fixed source runs rather than from one criticality calculation, is that only the negative, particle-competition mechanism survives, making its covariance structure, and hence the sign and magnitude of the diagonal approximation's bias, systematically different from the criticality-based results reported here. Directly evaluating such a case (Section \ref{sec:res-fixed-source}) shows that this expectation holds only in the analytic, infinite statistics sense (Equation \ref{eq:multinomial-cov}). The resulting covariance is intrinsically so small that it cannot be distinguished from zero at the statistics achievable with per-batch tallying, and the diagonal approximation's empirically observed bias for that case reflects that estimation noise rather than a genuine, oppositely signed covariance mechanism. In short, whether the full covariance treatment yields a larger or smaller uncertainty than the diagonal shortcut depends on the \gls{fm} generation strategy and the statistics regime, but for both strategies characterized in this work, every uncertainty estimator agrees closely enough in practice (Tables \ref{tab:res-keff-sigma} and \ref{tab:res-fixed-source-uq}) that this dependence does not change the practical recommendation of Section \ref{sec:res-fixed-source}, i.e., the use of a diagonal-only Delta method.

Despite that sign ambiguity, every estimator examined in this work (\textit{Covariances}, \textit{Independent}, \textit{Resample}, and \textit{Runs}) agrees closely in magnitude for the directly-evaluated \gls{aro} case: for \keff, the largest spread between any two uncertainty estimators is 5.7\% (Table \ref{tab:res-keff-sigma}); for the fission source RSU, it is under 1\% (Table \ref{tab:res-source-sigma}). The \gls{fm}-\gls{crd} case, evaluated at a finer axial discretization ($N_\text{ax}=15$) and under the additional linear interpolation and combination of the \gls{fm}-\gls{crd} method itself (Section \ref{sec:fm-crd}), shows a different pattern relative to \textit{Runs}, used there as the assumption-free reference in place of the unavailable \textit{Covariances}: while \textit{Resample} still agrees with \textit{Independent} to within 6.9\% for $\sigma_k$ and 5.8\% for the source RSU, both are roughly a factor of two more conservative than \textit{Runs} (Section \ref{sec:res-crd-uq}, Table \ref{tab:res-crd-uq}). This gap traces to Equation \ref{eq:crd-cov-diag}'s combination-weighted variance formula itself rather than to a failure of either estimator: combining $N_\text{CR}=4$ simultaneously-perturbed rods gives the \gls{aro} baseline's own coefficient variance a leading weight of $(1-N_\text{CR})^2=9$, rather than the $\pm1$ a directly-evaluated \gls{fm} would carry, and that single, ninefold-amplified term alone accounts for 79\% of \textit{Independent}'s total propagated variance for this test case (Section \ref{sec:res-crd-uq}). This is a structural property of the combination formula that \textit{Resample} reproduces independently, without any linearization. \textit{Runs}, a single direct calculation of the target configuration, carries no such amplification, so this gap is an expected, and favorable, consequence of correctly propagating the combination weights: since the real system \textit{Runs} samples does not itself combine or interpolate, having \textit{Independent} and \textit{Resample} overestimate it rather than underestimate it means the cheap, practical UQ estimate does not under-report the true uncertainty of the configuration it approximates. The \gls{fm}-\gls{crd} interpolation bias itself (60 pcm, Table \ref{tab:res-crd}) is a separate, deterministic error source from this propagated statistical uncertainty. It is about two standard deviations of \textit{Independent}'s own 29 pcm, a reasonable relative difference given that uncertainty (Section \ref{sec:res-crd-uq}). For a system with the \gls{fm} coefficient correlation structure characterized in Section \ref{sec:res-cov-structure} (near-unity individual pairwise correlations that nonetheless average out to a small net effect on the propagated variance), the diagonal-covariance approximation remains a safe choice even once carried through interpolation and combination, though a noticeably less tight one than for a directly-evaluated configuration. Since the \gls{aro} term's amplification grows with $N_\text{CR}$, this is expected to be less tight still for configurations combining more than four rods simultaneously. This does not mean the full covariance calculation is without value: performing it, at least once per system and \gls{fm} generation strategy as done in this work, is what establishes that the diagonal shortcut is safe to use and by how much it may be biased in either direction. But routinely repeating the full $\mathcal{O}(\nc^4)$ covariance calculation for every new configuration of an already-characterized system is not justified by the resulting gain in accuracy, especially for real-time applications in which the wall-clock time and memory requirements need to be minimized.

On computational cost grounds alone, the \textit{Independent} (diagonal-covariance) Delta method is the recommended default for propagating \gls{fm} statistical uncertainty in production use. It reduces both the memory and computational scaling of the covariance calculation from $\mathcal{O}(\nc^4)$ to $\mathcal{O}(\nc^2)$ (Section \ref{sec:simplified_methods}), leading to a reduction from a prohibitive 1.38 TB to a trivial 3.2 MB at $N_\text{ax}=15$ for this reactor alone (Section \ref{sec:fm-modeling-uq}), a gap that only grows more decisive for larger cores or finer discretizations, where the full covariance matrix would exceed even that limit and force the same kind of coarsening compromise made in this work solely to make it computable at all. This comparison is against the explicit covariance matrix. The full Delta method can also be evaluated from per-batch realizations in $\mathcal{O}(M\nc^2)$ time and $\mathcal{O}(\nc^2)$ memory (Equation \ref{eq:streaming}), but that requires per-batch tallies for every database entry. The diagonal approximation only needs the per-coefficient uncertainties that the Monte Carlo code already reports. \textit{Resample} offers no comparable advantage: it requires no covariance storage, but its accuracy is still subject to the resampling noise inherent to a finite number of Monte Carlo trials, and it requires re-solving the \gls{fm} eigenproblem $M$ times ($M=10{,}000$ in this work) rather than the single, cheap sandwich-product evaluation the Delta method needs once $\bm{\Sigma}_A$ (or its diagonal) is known.

The conclusions above pertain strictly to the steady-state, criticality-eigenvalue \gls{fm} formulation used throughout this work. A time-dependent \gls{fm} formulation, of the kind envisioned as the physics engine of a 3-D transport-based digital twin \cite{laureauTransientFissionMatrix2015,mascolinoNovelHybridDeterministic2024,mascolinoExperimentalValidation3D2024,mascolinoValidationTransientFission2019}, couples \gls{fm} coefficients evaluated at successive time steps or reactor configurations. The same near-total cancellation of off-diagonal covariance contributions to the propagated variance, established here for a single steady-state configuration (Section \ref{sec:res-cov-structure}), cannot simply be assumed to hold in that setting, particularly during fast transients where the fission source shape itself is evolving rather than converged. In that setting, \textit{Resample}'s core strategy (i.e., re-solving the \gls{fm} problem directly from perturbed coefficient realizations rather than linearizing about a single reference point) may generalize more naturally than the Delta method to configurations where the underlying linear perturbation assumption (Section \ref{sec:delta}) is less reliable, an approach already shown in this work to track the \textit{Independent} and \textit{Covariances} estimators closely even in the steady-state case (Table \ref{tab:res-keff-sigma}). Extending the covariance characterization of Section \ref{sec:covariances} to a time-dependent \gls{fm} database, and re-evaluating whether the diagonal approximation remains adequate under those conditions, is a natural next step building on both the present results and the fixed-source companion analysis of \cite{balouzaComparingFissionMatrix2026}. For the global fixed-source case characterized directly in this work (Section \ref{sec:res-fixed-source}), a possible next step could be to modify how OpenMC accumulates \gls{fm} tallies to collect per-particle (rather than per-batch) statistics, to confirm whether the small covariance between coefficients of the same column, $j$, is strictly negative as theoretically expected. This implementation, beyond the scope of this work, may not be particularly valuable given the extremely small positive or negative effect of the \gls{fm} coefficients' covariances on the \keff\ and fission source calculated using the \gls{fm} method.

\begin{figure}[p]
  \centering
  \includegraphics[width=\textwidth]{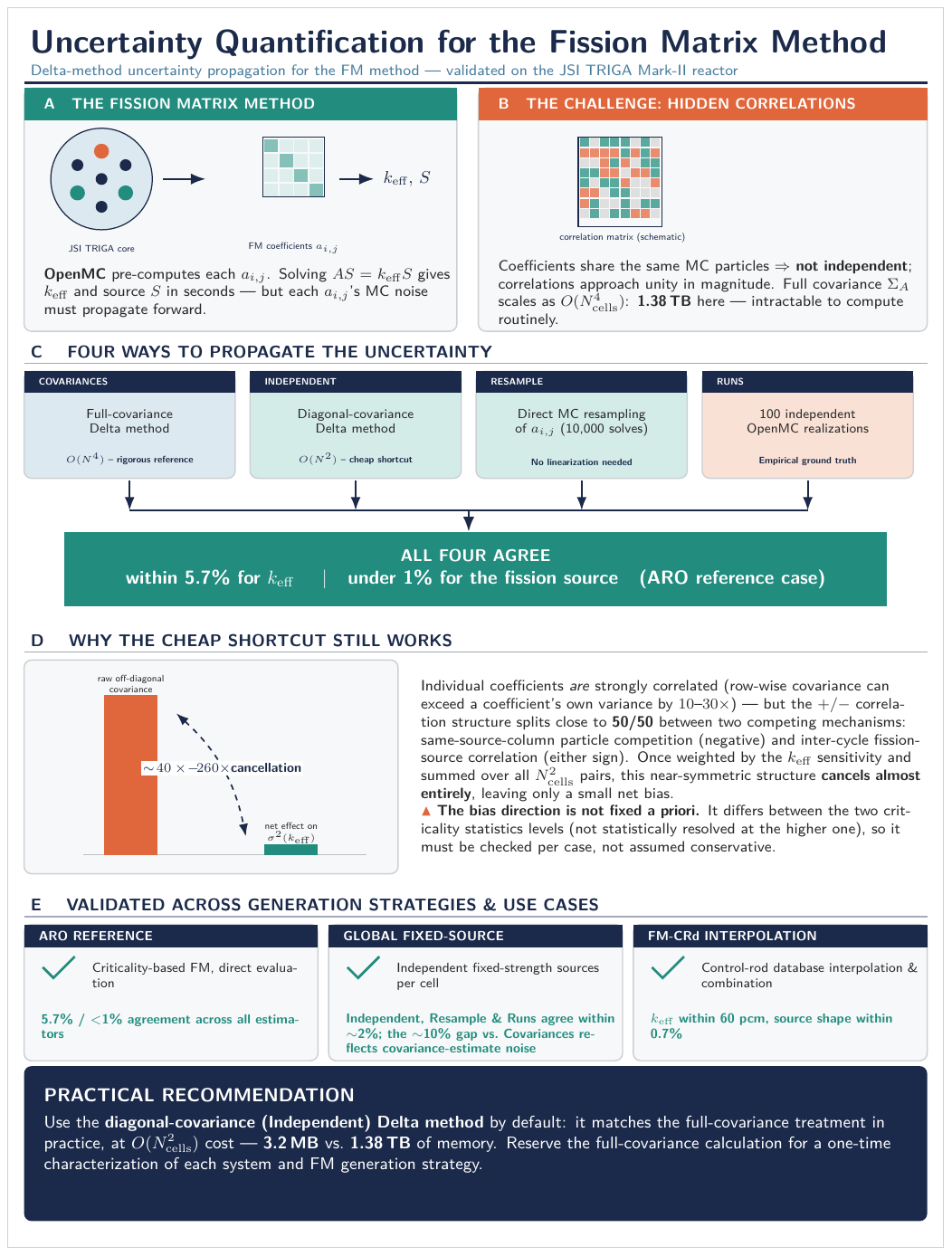}
  \caption{Graphical summary of the full-covariance Delta-method framework, its low-cost alternatives, and their validation against the JSI TRIGA Mark-II model.}
  \label{fig:graphical-summary}
\end{figure}

Figure~\ref{fig:graphical-summary} summarizes the full-covariance framework, its low-cost alternatives, and their validation against the JSI TRIGA Mark-II model discussed in this work.

\clearpage

\section*{Acknowledgments}
The author would like to acknowledge the precious help and parallel collaboration with researchers at the Jo\v{z}ef Stefan Institute, particularly Mr. Matic Kra\v{s}ovic, Dr. An\v{z}e Punger\v{c}i\v{c}, and Dr. Luka Snoj, for the modeling and validation of the JSI TRIGA Mark-II OpenMC model that was conducted prior to this work and without which this work could not have existed.

\section*{Declaration of generative AI and AI-assisted technologies in the manuscript preparation process}
During the preparation of this work, the author used the following genAI tools: Anthropic's Claude Code, models Sonnet 5 and Opus 5 (\url{https://claude.ai/code}); Google Gemini, model Flash 3.6 (\url{https://gemini.google.com/}); and FigureLabs (\url{https://figurelabs.ai/}). These genAI tools were used to generate flowcharts, schematics, and infographics; to aid in the production and debugging of analysis scripts; to draft portions of the narrative text from the author's own numerical results, outlines, and instructions; and to review the grammar and mathematical notation of the manuscript. After using these tools/services, the author reviewed, edited, and verified all AI-generated content against the underlying calculations and takes full responsibility for the content of the published article.

\section*{Declaration of Funding}
No funding was received for this work.

\section*{Disclosure Statement}
The author has no relevant financial or non-financial competing interests to report.

%% file: references.bib
@inproceedings{balouzaComparingFissionMatrix2026,
	address = {Phoenix, AZ, USA},
	title = {Comparing {Fission} {Matrix} {Approaches} using a {JSI} {TRIGA} {Mark}-{II} {Reactor} {Model}},
	booktitle = {Transactions of the {American} {Nuclear} {Society}},
	author = {Balouza, Samah and Krašovic, Matic and Pungerčič, Anže and Mascolino, Valerio},
	month = nov,
	year = {2026},
	note = {; accepted},
}

@misc{vitaFissionMatrixMethod2026a,
	address = {Rochester, NY},
	type = {{SSRN} {Scholarly} {Paper}},
	title = {The {Fission} {Matrix} {Method} for {Criticality} {Search} in {Heat}-{Pipe} {Micro} {Reactors}},
	url = {https://papers.ssrn.com/abstract=6843583},
	doi = {10.2139/ssrn.6843583},
	language = {en},
	urldate = {2026-09-14},
	publisher = {Social Science Research Network},
	author = {Vita, Christian and Abrate, Nicolo' and Dulla, Sandra and Mascolino, Valerio and Walters, William and Stauff, Nicolas E. and Franceschini, Fausto},
	month = may,
	year = {2026},
}

@inproceedings{paineNeuralNetworkEnabledThermal29,
	address = {Denver, CO, USA},
	title = {Neural {Network}-{Enabled} {Thermal} {Feedback} {Modeling} in the {Fission} {Matrix} {Method}},
	volume = {134},
	issn = {0003-18X},
	doi = {10.13182/T134-11738},
	booktitle = {Transactions of the {American} {Nuclear} {Society}},
	author = {Paine, Daris and Mascolino, Valerio},
	month = jun,
	year = {2026},
	pages = {652--655},
}

@article{strohDevelopmentPerformanceEvaluation2026,
	title = {Development and {Performance} {Evaluation} of {RAPID}-{AI} for {High}-{Fidelity} {Whole}-{Core} {Reactor} {Simulation}: {Analysis} of {Spatial} {Undersampling} in {Monte} {Carlo} {Eigenvalue} {Calculations}},
	issn = {0029-5639},
	shorttitle = {Development and {Performance} {Evaluation} of {RAPID}-{AI} for {High}-{Fidelity} {Whole}-{Core} {Reactor} {Simulation}},
	url = {https://doi.org/10.1080/00295639.2026.2710010},
	doi = {10.1080/00295639.2026.2710010},
	urldate = {2026-08-07},
	journal = {Nuclear Science and Engineering},
	publisher = {Taylor \& Francis},
	author = {Stroh, Brian and Haghighat, Alireza},
	month = aug,
	note = {Published online},
	year = {2026},
}

@article{penroseGeneralizedInverseMatrices1955,
	title = {A generalized inverse for matrices},
	volume = {51},
	number = {3},
	journal = {Mathematical Proceedings of the Cambridge Philosophical Society},
	publisher = {Cambridge University Press},
	author = {Penrose, Roger},
	year = {1955},
	pages = {406--413},
	doi = {10.1017/S0305004100030401},
}

@techreport{jcgmEvaluationMeasurementData2008,
	title = {Evaluation of measurement data --- {Guide} to the expression of uncertainty in measurement},
	number = {JCGM 100:2008},
	institution = {Bureau International des Poids et Mesures},
	author = {{Joint Committee for Guides in Metrology}},
	year = {2008},
}

@article{virtanenSciPy10Fundamental2020,
	title = {{SciPy} 1.0: fundamental algorithms for scientific computing in {Python}},
	volume = {17},
	issn = {1548-7105},
	shorttitle = {{SciPy} 1.0},
	url = {https://www.nature.com/articles/s41592-019-0686-2},
	doi = {10.1038/s41592-019-0686-2},
	language = {en},
	number = {3},
	urldate = {2026-07-05},
	journal = {Nature Methods},
	publisher = {Nature Publishing Group},
	author = {Virtanen, Pauli and Gommers, Ralf and Oliphant, Travis E. and Haberland, Matt and Reddy, Tyler and Cournapeau, David and Burovski, Evgeni and Peterson, Pearu and Weckesser, Warren and Bright, Jonathan and van der Walt, Stefan J. and Brett, Matthew and Wilson, Joshua and Millman, K. Jarrod and Mayorov, Nikolay and Nelson, Andrew R. J. and Jones, Eric and Kern, Robert and Larson, Eric and Carey, C. J. and Polat, Ilhan and Feng, Yu and Moore, Eric W. and VanderPlas, Jake and Laxalde, Denis and Perktold, Josef and Cimrman, Robert and Henriksen, Ian and Quintero, E. A. and Harris, Charles R. and Archibald, Anne M. and Ribeiro, Antonio H. and Pedregosa, Fabian and van Mulbregt, Paul},
	month = mar,
	year = {2020},
	pages = {261--272},
}

@article{harrisArrayProgrammingNumPy2020,
	title = {Array programming with {NumPy}},
	volume = {585},
	copyright = {2020 The Author(s)},
	issn = {1476-4687},
	url = {https://www.nature.com/articles/s41586-020-2649-2},
	doi = {10.1038/s41586-020-2649-2},
	language = {en},
	number = {7825},
	urldate = {2026-07-05},
	journal = {Nature},
	publisher = {Nature Publishing Group},
	author = {Harris, Charles R. and Millman, K. Jarrod and van der Walt, Stéfan J. and Gommers, Ralf and Virtanen, Pauli and Cournapeau, David and Wieser, Eric and Taylor, Julian and Berg, Sebastian and Smith, Nathaniel J. and Kern, Robert and Picus, Matti and Hoyer, Stephan and van Kerkwijk, Marten H. and Brett, Matthew and Haldane, Allan and del Río, Jaime Fernández and Wiebe, Mark and Peterson, Pearu and Gérard-Marchant, Pierre and Sheppard, Kevin and Reddy, Tyler and Weckesser, Warren and Abbasi, Hameer and Gohlke, Christoph and Oliphant, Travis E.},
	month = sep,
	year = {2020},
	pages = {357--362},
}

@book{magnusMatrixDifferentialCalculus2019,
	title = {Matrix {Differential} {Calculus} with {Applications} in {Statistics} and {Econometrics}},
	isbn = {978-1-119-54119-6},
	language = {en},
	publisher = {John Wiley \& Sons},
	author = {Magnus, Jan R. and Neudecker, Heinz},
	month = mar,
	year = {2019},
	note = {Google-Books-ID: 9jmNDwAAQBAJ},
}

@inproceedings{vitaCorrectionRatioApproach2026,
	address = {Turin, Italy},
	title = {A {Correction} {Ratio} {Approach} to {Fission} {Matrix} {Interpolation} in {Systems} with {Large} {Temperature} {Differences}},
	isbn = {979-12-81583-46-7},
	doi = {https://doi.org/10.5281/zenodo.20803590},
	booktitle = {Proceedings of {PHYSOR} 2026 {International} {Conference}},
	author = {Vita, Christian and Walters, William and Mascolino, Valerio and Roskoff, Nathan and Franceschini, Fausto and Abrate, Nicolo' and Dulla, Sandra},
	month = apr,
	year = {2026},
}

@article{mascolinoEvaluationRAPIDUNF2017,
	title = {Evaluation of {RAPID} for a {UNF} cask benchmark problem},
	volume = {153},
	doi = {10.1051/epjconf/201715305025},
	number = {ICRS-13 \& RPSD-2016},
	journal = {EPJ Web Conf.},
	author = {Mascolino, Valerio and Haghighat, Alireza and Roskoff, Nathan J},
	year = {2017},
	pages = {05025},
}

@article{carneyTheoryApplicationsFission2014,
	title = {Theory and applications of the fission matrix method for continuous-energy {Monte} {Carlo}},
	volume = {73},
	issn = {0306-4549},
	url = {https://www.sciencedirect.com/science/article/pii/S030645491400351X},
	doi = {10.1016/j.anucene.2014.07.020},
	urldate = {2025-11-20},
	journal = {Annals of Nuclear Energy},
	author = {Carney, Sean and Brown, Forrest and Kiedrowski, Brian and Martin, William},
	month = nov,
	year = {2014},
	pages = {423--431},
}

@article{snojHalfcenturyNuclearResearch2025,
	title = {A half-century of nuclear research, education and training: {Story} of the {JSI} {TRIGA} reactor},
	volume = {214},
	url = {https://www.sciencedirect.com/science/article/pii/S0306454924007850},
	doi = {10.1016/j.anucene.2024.111122},
	journal = {Annals of Nuclear Energy},
	author = {Snoj, Luka and Ambrožič, Klemen and Barbot, Loïc and Benedik, Ljudmila and Bratkič, Arne and Capan, Ivana and Reynard-Carette, Christelle and Cindro, Vladimir and Čalič, Dušan and Destouches, Christophe and Geslot, Benoit and Haghighat, Alireza and Henry, Romain and Horvat, Milena and Huseynov, Elchin M and de Izarra, Grégoire and Jaćimović, Radojko and Jazbec, Anže and Jenčič, Igor and Jeraj, Robert and Joyce, Malcom and Kotnik, Domen and Kramberger, Gregor and Lengar, Igor and Malec, Jan and Mandić, Igor and Mascolino, Valerio and Merljak, Vid and Mikuž, Marko and Noguère, Gilles and Peric, Julijan and Pungerčič, Anže and Radulović, Vladimir and Rupnik, Sebastjan and Smodiš, Borut and Šlejkovec, Zdenka and Štrok, Marko and Štancar, Ziga and Švajger, Ingrid and Thiollay, Nicolas and Tiselj, Iztok and Trkov, Andrej and Žefran, Bojan and Žerovnik, Gašper and Žiber, Ylenia Kogovšek and Goričanec, Tanja},
	year = {2025},
	pages = {111122--111122},
}

@article{leppanenStatusSerpentMonte2025,
	title = {Status of {Serpent} {Monte} {Carlo} code in 2024},
	volume = {11},
	copyright = {© J. Leppänen et al., Published by EDP Sciences, 2025},
	issn = {2491-9292},
	url = {https://www.epj-n.org/articles/epjn/abs/2025/01/epjn20240041/epjn20240041.html},
	doi = {10.1051/epjn/2024031},
	language = {en},
	urldate = {2026-06-23},
	journal = {EPJ Nuclear Sciences \& Technologies},
	publisher = {EDP Sciences},
	author = {Leppänen, Jaakko and Valtavirta, Ville and Rintala, Antti and Tuominen, Riku},
	year = {2025},
	pages = {3},
}

@techreport{kuleszaMCNPCodeVersion2024,
	title = {{MCNP}® {Code} {Version} 6.3.1 {Theory} \& {User} {Manual}},
	url = {https://www.osti.gov/biblio/2372634},
	doi = {10.2172/2372634},
	language = {English},
	number = {LA-UR--24-24602},
	urldate = {2026-06-23},
	institution = {Los Alamos National Laboratory (LANL), Los Alamos, NM (United States)},
	author = {Kulesza, Joel A. and Adams, Terry R. and Armstrong, Jerawan Chudoung and Bolding, Simon R. and Brown, Forrest B. and Bull, Jeffrey S. and Burke, Timothy Patrick and Clark, Alexander Rich and Forster III, Robert Arthur and Giron, Jesse Frank and Grieve, Tristan Sumner and Josey, Colin James and Martz, Roger Lee and McKinney, Gregg W. and Pearson, Eric J. and Rising, Michael Evan and Solomon Jr., Clell Jeffrey and Sriram, Swaminarayan and Trahan, Travis John and Weaver, Colin Andrew and Wilson, Stephen C. and Zukaitis, Anthony J.},
	month = may,
	year = {2024},
	doi = {10.2172/2372634},
}

@article{dufekFissionMatrixBased2009,
	title = {Fission matrix based {Monte} {Carlo} criticality calculations},
	volume = {36},
	issn = {0306-4549},
	url = {https://www.sciencedirect.com/science/article/pii/S0306454909001558},
	doi = {10.1016/j.anucene.2009.05.003},
	number = {8},
	urldate = {2026-06-23},
	journal = {Annals of Nuclear Energy},
	author = {Dufek, Jan and Gudowski, Wacław},
	month = aug,
	year = {2009},
	pages = {1270--1275},
}

@book{casellaStatisticalInference2024,
	title = {Statistical inference},
	isbn = {1-003-45628-6},
	publisher = {Chapman and Hall/CRC},
	author = {Casella, George and Berger, Roger},
	year = {2024},
}

@article{rauStrategiesFastFission2021,
	title = {Strategies for {Fast} {Fission} {Matrix} {Estimation} with {Fuel} {Temperature} and {Control} {Rod} {Feedback}},
	volume = {195},
	number = {10},
	journal = {Nuclear Science and Engineering},
	publisher = {Taylor \& Francis},
	author = {Rau, Adam J and Walters, William J},
	year = {2021},
	pages = {1017--1035},
}

@article{dalingerAnalysisFissionMatrix2025,
	title = {Analysis of {Fission} {Matrix} {Databases} with {Nonuniform} {Fuel} {Temperature} {Profiles} for {Static} and {Transient} {Calculations}},
	volume = {199},
	issn = {0029-5639},
	url = {https://doi.org/10.1080/00295639.2024.2328944},
	doi = {10.1080/00295639.2024.2328944},
	urldate = {2025-11-20},
	journal = {Nuclear Science and Engineering},
	publisher = {Taylor \& Francis},
	author = {Dalinger, Maximiliano and Walters, William},
	month = apr,
	year = {2025},
	note = {\_eprint: https://doi.org/10.1080/00295639.2024.2328944},
	pages = {S754--S764},
}

@article{waltersRAPIDFissionMatrix2018,
	title = {The {RAPID} {Fission} {Matrix} {Approach} to {Reactor} {Core} {Criticality} {Calculations}},
	volume = {192},
	issn = {0029-5639},
	url = {https://doi.org/10.1080/00295639.2018.1497395},
	doi = {10.1080/00295639.2018.1497395},
	number = {1},
	urldate = {2025-11-20},
	journal = {Nuclear Science and Engineering},
	publisher = {Taylor \& Francis},
	author = {Walters, William J. and Roskoff, Nathan J. and Haghighat, Alireza},
	month = oct,
	year = {2018},
	note = {\_eprint: https://doi.org/10.1080/00295639.2018.1497395},
	pages = {21--39},
}

@inproceedings{mascolinoValidationRAPIDsAlgorithm2021,
	address = {Raleigh, NC},
	title = {Validation of {RAPID}'s {Algorithm} for {Control} {Rods} {Movement} {Using} the {JSI} {TRIGA} {Reactor}},
	url = {https://www.ans.org/pubs/proceedings/article-50029/},
	booktitle = {Proceedings of {The} {International} {Conference} on {Mathematics} and {Computational} {Methods} {Applied} to {Nuclear} {Science} and {Engineering} ({M}\&{C2021})},
	author = {Mascolino, Valerio and Haghighat, Alireza and Pungerčič, Anže and Snoj, Luka},
	month = oct,
	year = {2021},
	pages = {572--582},
}

@phdthesis{mascolinoDevelopmentBenchmarkingAdvanced2021,
	address = {Washington DC, USA,},
	title = {Development and benchmarking of advanced {FM}-based particle transport algorithms for steady-state and transient conditions, implementation in {RAPID} and its {VRS} web-application},
	url = {https://vtechworks.lib.vt.edu/items/53cde524-7754-41c9-832c-7bdaee1fda19},
	school = {Virginia Polytechnic Institute and State University},
	author = {Mascolino, Valerio},
	year = {2021},
}

@article{mascolinoNovelHybridDeterministic2024,
	title = {A {Novel} {Hybrid} {Deterministic} and {Monte} {Carlo} {Neutron} {Transport} {Formulation} and {Algorithm} ({tRAPID}) for {Accurate} and {Fast} 3-{D} {Reactor} {Kinetics}},
	volume = {198},
	doi = {10.1080/00295639.2023.2197844},
	number = {3},
	journal = {Nuclear Science and Engineering},
	publisher = {Taylor \& Francis},
	author = {Mascolino, Valerio and Haghighat, Alireza},
	month = mar,
	year = {2024},
	pages = {592--627},
}

@inproceedings{mascolinoBenchmarkingRapidCode2018,
	address = {Cancun, Mexico, Apr. 22-26},
	title = {Benchmarking of the {Rapid} {Code} {System} {Using} the {GBC}-32 {Cask} {With} {Variable} {Burnups}},
	booktitle = {Proceedings of the {International} {Conference} on {Physics} of {Reactors} ({PHYSOR} 2018)},
	author = {Mascolino, Valerio and Roskoff, Nathan J and Haghighat, Alireza},
	year = {2018},
	pages = {697--708},
}

@article{haghighatMRTMethodologiesRealtime2016,
	title = {{MRT} methodologies for real-time simulation of nonproliferation and safeguards problems},
	volume = {87},
	url = {http://linkinghub.elsevier.com/retrieve/pii/S0306454915003151},
	doi = {10.1016/j.anucene.2015.06.004},
	journal = {Annals of Nuclear Energy},
	publisher = {Elsevier Ltd},
	author = {Haghighat, Alireza and Royston, Katherine and Walters, William},
	year = {2016},
	pages = {61--67},
}

@incollection{roskoffExperimentalComputationalValidation2018,
	address = {Santa Fe, NM, May 7-12},
	title = {Experimental and {Computational} {Validation} of {RAPID}},
	doi = {10.1520/STP160820170094},
	booktitle = {Reactor {Dosimetry}: 16th {International} {Symposium}},
	publisher = {ASTM International},
	author = {Roskoff, Nathan J and Haghighat, Alireza and Mascolino, Valerio},
	year = {2018},
	pages = {544--554},
}

@book{bellNuclearReactorTheory1970,
	title = {Nuclear reactor theory},
	publisher = {US Atomic Energy Commission, Washington, DC (United States)},
	author = {Bell, George I and Glasstone, Samuel},
	year = {1970},
}

@article{jerajTRIGAMarkII1999,
	title = {{TRIGA} {Mark} {II} {Reactor} {U} (20)–{Zirconium} {Hydride} {Fuel} {Rods} {In} {Water} {With} {Graphite} {Reflector}, {International} {Handbook} of {Evaluated} {Criticality} {Safety} {Benchmark} {Experiments}},
	journal = {NEA/NSC/DOC/(95) 03/III, Tech. Rep.},
	author = {Jeraj, R and Ravnik, M},
	year = {1999},
}

@article{laureauTransientFissionMatrix2015,
	title = {Transient {Fission} {Matrix}: {Kinetic} calculation and kinetic parameters {$\beta_\mathrm{eff}$} and {$\Lambda_\mathrm{eff}$} calculation},
	volume = {85},
	journal = {Annals of Nuclear Energy},
	publisher = {Elsevier},
	author = {Laureau, A and Aufiero, M and Rubiolo, P R and Merle-Lucotte, E and Heuer, D},
	year = {2015},
	pages = {1035--1044},
}

@article{mervinUncertaintyUnderpredictionMonte2014,
	title = {Uncertainty {Underprediction} in {Monte} {Carlo} {Eigenvalue} {Calculations}},
	volume = {173},
	doi = {10.13182/nse11-104},
	number = {3},
	journal = {Nuclear Science and Engineering},
	author = {Mervin, Brenden T. and Mosher, Scott W. and Wagner, John C. and Maldonado, G. I.},
	year = {2014},
	pages = {276--292},
}

@inproceedings{mascolinoExperimentalComputationalBenchmarking2019,
	address = {Portland, OR},
	title = {Experimental and {Computational} {Benchmarking} of {RAPID} using the {JSI} {TRIGA} {MARK}-{II} {Reactor}},
	url = {https://www.ans.org/pubs/proceedings/article-46717/},
	booktitle = {Proceedings of the {International} {Conference} on {Mathematics} and {Computational} {Methods} applied to {Nuclear} {Science} and {Engineering} ({M}\&{C} 2019)},
	author = {Mascolino, Valerio and Pungerčič, Anže and Haghighat, Alireza and Snoj, Luka},
	year = {2019},
	pages = {1328--1337},
}

@article{romanoOpenMCMonteCarlo2013,
	title = {The {OpenMC} monte carlo particle transport code},
	volume = {51},
	journal = {Annals of Nuclear Energy},
	publisher = {Elsevier},
	author = {Romano, Paul K and Forget, Benoit},
	year = {2013},
	pages = {274--281},
}

@inproceedings{mascolinoValidationTransientFission2019,
	address = {Portland, OR},
	title = {Validation of the {Transient} {Fission} {Matrix} {Code} {tRAPID} against the {Flattop}-{Pu} {Benchmark}},
	url = {https://www.ans.org/pubs/proceedings/article-46718/},
	booktitle = {Proceedings of the {International} {Conference} on {Mathematics} and {Computational} {Methods} applied to {Nuclear} {Science} and {Engineering} ({M}\&{C} 2019)},
	author = {Mascolino, Valerio and Haghighat, Alireza},
	year = {2019},
	pages = {1338--1347},
}

@article{tophamIterativeFissionMatrix2020,
	title = {An iterative fission matrix scheme for calculating steady-state power and critical control rod position in a {TRIGA} reactor},
	volume = {135},
	journal = {Annals of Nuclear Energy},
	publisher = {Elsevier},
	author = {Topham, Tyler J and Rau, Adam and Walters, William J},
	year = {2020},
	pages = {106984--106984},
}

@book{haghighatMonteCarloMethods2020,
	edition = {2nd},
	title = {Monte {Carlo} {Methods} for {Particle} {Transport}},
	isbn = {0-429-58410-5},
	publisher = {CRC Press Taylor \& Francis Group},
	author = {Haghighat, Alireza},
	year = {2020},
	note = {Pages: 213},
}

@article{heCorrectionMethodRAPID2020,
	title = {A correction method for {RAPID} fission matrix calculations with control rod movement},
	volume = {121},
	journal = {Progress in Nuclear Energy},
	publisher = {Elsevier},
	author = {He, Donghao and Walters, William J},
	year = {2020},
	pages = {103226--103226},
}

@article{pungercicVerificationNovelFuel2023,
	title = {Verification of a novel fuel burnup algorithm in the {RAPID} code system based on {Serpent}-2 simulation of the {TRIGA} {Mark} {II} research reactor},
	doi = {10.1016/j.net.2023.06.040},
	journal = {Nuclear Engineering and Technology},
	author = {Pungerčič, Anže and Mascolino, Valerio and Haghighat, Alireza and Snoj, Luka},
	year = {2023},
}

@article{mascolinoDevelopmentValidationNew2022,
	title = {Development and validation of new algorithms for control rods insertion modeling in the {RAPID} code system using the {JSI} {TRIGA} {Mark}-{II} reactor},
	volume = {166},
	doi = {10.1016/J.ANUCENE.2021.108711},
	journal = {Annals of Nuclear Energy},
	publisher = {Pergamon},
	author = {Mascolino, Valerio and Haghighat, Alireza and Snoj, Luka},
	month = feb,
	year = {2022},
	pages = {108711--108711},
}

@article{mascolinoVerificationValidationRAPID2021,
	title = {Verification and {Validation} of {RAPID} {Formulations} and {Algorithms} {Based} on {Dosimetry} {Measurements} at the {JSI} {TRIGA} {Mark}-{II} {Reactor}},
	volume = {195},
	doi = {10.1080/00295639.2021.1890321},
	number = {9},
	journal = {Nuclear Science and Engineering},
	publisher = {Taylor \& Francis},
	author = {Mascolino, Valerio and Haghighat, Alireza and Snoj, Luka},
	month = sep,
	year = {2021},
	pages = {937--953},
}

@article{heDevelopmentUncertaintyAnalysis2022,
	title = {Development of {Uncertainty} {Analysis} {Techniques} for the {Fission} {Matrix}–{Based} {Neutron} {Transport} {Code} {RAPID}},
	volume = {196},
	number = {9},
	journal = {Nuclear Science and Engineering},
	publisher = {Taylor \& Francis},
	author = {He, Donghao and Walters, William},
	year = {2022},
	pages = {1101--1113},
	doi = {10.1080/00295639.2022.2049991},
}

@inproceedings{pungercicApplicationBRAPIDFission2022,
	address = {Pittsburgh, PA, May 15-20, 2022},
	title = {Application of the {bRAPID} {Fission} {Matrix} {Burnup} {Methodology} to the {JSI} {TRIGA} {Mark} {II} {Research} {Reactor}},
	isbn = {978-0-89448-787-3},
	doi = {10.13182/PHYSOR22-37787},
	booktitle = {Proceedings of the {International} {Conference} on {Physics} of {Reactors} ({PHYSOR} 2022)},
	author = {Pungerčič, Anže and Haghighat, Alireza and Snoj, Luka},
	year = {2022},
	pages = {2275--2284},
}

@article{mascolinoExperimentalValidation3D2024,
	title = {Experimental {Validation} of the 3-{D} {Neutron} {Kinetics} {Algorithm} of {RAPID} using the {JSI} {TRIGA} {Mark}-{II} reactor},
	volume = {176},
	doi = {10.1016/j.pnucene.2024.105391},
	number = {November 2024},
	journal = {Progress in Nuclear Energy},
	author = {Mascolino, Valerio and Pungerčič, Anže and Snoj, Luka and Haghighat, Alireza},
	year = {2024},
	pages = {105391--105391},
}

@article{romanoOpenMCStateoftheartMonte2015,
	title = {{OpenMC}: {A} state-of-the-art {Monte} {Carlo} code for research and development},
	volume = {82},
	journal = {Annals of Nuclear Energy},
	publisher = {Elsevier},
	author = {Romano, Paul K and Horelik, Nicholas E and Herman, Bryan R and Nelson, Adam G and Forget, Benoit and Smith, Kord},
	year = {2015},
	pages = {90--97},
}

@book{johnsonDiscreteMultivariateDistributions1997,
	title = {Discrete multivariate distributions},
	isbn = {978-0-471-12844-1},
	publisher = {Wiley},
	author = {Johnson, Norman L. and Kotz, Samuel and Balakrishnan, N.},
	series = {Wiley Series in Probability and Statistics},
	year = {1997},
}
